\documentclass[aps,prd,10pt,a4paper,twocolumn,amsfonts,amsmath,amssymb,superscriptaddress,notitlepage,preprintnumbers]{revtex4-2}

\usepackage{hyperref}
\usepackage{bm}
\usepackage{physics}
\usepackage{graphicx}
\usepackage{comment}
\usepackage[dvipsnames]{xcolor}

\newcommand{\hMpc}{h^{-1}\,\mathrm{Mpc}}

\newcommand{\hMpcinv}{h\,\mathrm{Mpc}^{-1}}

\newcommand{\rdfid}{r_d^\mathrm{fid}}
\newcommand{\DHfid}{D_H^\mathrm{fid}}
\newcommand{\DMfid}{D_M^\mathrm{fid}}
\newcommand{\sigmav}{\sigma_\mathrm{v}}
\newcommand{\sigmad}{\sigma_\mathrm{d}}

\newcommand{\kmax}{k_\mathrm{max}}

\newcommand{\mnras}{Mon. Not. Roy. Astron. Soc.}

\newcommand{\jcap}{J. Cosmol. Astropart. Phys.}
\newcommand{\physrep}{Phys. Rept.}
\newcommand{\aap}{Astron. Astrophys.}

\newcommand{\pasj}{Publ. Astron. Soc. Jpn.}
\newcommand{\pasp}{Publ. Astron. Soc. Pac.}

\begin{document}
\preprint{YITP-23-32}

\title{Perturbation theory challenge for cosmological parameters estimation II.:
Matter power spectrum in redshift space}

\author{Ken Osato}
\email[]{ken.osato@chiba-u.jp}
\affiliation{Center for Frontier Science, Chiba University, Chiba 263-8522, Japan}
\affiliation{Department of Physics, Graduate School of Science, Chiba University, Chiba 263-8522, Japan}
\affiliation{Kavli Institute for the Physics and Mathematics of the Universe,
The University of Tokyo Institutes for Advanced Study (UTIAS), The University of Tokyo,
Chiba 277-8583, Japan}
\affiliation{Center for Gravitational Physics and Quantum Information,
Yukawa Institute for Theoretical Physics,
Kyoto University, Kyoto 606-8502, Japan}

\author{Takahiro Nishimichi}
\affiliation{Department of Astrophysics and Atmospheric Sciences, Faculty of Science,
Kyoto Sangyo University, Kyoto 603-8555, Japan}
\affiliation{Kavli Institute for the Physics and Mathematics of the Universe,
The University of Tokyo Institutes for Advanced Study (UTIAS), The University of Tokyo,
Chiba 277-8583, Japan}
\affiliation{Center for Gravitational Physics and Quantum Information,
Yukawa Institute for Theoretical Physics,
Kyoto University, Kyoto 606-8502, Japan}

\author{Atsushi Taruya}
\affiliation{Center for Gravitational Physics and Quantum Information,
Yukawa Institute for Theoretical Physics,
Kyoto University, Kyoto 606-8502, Japan}
\affiliation{Kavli Institute for the Physics and Mathematics of the Universe,
The University of Tokyo Institutes for Advanced Study (UTIAS), The University of Tokyo,
Chiba 277-8583, Japan}

\author{Francis Bernardeau}
\affiliation{Institut de Physique Th\'eorique, Universit\'e Paris-Saclay, CEA, CNRS, URA 2306,
91191 Gif-sur-Yvette, France}
\affiliation{Institut d'Astrophysique de Paris, Sorbonne Universit\'e, CNRS, UMR 7095,
75014 Paris, France}

\begin{abstract}
Constraining cosmological parameters from large-scale structure observations requires
precise and accurate tools to compute its properties.
While perturbation theory (PT) approaches can serve this purpose, exploration of large parameter space is challenging
due to the potentially large computational cost of such calculations.
In this study, we show that a response function approach applied to the regularized PT (RegPT) model
at 2-loop order,
plus correction terms induced by redshift space distortion effects, can reduce the runtime
by a factor of 50 compared to direct integration.
We illustrate the performance of this approach by
performing the parameter inference of five fundamental cosmological parameters
from the redshift space power spectrum measured from $N$-body simulations as mock measurements,
and inferred cosmological parameters are directly compared with
parameters used to generate initial conditions of the simulations.
From this \textit{PT challenge} analysis, the constraining power of cosmological parameters
and parameter biases are quantified with the survey volume and galaxy number density
expected for the \textit{Euclid} mission at the redshift $z = 1$
as a function of the maximum wave-number of data points $\kmax$.
We find that RegPT with correction terms reproduces the input cosmological parameters
without bias up to maximum wave-number $\kmax = 0.18 \, \hMpcinv$.
Moreover, RegPT+, which introduces one free parameter
to RegPT to handle the damping feature on small scales,
delivers the best performance among the examined models and
achieves tighter constraints without significant parameter bias
for higher maximum wave-number $\kmax = 0.21 \, \hMpcinv$.

\end{abstract}

\date{\today}

\maketitle

\section{Introduction}
\label{sec:introduction}
The widely accepted scenario in modern cosmology is that
quantum fluctuations generated by inflation in the earliest stage of the Universe
results in primordial density perturbation and the fluctuations
lead to the formation of self-gravitating bound objects called dark halos
by gravitational instability.
As the Universe evolves, dark halos successively undergo mergers
and form structures with a wide range of spatial scales \cite{White1991}.
In the hierarchy of structures, the largest structures are referred to as
the large-scale structures (LSS) of the Universe.
The spatial clustering of LSS, in particular baryon acoustic oscillation (BAO) \cite{Aubourg2015},
is driven by dark matter and is sensitive to the accelerated expansion induced by dark energy.
Thus, observations of LSS are the key to understanding the nature of the dark sector
\cite[for a review, see][]{Weinberg2013}.

One widely employed probe to map the LSS is the use of the galaxies
as tracers of the matter distribution at large cosmological scales.
Over the last decade, many such surveys have been built:
6dF Galaxy Survey \cite{Beutler2011}, WiggleZ \cite{Blake2011}, VIPERS \cite{delaTorre2017},
and extended Baryon Acoustic Oscillation Survey \cite[eBOSS;][]{Alam2021}.
In the coming era, a new generation of spectroscopic surveys is under preparation:
the Subaru Prime Focus Spectrograph \cite[PFS;][]{Takada2014},
Dark Energy Spectrograph Instrument \cite[DESI;][]{DESI2016a,DESI2016b},
\textit{Euclid} \cite{Laureijs2011,Amendola2018},
and Nancy Grace Roman Space Telescope \footnote{\url{https://roman.gsfc.nasa.gov/}},
which will probe the universe with wider and deeper coverage.
These next-generation surveys are promised to advance our understanding of the Universe
through precise and accurate measurements of galaxy clustering.

In the practical analysis of galaxy clustering, we rely on statistics to summarize
the information of observed galaxy distribution.
The most fundamental and widely used statistics is the two-point correlation function
or its Fourier space counterpart, the power spectrum.
The accurate and precise theoretical model to predict these statistics given a cosmological model
is an essential component in the statistical analysis to constrain cosmological models.
For this purpose, various approaches have been developed so far.
Among such methods, the perturbation theory (PT) of LSS \cite[for a review, see][]{Bernardeau2002}
has been commonly employed in practical analysis.
In the PT framework, the cosmic matter is approximated as single-stream fluid
and the evolution is governed by continuity equation, Euler equation, and Poisson equation.
These equations can be expanded with respect to the linear density contrast
and this naive approach is called as standard perturbation theory (SPT).
SPT is fast enough to apply for statistical inference
and the sub-percent level accuracy is achieved up to the mildly non-linear regime.
However, SPT is known to exhibit poor convergence of PT expansion and
the UV-sensitive behaviors \cite{Blas2014,Bernardeau2014}.
These severely restrict the applicable range of SPT predictions,
and a better control of the convergence as well as UV sensitivity needs to be exploited.

In order to circumvent these problems and realize better convergence and accuracy,
approaches beyond SPT have been developed based on resummation technique
in Lagrangian space \cite{Matsubara2008,Okamura2011} and in Eulerian space \cite{Crocce2006,Bernardeau2008,Bernardeau2012,Taruya2008,Taruya2009,Pietroni2008}.
the UV sensitivity has been shown to be mitigated,
and the applicable range of PT treatment can be extended to small scales.
The latter approach is especially referred to as the effective field theory (EFT) of LSS \cite{Baumann2012,Carrasco2012,Carrasco2014,Cabass2023}.
The main concept of this treatment is to filter out small-scale uncertainties,
including the galaxy bias, and to absorb their impacts on large-scale modes by adding an effective stress tensor in the single-stream fluid equations.
As a trade-off, in EFT, controlling UV-sensitive behaviors needs free parameters that characterize the amplitude of counter terms, and these must be marginalized over
when comparing the predictions with observations or simulations,
although each coefficient of counter terms has a clear physical meaning such as the effective sound speed.

In all cases, the statistical inference of
cosmological parameters from the power spectrum involves theoretical calculations of power spectra
for a large number of sets of cosmological parameters.
For example, the standard cosmological model, where the Universe is composed of ordinary matter,
cold dark matter (CDM) and the cosmological constant and the geometry is flat, i.e., flat $\Lambda$CDM model,
contains at least five cosmological parameters.
This number is significantly augmented when one takes into account nuisance parameters for observational systematics,
such as the galaxy bias, and the impact of redshift space distortion (RSD) effect \cite{Jackson1972,Kaiser1987}.
For instance, for a parameter space  of three cosmological parameters \cite{Osato2019},
$\mathcal{O} (10^5)$ evaluations of statistics were required to perform the parameter inference.
A full exploration of the parameter space will be obviously more demanding.

So far, a variety of approaches to predict the power spectrum
have been used to explore such large parameter space.
Previous works \cite{Schmittfull2016a,Schmittfull2016b,McEwen2016,Fang2017}
showed that the computational cost of loop correction terms in the SPT expansion
could be reduced with the \texttt{FFTLog} algorithm \cite{Hamilton2000}.
These methods are formulated up to next-to-leading order (NLO)
but have not been implemented at next-to-next-leading order (NNLO).
NNLO direct computations have so far been hampered by their computational cost.
In order to gain accuracy, EFT approaches have been widely advocated as an alternative to NNLO computations,
and applied to full-shape analysis of galaxy power spectrum \cite{dAmico2020,Colas2020,Ivanov2020}.
The advantage is that their computational cost is indeed comparable to that of NLO computations.
The price to pay is the introduction of many extra nuisance parameters to control the impact of the small-scale fluctuations that
potentially degrade their constraining power \cite{Osato2019}.
Finally, it should be noted that simulation-based approaches, e.g.,
emulator approach \cite{Heitmann2010,Heitmann2009,Nishimichi2019,Euclid2021},
are successful in predicting power spectrum down to small scales in a fast manner \cite{Kobayashi2020} and
have been successfully employed to constrain cosmological parameters from galaxy power spectrum \cite{Kobayashi2022}.
This is not however the path we choose to follow here.

In general, analytical PT treatments mentioned above have their own limitation,
and the robustness of these approaches has to be tested against numerical simulations,
in which several non-linear systematics, including gravitational clustering, RSD, and galaxy bias,
are properly accounted for.
However, most of the previous studies have restricted their analysis to
the case where cosmological parameters are fixed. 
For a more practical setup, one would be interested in deriving cosmological constraints,
allowing all the parameters in the theoretical models to be free.
Nevertheless, in the presence of parameter degeneracies,
even if an analytical method succeeds in reproducing the observed power spectra,
an unbiased parameter estimation is not always guaranteed.  

The main goal of this paper is to conduct
a cosmology challenge analysis of the perturbation theory, a \textit{PT Challenge}, extending to redshift space the studies done previously in real space
\cite{Osato2019,Nishimichi2020,Eggemeier2020,Eggemeier2021,Pezzotta2021}.
Since cosmological parameters used to generate the initial conditions of the simulation are known, we can directly compare between the inferred and true values.
In this analysis, we mainly focus on the regularized PT approach RegPT \cite{Taruya2012}
with acceleration by the response function expansion fast-RegPT \cite{Osato2021}.
With the help of the response function, we can significantly reduce the computational cost
of the calculation of power spectrum and bispectrum.
Through this analysis, we can discuss which model serves the best to give precise and accurate estimates. 
Further, we determine the extent to which each of PT models/methods
can be used as a reliable theoretical template to reproduce safely the cosmological parameters without any systematic bias.
Note that the target power spectrum in this analysis is the \textit{matter} power spectrum.
Indeed, the matter power spectrum differs from the galaxy power spectrum, which is measured in real galaxy clustering measurements.
Since there are various uncertainties in modeling of the galaxy power spectrum
and construction of galaxy mock samples,
we focus only on the matter power spectrum to evade such uncertainties.
Our scope in this work is to assess the performance of PT models at the matter level,
and the analysis with the galaxy power spectrum will be carried out
in the subsequent paper of the PT Challenge series.

In Section~\ref{sec:theory}, we briefly overview the basics of analytical treatments
on redshift space power spectrum: SPT, RegPT, IR-resummed EFT.
In Section~\ref{sec:fast}, we present the fast scheme of RegPT calculation
which realizes full fundamental cosmological parameter inference.
In Section~\ref{sec:challenge}, we present details of PT challenge:
statistical analysis and $N$-body simulations,
and in Section~\ref{sec:results}, the results of PT challenge are presented.
We make concluding remarks in Section~\ref{sec:conclusions}.

Throughout this paper, we assume flat $\Lambda$CDM Universe.
Though we will perform the statistical inference of cosmological parameters,
the fiducial cosmological parameters to generate the initial conditions
of $N$-body simulations for mock measurements are based on the results of
temperature and polarization anisotropies (TT,TE,EE+lowP dataset)
measured in \textit{Planck} 2015 results \cite{Planck2015XIII}.
The CDM density parameter is $\omega_\mathrm{cdm} = 0.1198$,
the baryon density parameter is $\omega_\mathrm{b} = 0.02225$, and
the massive neutrino density parameter is $\omega_\nu = 0.00064$.
For neutrinos, we assume that one of three generations is massive
with mass $m_\nu = 0.06\,\mathrm{eV}$ and the other two are massless.
The dark energy density parameter $\Omega_\mathrm{de} = 0.6844$
and the dark energy is the cosmological constant
with the equation of state parameter $w_\mathrm{de} = -1$.
The Hubble parameter at the present Universe is
$H_0 /  (\mathrm{km}\,\mathrm{s}^{-1}\,\mathrm{Mpc}^{-1}) = 100 h = 67.27$,
which is determined through the flatness
$(\omega_\mathrm{cdm} + \omega_\mathrm{b} + \omega_\nu)/h^2 + \Omega_\mathrm{de} = 1$.
The amplitude and tilt of the primordial scalar perturbation is
$\ln (10^{10} A_\mathrm{s}) = 3.094$ and $n_\mathrm{s} = 0.9645$, respectively,
with the pivot scale $k_\mathrm{piv} = 0.05 \, \mathrm{Mpc}^{-1}$.

\section{Theory}
\label{sec:theory}
In this section, we briefly review the basics of SPT and RegPT
and the calculations of power spectrum.
In addition to these PT schemes, we present one of EFT treatments: IR-resummed EFT.
In the redshift space, non-linear coupling with density and velocity fields has appreciable contributions
to the power spectrum. We overview how we can incorporate
the effect specific to the redshift space based on the PT framework.

\subsection{Standard perturbation theory}
Since the density fluctuation is small on large scales and/or at early times,
the cosmic matter field can be approximated as a single-stream fluid.
Then, the evolution of the cosmic fluid is described by continuity, Euler, and Poisson equations,
and the density and velocity divergence field can be expanded in a perturbative manner
with respect to the linear density field at the present Universe:
\begin{equation}
  \bm{\Psi} (\bm{k}; \eta) = \sum_{n = 1}^{\infty}e^{n \eta} \bm{\Psi}^{(n)} (\bm{k}) ,
  \label{eq:PT_Psi}
\end{equation}
where $\bm{\Psi} (\bm{k}; \eta) = (\delta (\bm{k}; \eta) , \theta (\bm{k}; \eta))$
is the density-velocity doublet,
$\delta (\bm{k}; \eta)$ and $\theta (\bm{k}; \eta)$ are the Fourier transform of
the density and velocity divergence field, respectively.
The velocity divergence field is normalized as $\theta = - \nabla \cdot \bm{v} /(aHf)$,
where $\bm{v}$ is the velocity field, $a$ is the scale factor, $H$ is the Hubble parameter,
$f = \dd \ln D_+ / \dd \ln a$ is the linear growth rate,
and $D_+ (a)$ is the linear growth factor normalized as $D_+ (a = 1) = 1$.
We have considered the fastest growing mode and
adopted Einstein--de Sitter Universe in the expansion,
and thus, the time-dependence of the kernel functions is factorizable
as $e^{n \eta}$, where $\eta \equiv \ln D_+$.
The doublet at linear order ($n = 1$)
is $\bm{\Psi}^{(1)} (\bm{k}) = (\delta_0 (\bm{k}), \delta_0 (\bm{k}))$,
where $\delta_0 (\bm{k})$ is the Fourier transform of the linear density field
at the present Universe.
The $n$-th order term of the doublet $\bm{\Psi}^{(n)}$ is expressed as
mode coupling of the Fourier transform of the linear density field
due to non-linear gravitational evolution:
\begin{align}
  \Psi_a^{(n)} (\bm{k}) =& \int \frac{\dd[3]{q_1}}{(2\pi)^3} \cdots
  \int \frac{\dd[3]{q_n}}{(2\pi)^3}
  (2\pi)^3 \delta_\mathrm{D} (\bm{k} - \bm{q}_{1 \cdots n})
  \nonumber \\
  & \times \ F_a^{(n)} (\bm{q}_1, \ldots, \bm{q}_n)
  \delta_0 (\bm{q}_1) \cdots \delta_0 (\bm{q}_n) ,
\end{align}
where $\delta_\mathrm{D}$ is the Dirac delta function,
$\bm{q}_{1 \cdots n} \equiv \bm{q}_1 + \cdots + \bm{q}_n$, and
$F_a^{(n)}$ are symmetrized kernel functions and
the explicit expression is derived via the recursion relation \cite[see, e.g.,][]{Crocce2006}.

Based on the PT framework, one can also compute the statistics of the density and velocity divergence field
in a perturbative manner.
As a working example, let us consider the most fundamental statistics to characterize the cosmic field:
power spectrum $P_{ab}$ and bispectrum $B_{abc}$,
which is the Fourier space version of 2-point and 3-point correlation functions, respectively.
The definitions are given as
\begin{equation}
  \langle \Psi_a (\bm{k}; \eta) \Psi_b (\bm{k}'; \eta) \rangle \equiv
  (2 \pi)^3 \delta_\mathrm{D} (\bm{k} + \bm{k}') P_{ab} (k; \eta),
\end{equation}
\begin{multline}
  \langle \Psi_a (\bm{k}_1; \eta) \Psi_b (\bm{k}_2; \eta) \Psi_c (\bm{k}_3; \eta) \rangle \\
  \equiv (2 \pi)^3 \delta_\mathrm{D} (\bm{k}_{123}) B_{abc} (\bm{k}_1, \bm{k}_2, \bm{k}_3; \eta),
\end{multline}
where the subscript ($a, b, c$) denotes the density $\delta$ or the velocity divergence $\theta$.
Based on the perturbative expressions (Eq.~\ref{eq:PT_Psi}),
the power spectrum and the bispectrum can be calculated as the loop expansion:
\begin{equation}
  P_{ab} (k; \eta) = P_{ab,\text{tree}} (k, \eta) + P_{ab,\text{1-loop}} (k, \eta) + \cdots,
\end{equation}
\begin{multline}
  B_{abc} (\bm{k}_1, \bm{k}_2, \bm{k}_3; \eta)
  = B_{abc,\text{tree}} (\bm{k}_1, \bm{k}_2, \bm{k}_3; \eta) \\
  + B_{abc,\text{1-loop}} (\bm{k}_1, \bm{k}_2, \bm{k}_3; \eta) + \cdots .
\end{multline}
These expansions are ordered with respect to
the linear power spectrum at the present Universe $P_0 (k)$, which is defined as
\begin{equation}
  \langle \delta_0 (\bm{k}) \delta_0 (\bm{k}') \rangle \equiv
  (2\pi)^3 \delta_\mathrm{D} (\bm{k} + \bm{k}') P_0 (k).
\end{equation}
The tree level (LO) terms of power spectrum and bispectrum ($P_{ab,\text{tree}}$ and $B_{abc,\text{tree}}$)
are proportional to the linear power spectrum and the square of the linear power spectrum, respectively.
Next, the 1-loop level (NLO) terms ($P_{ab,\text{1-loop}}$ and $B_{abc,\text{1-loop}}$)
require the loop integral with the external wave-vector and the integrand for power spectrum (bispectrum)
contains the square (cubic) power of the linear power spectrum.
In general, $n$-loop terms involve $3n-1$ dimensional integrals and the power of the linear power spectrum in the integrands
is $n+1$ for power spectrum and $n+2$ for bispectrum.
In practice, one needs to truncate the infinite series of the expansion to compute the power spectrum and the bispectrum,
and in this paper, terms up to 2-loop orders are considered for the power spectrum calculations.

\subsection{Regularized perturbation theory}
In order to improve the convergence of the SPT expansion,
the resummation technique, which reorganizes the terms of the SPT expansion,
has been developed.
Among the resummed PT frameworks, we consider the regularized perturbation theory (RegPT) \cite{Taruya2012},
which utilizes the multi-point propagators to reorganize the SPT expansion,
i.e., $\Gamma$-expansion \cite{Bernardeau2008}.

First, we introduce the expressions of multi-point propagators,
which are defined as the response of the density and velocity divergence
fields with respect to the linear density field.
The $(n+1)$-point propagator $\Gamma_a^{(n)}$ is defined as
\begin{multline}
  \frac{1}{n!} \left\langle
  \frac{\delta^n \Psi_a (\bm{k} ; \eta)}{\delta \delta_0 (\bm{k}_1) \cdots \delta \delta_0 (\bm{k}_n)}
  \right\rangle \\
  \equiv \delta_\mathrm{D} (\bm{k} - \bm{k}_{1 \cdots n}) (2 \pi)^{-3(n-1)}
  \Gamma_a^{(n)} (\bm{k}_1 , \ldots \bm{k}_n; \eta) .
\end{multline}
The expression of the propagator is given as
\begin{align}
  & \Gamma_a^{(n)} (\bm{k}_1 , \ldots, \bm{k}_n; \eta) \nonumber \\
  & = \Gamma_{a,\text{tree}}^{(n)} (\bm{k}_1 , \ldots, \bm{k}_n ; \eta) +
  \sum_{p = 1}^{\infty} \Gamma_{a,p\text{-loop}}^{(n)} (\bm{k}_1 , \ldots, \bm{k}_n ; \eta), \\
  & \Gamma_{a,\text{tree}}^{(n)} (\bm{k}_1 , \ldots, \bm{k}_n ; \eta)
  \equiv e^{n \eta} F_{a}^{(n)} (\bm{k}_1 , \ldots, \bm{k}_n) \\
  & \Gamma_{a,p\text{-loop}}^{(n)} (\bm{k}_1 , \ldots, \bm{k}_n ; \eta) \equiv \nonumber \\
  & e^{(n+2p)\eta} c_p^{(n)} \int \frac{\dd[3]{q_1}}{(2\pi)^3} \cdots
  \int \frac{\dd[3]{q_p}}{(2\pi)^3} \nonumber \\
  & \times F_a^{(n)} (\bm{q}_1, -\bm{q}_1, \ldots, \bm{q}_p, -\bm{q}_p,
  \bm{k}_1 , \ldots, \bm{k}_n) P_0 (q_1) \cdots P_0 (q_p) \\
  & \equiv e^{(n+2p) \eta} \bar{\Gamma}^{(n)}_{a,p\text{-loop}} (\bm{k}_1 , \ldots, \bm{k}_n) ,
\end{align}
where $c_p^{(n)} \equiv \binom{n+2p}{n} (2n-1)!!$.
There is a notable feature that the propagator has an asymptotic form at high-$k$ limit \cite{Bernardeau2008}:
\begin{multline}
  \lim_{k \to \infty} \Gamma_a^{(n)} (\bm{k}_1 , \ldots, \bm{k}_n; \eta) = \\
  \exp \left[ -\frac{k^2 e^{2\eta} \sigmad^2}{2} \right]
  \Gamma_{a,\text{tree}}^{(n)} (\bm{k}_1 , \ldots, \bm{k}_n; \eta),
\end{multline}
where $k \equiv |\bm{k}_1 + \cdots + \bm{k}_n|$,
$\sigmad^2$ is the dispersion of the displacement field:
\begin{equation}
  \sigmad^2 (k) \equiv \int_0^{k_\Lambda (k)} \frac{\dd q}{6 \pi^2} P_0 (q) ,
  \label{eq:def_sigmad}
\end{equation}
and we have introduced the UV cutoff scale $k_\Lambda (k)$ and adopted $k_\Lambda (k) = k/2$
to match the result of $N$-body simulations \cite{Taruya2012}.
This displacement dispersion controls the damping feature on small scales and
it is known that making $\sigmad$ a free parameter improves the fit to the $N$-body simulation results \cite{Osato2019}
though the generalized Galilean invariance is broken
(see e.g., \cite{Peloso2013,Bernardeau2014,Peloso2017}).
We refer to this one-parameter extended RegPT model as RegPT+.

It is possible to express the statistics of density and velocity fields with the propagator expansion.
The power spectrum and the bispectrum based on RegPT are formally expressed as the infinite series
\cite{Bernardeau2008}:
\begin{widetext}
\begin{align}
  \langle \Psi_a (\bm{k} ; \eta) \Psi_b (\bm{k}' ; \eta) \rangle =&
  \delta_\mathrm{D} (\bm{k} + \bm{k}')
  \sum_{p=1}^\infty p! \int \frac{\dd[3]{q_1} \cdots \dd[3]{q_p}}{(2 \pi)^{3p}}
  \delta_\mathrm{D} (\bm{k} - \bm{q}_{1 \ldots p}) \nonumber \\
  & \times \Gamma_a^{(p)} (\bm{q}_1, \ldots \bm{q}_p ; \eta)
  \Gamma_b^{(p)} (\bm{q}_1, \ldots \bm{q}_p ; \eta) P_0 (q_1) \cdots P_0 (q_p) ,
\end{align}
\begin{align}
  \langle \Psi_a (\bm{k}_1 ; \eta) \Psi_b (\bm{k}_2 ; \eta) \Psi_c (\bm{k}_3 ; \eta) \rangle =&
  \sum_{\alpha, \beta, \gamma} \binom{\alpha + \beta}{\alpha} \binom{\beta + \gamma}{\beta}
  \binom{\gamma + \alpha}{\gamma} \alpha! \beta! \gamma!
  \int \frac{\dd[3]{p_1} \cdots \dd[3]{p_\alpha}}{(2\pi)^{3(\alpha -1)}}
  \frac{\dd[3]{q_1} \cdots \dd[3]{q_\beta}}{(2\pi)^{3(\beta -1)}}
  \frac{\dd[3]{r_1} \cdots \dd[3]{r_\gamma}}{(2\pi)^{3(\gamma -1)}} \nonumber \\
  & \times \delta_\mathrm{D} (\bm{k}_1 - \bm{p}_{1 \ldots \alpha} - \bm{q}_{1 \ldots \beta})
  \delta_\mathrm{D} (\bm{k}_2 + \bm{q}_{1 \ldots \beta} - \bm{r}_{1 \ldots \gamma})
  \delta_\mathrm{D} (\bm{k}_3 + \bm{r}_{1 \ldots \gamma} + \bm{p}_{1 \ldots \alpha}) \nonumber \\
  & \times \Gamma_a^{(\alpha+\beta)} (\bm{p}_1,\ldots,\bm{p}_\alpha, \bm{q}_1,\ldots,\bm{q}_\beta ; \eta)
  \Gamma_b^{(\beta+\gamma)} (-\bm{q}_1,\ldots,-\bm{q}_\beta, \bm{r}_1,\ldots,\bm{r}_\gamma ; \eta) \nonumber \\
  & \times \Gamma_c^{(\gamma+\alpha)} (-\bm{r}_1,\ldots,-\bm{r}_\gamma, -\bm{p}_1,\ldots,-\bm{p}_\alpha ; \eta)
  \nonumber \\
  & \times P_0 (p_1) \cdots P_0 (p_\alpha) P_0 (q_1) \cdots P_0 (q_\beta) P_0 (r_1) \cdots P_0 (r_\gamma),
  \label{eq:bispectrum}
\end{align}
\end{widetext}
where the summation in Eq.~\eqref{eq:bispectrum} runs over non-negative integers
with the constraint that at most one of the indices ($\alpha, \beta, \gamma$) is zero.
Similarly to the SPT expansion, one needs to truncate the expansion.
Hereafter, we suppress the time variable $\eta$ for simplicity.
For more explicit construction of the power spectrum and bispectrum calculations respectively at 2- and 1-loop orders,
refer to Refs.~\cite{Taruya2012,Taruya2013}.

\subsection{Redshift space power spectrum}
The peculiar motion of galaxies along the line-of-sight direction
is known to induce the anisotropy onto the power spectrum observed via spectroscopic surveys,
referred to as the RSD effect,
and this effect must be taken into account for the analysis of the power spectrum measurements.
The redshift space power spectrum is enhanced on large scales due to a 
coherent infall toward the gravitational potential well, which is referred to as the Kaiser effect \cite{Kaiser1987}.
The non-linear version of the Kaiser formula is given as
\begin{equation}
  P_\mathrm{Kaiser}^{(\mathrm{S})} (k, \mu) = P_{\delta \delta} (k)
  + 2 f \mu^2 P_{\delta \theta} (k) + f^2 \mu^4 P_{\theta \theta} (k) ,
\end{equation}
where $P_{\delta \delta}$, $P_{\delta \theta}$, and $P_{\theta \theta}$
is the density auto-, the density-velocity cross-, and the velocity auto-spectra in the real space,
respectively, and $\mu$ is the directional cosine with respect to the line-of-sight direction.
Hereafter, we take the $z$-axis as the line-of-sight direction.
Furthermore, the non-linear coupling of density and velocity fields
give rise to additional contributions at small scales \cite{Scoccimarro2004}.
This effect can be modeled in a perturbative manner using the cumulant expansion
and the formalism is presented in Ref.~\cite{Taruya2010},
which is referred to as the Taruya--Nishimichi--Saito (TNS) correction,
and further extended with RegPT \cite{Taruya2013}.
The resultant power spectrum in the redshift space based on RegPT is given as
\begin{align}
  P^{(\mathrm{S})}_\text{RegPT} (k, \mu) =& D_\mathrm{FoG} (k \mu f \sigma_\mathrm{v})
  \nonumber \\
  & \times \left[ P_{\delta \delta} (k) + 2 f \mu^2 P_{\delta \theta} (k) +
  f^2 \mu^4 P_{\theta \theta} (k) \right.
  \nonumber \\
  & \left. + A(k, \mu) + B (k, \mu)  \right] ,
  \label{eq:RegPT_Pk_zspace}
\end{align}
where $D_\mathrm{FoG}$ is the damping function due to the Finger-of-God (FoG) effect
and will be discussed later,
and two correction terms denoted as $A (k, \mu)$ and $B(k, \mu)$ are introduced.
These correction terms are referred to as TNS terms.
The explicit expressions of $A$- and $B$-terms are given as
\begin{align}
  A (k, \mu) =& (k \mu f) \int \frac{\dd[3] p}{(2\pi)^3} \frac{\mu_p}{p}
  \nonumber \\
  & \times \left[ B_\sigma (\bm{p}, \bm{k}-\bm{p}, -\bm{k}) -
  B_\sigma (\bm{p}, \bm{k}, -\bm{k}-\bm{p}) \right], \\
  B (k, \mu) =& (k \mu f)^2 \int \frac{\dd[3] p}{(2\pi)^3}
  F(\bm{p}) F(\bm{k}-\bm{p}), \\
  F(\bm{p}) \equiv& \frac{\mu_p}{p} \left[ P_{\delta \theta} (p) +
  f \frac{\mu_p}{p} P_{\theta \theta} (p) \right] ,
\end{align}
where the cross-bispectrum $B_\sigma$ is defined as
\begin{align}
  \left\langle \theta (\bm{k}_1)
  \left\{ \delta (\bm{k}_2) + f \frac{\mu_2}{k_2} \theta(\bm{k}_2) \right\}
  \left\{ \delta (\bm{k}_3) + f \frac{\mu_3}{k_3} \theta(\bm{k}_3) \right\}
  \right\rangle
  \nonumber \\
  \equiv (2 \pi)^3 \delta_\mathrm{D} (\bm{k}_1 + \bm{k}_2 + \bm{k}_3)
  B_\sigma (\bm{k}_1, \bm{k}_2, \bm{k}_3) .
\end{align}
These expressions do not depend on the direction of the wave-vector $\bm{k}$,
and thus, we take $\bm{k} = (0, 0, k)$ and the directional cosine as
$\mu_p = (\bm{k} \cdot \bm{p}) / (k p) = p_z / p$.
The expression of the $A$-term is written as the integrals of density and velocity bispectra:
\begin{align}
  A(k, \mu) =& \sum_{n=1}^{3} \sum_{a, b=1}^{2} \mu^{2n} f^{a+b-1} \frac{k^3}{(2 \pi)^2}
  \int_0^{\infty} \!\! \dd r \int_{-1}^{+1} \!\! \dd x
  \nonumber \\
  & \times \left[ A^n_{ab} (r, x) B_{2ab} (\bm{p}, \bm{k} - \bm{p}, -\bm{k}) \right.
  \nonumber \\
  & \left. +\tilde{A}^n_{ab} (r, x) B_{2ab} (\bm{k} - \bm{p}, \bm{p}, -\bm{k} ) \right]
  \nonumber \\
  \equiv & \sum_{n=1}^{3} \mu^{2n} A_{2n} (k) ,
\label{eq:Aterm}
\end{align}
where $r$ and $x$ are the dimensionless variables defined as $r = p/k$ and $x = (\bm{p} \cdot \bm{k})/(p k)$.
As a notation, we also assign an integer to the subscript:
$a = 1$ denotes the density field $\delta$ and $a = 2$ denotes the velocity divergence field $\theta$.
Next, the $B$-term is expressed as the integrals of the power spectra:
\begin{align}
  B(k, \mu) =& \sum_{n=1}^{4} \sum_{a, b=1}^{2} \mu^{2n} (-f)^{a+b} \frac{k^3}{(2 \pi)^2}
  \int_0^{\infty} \!\! \dd r \int_{-1}^{+1} \!\! \dd x
  \nonumber \\
  & \times B^n_{ab} (r, x) \frac{P_{a2} (k\sqrt{1+r^2-2rx}) P_{b2}(kr)} {(1+r^2-2rx)^a}
  \nonumber \\
  \equiv & \sum_{n=1}^{4} \mu^{2n} B_{2n} (k) .
\label{eq:Bterm}
\end{align}
The explicit expressions of auxiliary functions ($A_{ab}^n, \tilde{A}_{ab}^n, B_{ab}^n$)
are found in Ref.~\cite{Taruya2010}.
Hereafter, we consider the power spectrum in the redshift space at 2-loop order,
and thus, density auto-, velocity auto-, and density-velocity cross-spectra in Eq.~\eqref{eq:RegPT_Pk_zspace}
are computed at 2-loop order.
For TNS terms, the bispectra in the $A$-term and the power spectra in the $B$-term
should be computed at 1-loop order to be consistent.
It is possible to derive the expression of the redshift space power spectrum
based on SPT at 1-loop order in the same manner
and the derivation is described in Appendix~\ref{sec:1loop_SPT}.

At small scales, the random motion of galaxies suppresses the structures,
which is referred to as the FoG effect \cite{Jackson1972}.
The damping effect is well explained by the phenomenological model,
which introduces the damping function $D_\mathrm{FoG} (k \mu f \sigmav)$ to control the overall amplitude
of the power spectrum and the functional form is taken as Gaussian or Lorentzian form:
\begin{equation}
D_\mathrm{FoG} (x) =
\begin{cases}
  \exp (-x^2) & (\text{Gaussian}) , \\
  (1+ x^2/2)^{-2} & (\text{Lorentzian}) .
\end{cases}
\end{equation}
In most of the models, the velocity dispersion $\sigma_\mathrm{v}$ is treated as a free parameter
and can be calculated at linear order as
\begin{equation}
  \label{eq:sigmav_L}
  \sigma^2_{\mathrm{v, L}} = \int_0^\infty \frac{\dd q}{6 \pi^2} P_\mathrm{L} (q; z) ,
\end{equation}
where $P_\mathrm{L} (k; z) = e^{2 \eta(z)} P_0 (k)$ is the linear power spectrum at the redshift $z$.
We also consider another functional form with one free parameter $\gamma$ \cite{Okumura2012}:
\begin{equation}
  D_\mathrm{FoG} (x) = \left(1 + \frac{x^2}{\gamma} \right)^{-\gamma} .
\end{equation}
This FoG model becomes Lorentzian FoG for $\gamma = 2$ and
asymptotes to Gaussian FoG with $\gamma \to \infty$.
Thus, this model contains the Gaussian and Lorentzian FoGs and has more flexibility.
Hereafter, we refer to this FoG model as ``$\gamma$ FoG''.

\subsection{IR-resummed effective field theory}
The EFT provides a practical way to compute the power spectrum
down to small scales ($\gtrsim 0.3 \, \hMpcinv$)
once free parameters are calibrated with $N$-body simulations or marginalized over
as nuisance parameters in comparing its predictions with simulations/observations.
In order to examine the performance of the EFT against PT models, we adopt the latter approach
and consider specifically the IR-resummed EFT model, where the damping of BAO feature due to the large-scale bulk motion
is corrected in the perturbative approach \cite{Baldauf2015b,Vlah2016,Blas2016}.
Here, we describe the power spectrum based on IR-resummed EFT at 1-loop order.
Although our primary focus is to test the 2-loop PT predictions,
the 1-loop EFT is now being used for parameter inference on observational data.
Thus, comparing its performance with those of the 2-loop PT would provide a useful guideline
for future applications to observations.

First, we divide the linear power spectrum $P_\mathrm{L} (k)$ into the wiggle and smooth parts.
The smooth part $P_\mathrm{L}^\mathrm{nw} (k)$
is obtained by smoothing the linear power spectrum with no-wiggle power spectrum:
\begin{align}
  P_\mathrm{L}^\mathrm{nw} (k) =& P_\mathrm{EH} (k) \frac{1}{\sqrt{2\pi} \log_{10} \lambda}
  \int \dd (\log_{10} q) \frac{P_\mathrm{L} (q)}{P_\mathrm{EH} (q)}
  \nonumber \\
  & \times \exp \left[ -\frac{(\log_{10} k - \log_{10} q)^2}{2 (\log_{10} \lambda)^2} \right] ,
\end{align}
where we adopt the smoothing scale $\lambda = 10^{0.25} \, \hMpcinv$ \cite{Vlah2016} and
$P_\mathrm{EH}$ is no-wiggle power spectrum from the fitting formula in Ref.~\cite{Eisenstein1998}.
The wiggle part $P_\mathrm{L}^\mathrm{w} (k)$ is obtained by taking the residual:
\begin{equation}
  P_\mathrm{L}^\mathrm{w} (k) = P_\mathrm{L}(k) - P_\mathrm{L}^\mathrm{nw} (k) .
\end{equation}
The expression for IR-resummed EFT power spectrum \cite{Ivanov2018} is given as
\begin{align}
  P^{(\mathrm{S})}_\text{EFT} (k, \mu) =& P^\mathrm{nw}_\mathrm{L} (k, \mu)
  + P^\mathrm{nw}_\text{1-loop} (k, \mu)
  \nonumber \\
  & + P^\mathrm{nw}_\mathrm{ctr} (k, \mu) + P_\mathrm{shot}
  \nonumber \\
  & + e^{-k^2 \Sigma^2_\mathrm{tot} (\mu)} \left[
  (1+k^2 \Sigma^2_\mathrm{tot} (\mu)) P^\mathrm{w}_\mathrm{L} (k, \mu) \right.
  \nonumber \\
  & \left. + P^\mathrm{w}_\text{1-loop} (k, \mu) + P^\mathrm{w}_\mathrm{ctr} (k, \mu) \right] ,
\end{align}
where $P_\mathrm{shot}$ is the constant shot noise term.
The total damping function $\Sigma^2_\mathrm{tot} (\mu)$ is defined as
\begin{align}
  \Sigma^2_\mathrm{tot} (\mu) &= [1+f \mu^2 (2+f) ] \Sigma^2
  + f^2 \mu^2 (\mu^2 - 1) \delta \Sigma^2 , \\
  \Sigma^2 &= \frac{1}{6 \pi^2} \int_0^{k_S} \!\! \dd q \,
  P^\mathrm{nw}_\mathrm{L} (q) \left[ 1 - j_0 ( q r_d ) + 2 j_2 ( q r_d ) \right] , \\
  \delta \Sigma^2 &= \frac{1}{2 \pi^2} \int_0^{k_S} \!\! \dd q \,
  P^\mathrm{nw}_\mathrm{L} (q) j_2 ( q r_d ) ,
\end{align}
where $r_d$ is the sound horizon scale at the drag epoch,
and $j_0$ and $j_2$ are 0th and 2nd order spherical Bessel functions.
We adopt $k_S = 0.2 \, h^{-1} \, \mathrm{Mpc}$ \cite{Ivanov2018}.
For the counter term $P_\mathrm{ctr} (k, \mu)$, we adopt the following form:
\begin{align}
  P_\mathrm{ctr} (k, \mu) =& -2 \tilde{c}_0 k^2 P_\mathrm{L} (k) - 2 \tilde{c}_2 f\mu^2 k^2 P_\mathrm{L} (k)
  \nonumber \\
  & - 2 \tilde{c}_4 f^2 \mu^4 k^2 P_\mathrm{L} (k)
  \nonumber \\
  & + \tilde{c}_{\nabla^4_z \delta} f^4 \mu^4 k^4 (1+f \mu^2)^2 P_\mathrm{L} (k),
\end{align}
where four parameters ($\tilde{c}_0$, $\tilde{c}_2$, $\tilde{c}_4$, $\tilde{c}_{\nabla^4_z \delta}$) are introduced.
For 1-loop correction and counter terms,
smooth and wiggle parts are obtained by regarding the expressions
as the functional of the linear power spectrum and plugging the smoothed and wiggly linear power spectra
into SPT expressions:
\begin{align}
P^\mathrm{nw}_\text{1-loop} & \equiv P_\text{1-loop} [P^\mathrm{nw}_\mathrm{L}] , \\
P^\mathrm{w}_\text{1-loop} & \equiv
P_\text{1-loop} [P_\mathrm{L}] -
P_\text{1-loop} [P^\mathrm{nw}_\mathrm{L}] , \\
P^\mathrm{nw}_\mathrm{ctr} & \equiv P_\mathrm{ctr} [P^\mathrm{nw}_\mathrm{L}] , \\
P^\mathrm{w}_\mathrm{ctr} & \equiv
P_\mathrm{ctr} [P_\mathrm{L}] -
P_\mathrm{ctr} [P^\mathrm{nw}_\mathrm{L}] .
\end{align}

\subsection{Alcock--Paczy\'nski effect}
When converting the galaxy position and redshift into the comoving coordinate,
one needs to assume cosmological parameters relevant to the geometry of the Universe.
The cosmological parameters are varied for inference but
due to the high computational cost, the redshift-distance conversion is performed only once
with fiducial cosmological parameter set.
The wrong fiducial cosmological parameters induce spurious anisotropy to the power spectrum.
This effect is called as Alcock--Paczy\'nski (AP) effect \cite{Alcock1979}.
The distortion is proportional to the Hubble distance for the line-of-sight direction
and the angular diameter distance for the direction perpendicular to the line-of-sight direction.
Then, we define AP parameters:
\begin{equation}
  \alpha_\parallel = \frac{D_H}{\DHfid} \frac{\rdfid}{r_d} ,\
  \alpha_\perp = \frac{D_M}{\DMfid} \frac{\rdfid}{r_d} ,
\end{equation}
where $r_d$ is the sound horizon at the drag epoch,
$D_M (z) = (1+z) D_A (z)$ is the proper motion distance,
$D_H (z) = c/H (z)$ is the Hubble distance,
$D_A (z)$ is the angular diameter distance,
$c$ is the speed of light,
$H (z)$ is the Hubble parameter,
and the superscript ``fid'' represents the value
computed with the fiducial cosmology.
Accordingly, the distorted power spectrum is given as
\begin{align}
P^{(\mathrm{S})}_{\mathrm{AP}} (k, \mu) &= \left( \frac{\rdfid}{r_d} \right)^3
\frac{1}{\alpha_\parallel \alpha_\perp^2} P^{(\mathrm{S})} (q, \nu), \\
q &= \frac{k}{\alpha_\perp} \left[ 1 + \mu^2 \left(
\frac{\alpha_\perp^2}{\alpha_\parallel^2} - 1 \right)
\right]^{\frac{1}{2}} , \\
\nu &= \frac{\mu \alpha_\perp}{\alpha_\parallel}
\left[ 1 + \mu^2 \left(
\frac{\alpha_\perp^2}{\alpha_\parallel^2} - 1 \right)
\right]^{\frac{1}{2}} .
\end{align}
This power spectrum is an observable when the fiducial cosmology is fixed to
derive the redshift-distance relation.

The information on anisotropic clustering signal
is expressed by the pair of $\alpha_\parallel$ and $\alpha_\perp$.
On the other hand, different combinations are introduced in the literature.
Among such pairs, we consider the dilation (warping) parameter $\alpha$ ($\epsilon$) \cite{Padmanabhan2008},
the volume-averaged distance $D_V (z)$, and the AP parameter $F_\mathrm{AP} (z)$:
\begin{align}
  \alpha & \equiv \alpha_\parallel^{\frac{1}{3}} \alpha_\perp^{\frac{2}{3}} , \\
  \epsilon & \equiv \left( \frac{\alpha_\parallel}{\alpha_\perp} \right)^{\frac{1}{3}} - 1 ,\\
  D_V (z) & \equiv (z D_M^2 (z) D_H (z))^{\frac{1}{3}}, \\
  F_\mathrm{AP} (z) & \equiv D_M (z) / D_H (z) .
\end{align}
The dilation parameter $\alpha$ corresponds to the isotropic deformation
and thus can be constrained from the BAO scale.
On the other hand, the warping parameter $\epsilon$ describes the anisotropic deformation and
the change of this parameter has more impact on the anisotropic moments.
Another pair is the volume-averaged distance $D_V (z)$ and the AP parameter $F_\mathrm{AP} (z)$,
which are widely employed in previous galaxy clustering analyses.

\section{Fast schemes for the redshift space power spectrum}
\label{sec:fast}
In the practical parameter inference from the power spectrum measurements,
a fast methodology is critical to explore the large parameter space.
Typically, a runtime of less than a few minutes for each cosmological model is required.
In order to realize fast calculations of the redshift space power spectrum at 2-loop order,
we employ the response function approach to speed up the computations
of the density and velocity power spectra at 2-loop order and the TNS correction terms.

The codes to compute the power spectrum with TNS correction terms are implemented in the framework of
\texttt{Eclairs} \cite{Eclairs} and will be publicly available
at the github repositry \footnote{\url{https://github.com/0satoken/Eclairs}}.

\subsection{Response function approach}
In this section, we describe the response function approach to speed up the calculations of
the power spectrum and the bispectrum \cite[for details, see Ref.][]{Osato2021}
and implement this approach for calculations of the TNS correction terms.
In general, the expressions of the power spectrum based on PT approaches can be
regarded as the functionals with respect to the linear power spectrum.
In the response function approach, the PT expressions are expanded with functional derivatives and
if the fiducial cosmology is close to the target cosmology, the power spectrum for the target cosmology
is well approximated with the expansion up to the leading order:
\begin{multline}
  P_{ab} (k; \bm{\theta}^\mathrm{tar}, \sigmad^\mathrm{tar})
  \simeq P_{ab} (k; \bm{\theta}^\mathrm{fid}, \sigmad^\mathrm{tar}) \\
   + \int \dd q \frac{\delta P_{ab} [P_0 ; \bm{\theta}^\mathrm{fid}, \sigmad^\mathrm{tar}]}{\delta P_0 (q)}
  \delta P_0 (q),
  \label{eq:resp_Pk}
\end{multline}
\begin{multline}
  B_{abc} (\bm{k}_1, \bm{k}_2, \bm{k}_3 ; \bm{\theta}^\mathrm{tar}, \sigmad^\mathrm{tar}) \simeq
  B_{abc} (\bm{k}_1, \bm{k}_2, \bm{k}_3 ; \bm{\theta}^\mathrm{fid}, \sigmad^\mathrm{tar}) \\
  + \int \dd q \frac{\delta B_{abc} [P_0 ; \bm{\theta}^\mathrm{fid}, \sigmad^\mathrm{tar}]}{\delta P_0 (q)}
  \delta P_0 (q),
  \label{eq:resp_Bk}
\end{multline}
where $\delta P_{ab} / \delta P_0$ and $\delta B_{abc} / \delta P_0$ are
the response functions of the power spectrum and the bispectrum, respectively,
$\bm{\theta}^\mathrm{tar}$ and $\bm{\theta}^\mathrm{fid}$ are sets of cosmological parameters
for target and fiducial cosmologies, respectively,
and $\delta P_0 (q) = P_0^\mathrm{tar} (q) - P_0^\mathrm{fid} (q)$
is the difference of linear power spectra between the fiducial and target cosmologies.
Once the fiducial spectra and response functions are precomputed,
the computational cost to evaluate spectra at target cosmologies is just to
perform the one-dimensional integrations
(second terms of the right hand sides in Eqs.~\ref{eq:resp_Pk} and \ref{eq:resp_Bk}).
Since the original expressions at 2-loop order involve five-dimensional integrations,
this response function approach considerably reduces the computational costs.
Note that the displacement dispersion $\sigmad$ is always evaluated at the target cosmology
because the cosmology dependence of the displacement dispersion
is incorporated without cost after the precomputations.

The redshift space power spectrum (Eq.~\ref{eq:RegPT_Pk_zspace})
has contributions from three 2-loop spectra
($P_{\delta \delta}$, $P_{\delta \theta}$, $P_{\theta \theta}$) and TNS corrections,
and all of them involve five-dimensional integrations, which hamper fast predictions
essential to the statistical inference.
For 2-loop spectra, we can make use of the response function approach
and the computations require only one-dimensional integrations, which can be computed
within a few seconds in general.
The most time-consuming part is five-dimensional integrations
in the $A$-terms; the 1-loop bispectrum contains three-dimensional integrations
and for the $A$-term, the bispectra are further integrated twice.
The response function approach is also applicable to TNS correction terms.
With the response function expansion,
the bispectra in the $A$-terms and the power spectra in the $B$-terms
can be calculated in a fast manner, and both of $A$- and $B$-terms
become three-dimensional integrations.
The response function approach is not applied to the entire $A$- and $B$-terms
because the variation of $\sigmad$ is not easily incorporated.
This reduction of computational time makes the analytical PT scheme feasible
for cosmological parameter estimation with the redshift space power spectrum.

\subsection{Validation}
In order to validate our fast scheme, we compare the results with the fast approach
and direct integrations.
We follow the validation procedure in Ref.~\cite{Osato2021}.
First, we generate $10$ fiducial and $10$ target cosmological parameter sets
around \textit{Planck} 2015 best-fit cosmological parameters,
which are homogeneously sampled with the latin hypercube design \cite{Zhang2019}.
The parameter ranges of fiducial and target cosmological parameters
are described in Section~III.B of Ref.~\cite{Osato2021}.
For fiducial sets, we precompute power spectra, bispectra, and response functions
and for each target set, we select the nearest fiducial set according to the distance
between fiducial and target linear power spectra, which is denoted as $d$
(see Eq.~22 of Ref.~\cite{Osato2021} for definition).
In Ref.~\cite{Osato2021}, we have found that the response function approach
reproduces the results with direct integrations with the accuracy of $0.5\%$ for power spectra
and $2\%$ for bispectra up to $0.3 \, \hMpcinv$.

Here, we apply the response function technique to the TNS correction terms
and investigate the accuracy.
For power spectra, the number of sampling for wave-numbers should be small but
sufficient to reconstruct the full shape of spectra.
We employ the following adaptive sampling scheme:
\begin{equation}
\begin{cases}
[10^{-3}, 10^{-2}] \, \hMpcinv & (\Delta \log k = \mathrm{const.}, n_k = 10), \\
[10^{-2}, 0.3] \, \hMpcinv & (\Delta k = \mathrm{const.}, n_k = 80), \\
[0.3, 1] \, \hMpcinv & (\Delta \log k = \mathrm{const.}, n_k = 30), \\
\end{cases}
\end{equation}
where $n_k$ is the number of sampling.
For the intermediate range, we use linear spacing to capture the BAO feature,
otherwise the log-equally spaced sampling is employed.
If the spectra with arbitrary wave-numbers are necessary, we employ cubic spline.
In $A$-term calculations, bispectra with specific configurations
$(\bm{k}_1, \bm{k}_2, \bm{k}_3) = (\bm{p}, \bm{k}-\bm{p}, -\bm{k})$ are required.
The range of the wave-number $k$ is $(10^{-3}, 1) \, \hMpcinv$ with
log-equally spaced $300$ wave-numbers.
For integration with respect to the wave-vector $\bm{p}$,
numerical integrations are performed with Gaussian quadrature
for $r = p/k$ and $x = (\bm{p} \cdot \bm{k})/(p k)$
and the number of sampling is $n_r = 600$ and $n_x = 10$, respectively.

\begin{figure*}
\includegraphics[width=\textwidth]{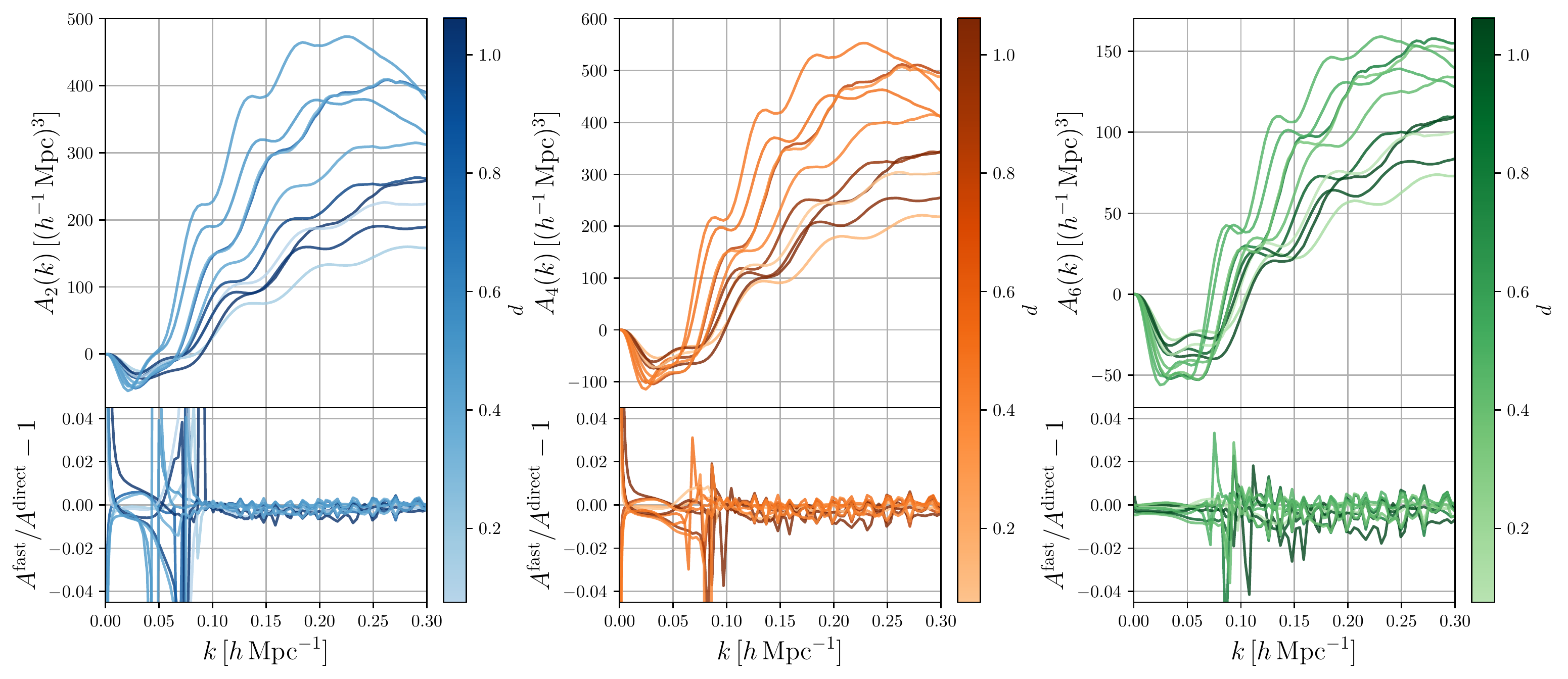}
\caption{The $A$-terms at the redshift $z = 0.9$ computed with the response function approach.
The upper panels show $A$-terms for the 10 target models
and the color corresponds to the distance from the nearest fiducial model (Eq.~\ref{eq:distance}).
The lower panels show the fractional difference
with respect to the results with direct integration.}
\label{fig:Aterm_validation}
\end{figure*}

\begin{figure*}
\includegraphics[width=0.8\textwidth]{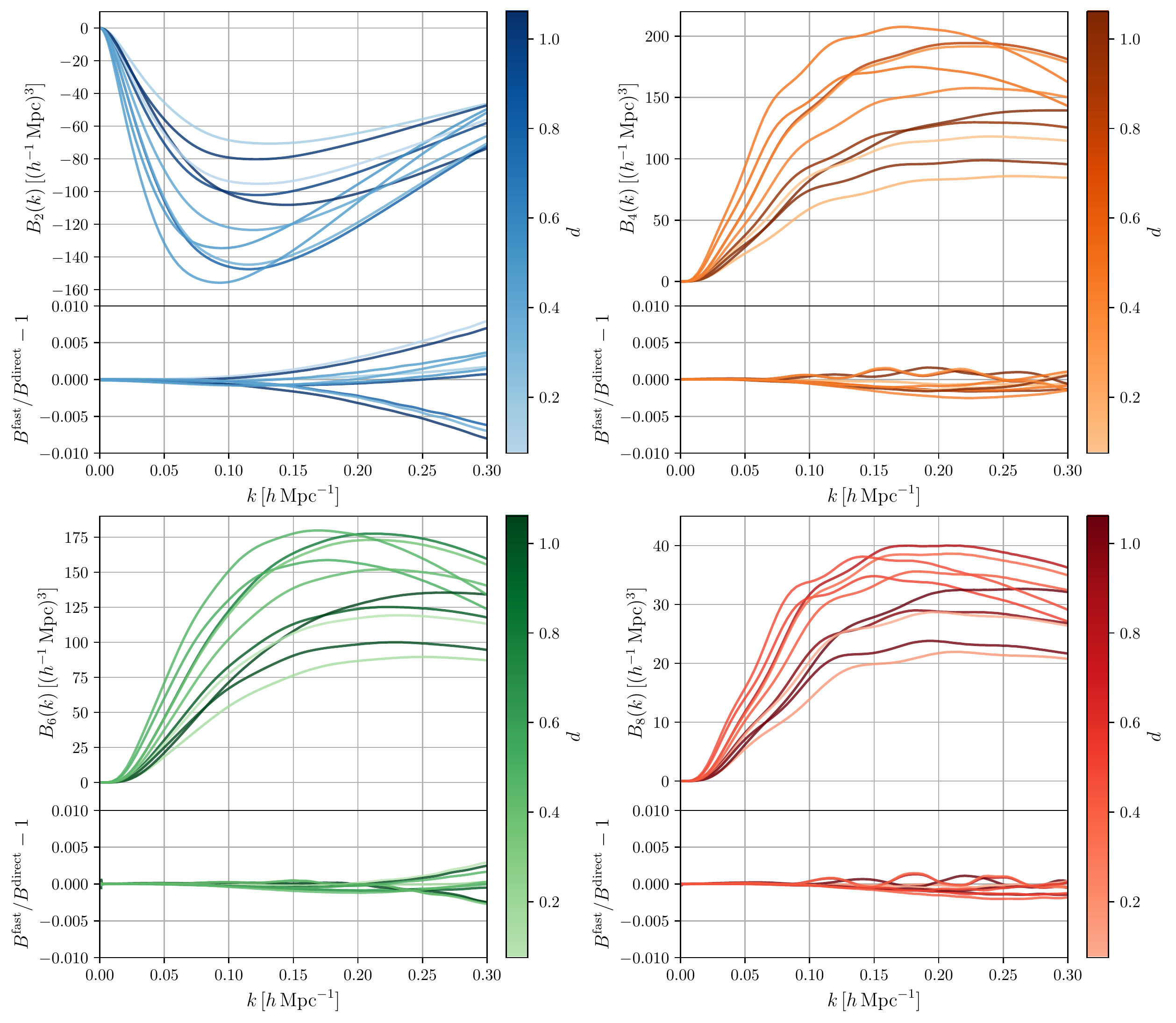}
\caption{The $B$-terms at the redshift $z = 0.9$ computed with the response function approach.
The configuration of the lower panels and the color is the same as Figure~\ref{fig:Aterm_validation}.}
\label{fig:Bterm_validation}
\end{figure*}

Figure~\ref{fig:Aterm_validation} shows the results with the response function approach
for $10$ target models and the ratios with respect to the direct integration results.
Each result is color-coded by the distance between target and fiducial cosmologies $d$,
which is defined as
\begin{equation}
  \label{eq:distance}
  d^2 = \frac{1}{n_k} \sum_{i=1}^{n_k}
  \frac{[\log (P_0^\text{tar} (k_i)) - \log (c P_0^\text{fid} (k_i))]^2}{\sigma^2_{k_i}},
\end{equation}
where $\sigma_{k_i} = k_i / (1\, \hMpcinv)$, $n_k = 20$, and $c$ is the scaling factor.
The scaling factor and fiducial cosmological model are determined to minimize the distance $d$.
The detailed procedure to select the fiducial model is described in Section~III.B of Ref.~\cite{Osato2021}.
The accuracy of $A$-term calculations is about $1\%$ at all scales.
There is a noisy feature around $k = 0.07 \, \hMpcinv$ but this originates from zero-crossing
of direct integration results.
Similarly, Figure~\ref{fig:Bterm_validation} shows the results for $B$-terms.
The accuracy for the $B$-terms is more stable and well within $1\%$ for all scales.

\begin{figure*}
\includegraphics[width=\textwidth]{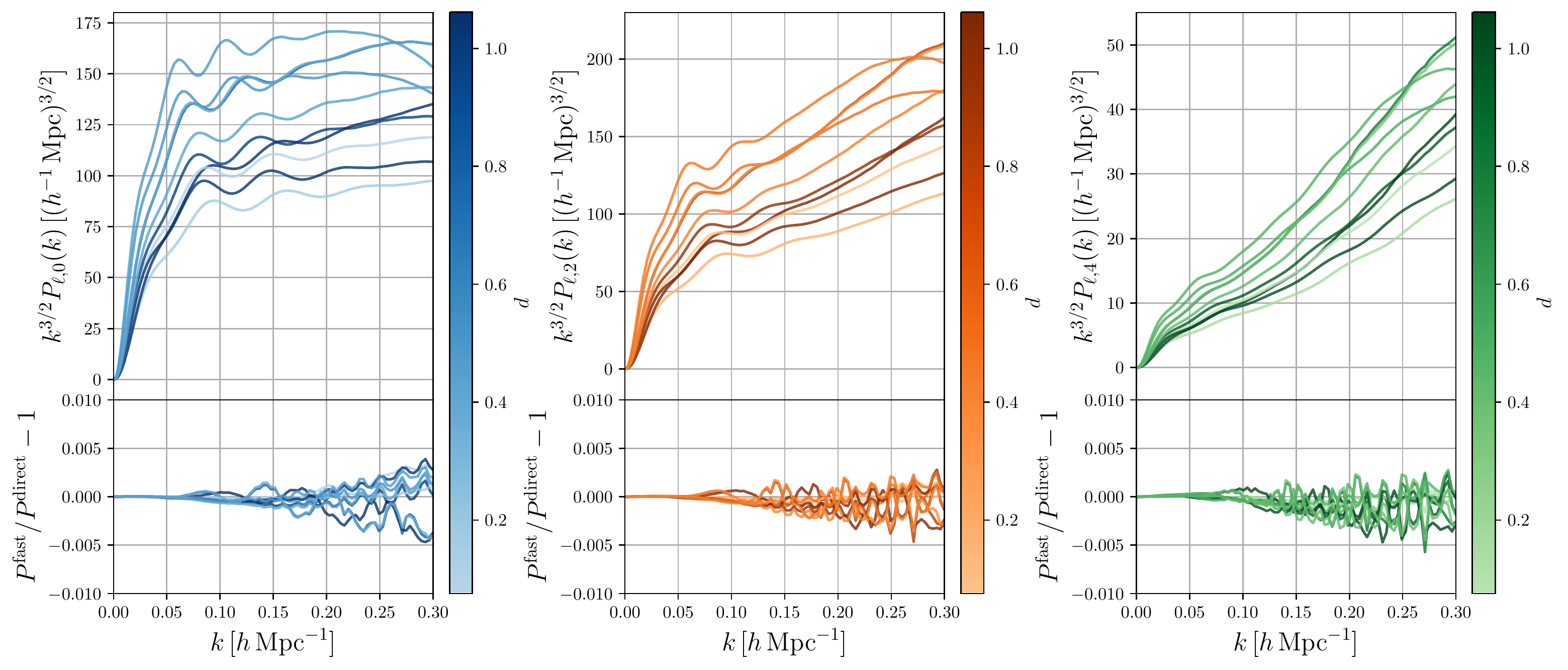}
\caption{The monopole, quadrupole, and hexadecapole moments at the redshift $z = 0.9$ computed with the response function approach.
The configuration of the lower panels and the color is the same as Figure~\ref{fig:Aterm_validation}.}
\label{fig:multipoles_validation}
\end{figure*}

\begin{figure*}
\includegraphics[width=\textwidth]{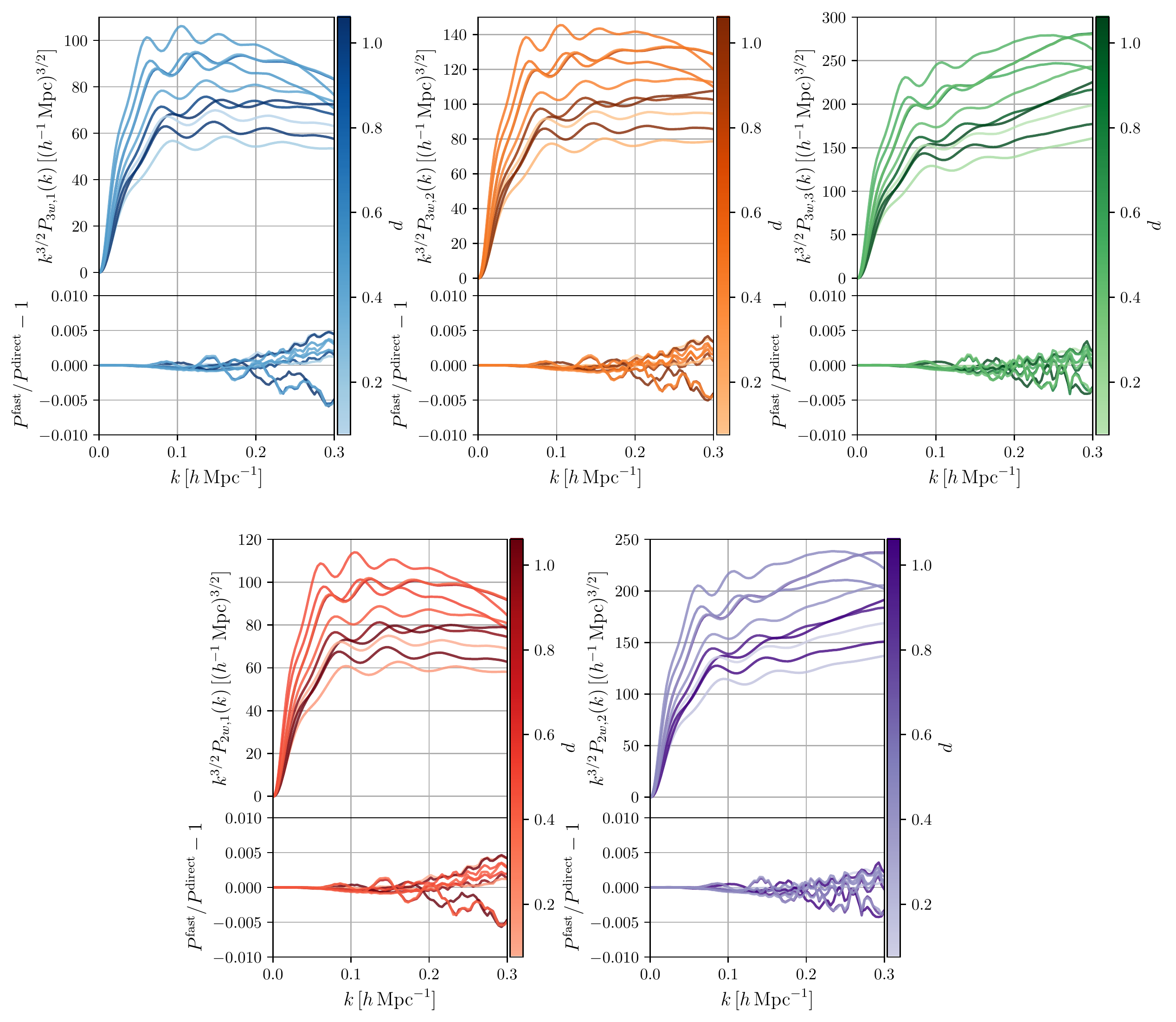}
\caption{The wedge power spectra at the redshift $z = 0.9$ computed with the response function approach.
The upper three figures show the results with 3 wedges $[0, 1/3], [1/3, 2/3], [2/3, 1]$
and the lower two figures show the results with 2 wedges  $[0, 1/2], [1/2, 1]$.
The configuration of the lower panels and the color is the same as Figure~\ref{fig:Aterm_validation}.}
\label{fig:wedges_validation}
\end{figure*}

In the practical analysis, instead of two-dimensional power spectrum $P^\mathrm{(S)} (k,\mu)$,
the multipole expansion is widely used to characterize the anisotropy.
The $\ell$-th order multipole moment is defined as
\begin{equation}
  P_\ell (k) \equiv \frac{2 \ell + 1}{2} \int_{-1}^{+1} \!\!
  \dd \mu \, \mathcal{P}_\ell (\mu) P^\mathrm{(S)} (k, \mu) ,
\end{equation}
where $\mathcal{P}_\ell$ is the Lengendre polynomial of order $\ell$.
Another representation of anisotropic power spectrum is wedges \cite{Kazin2012}
which are mean power spectrum with a given range of the directional cosine:
\begin{equation}
  P_w (k) \equiv \frac{1}{\mu_\mathrm{max}^w - \mu_\mathrm{min}^w}
  \int_{\mu_\mathrm{min}^w}^{\mu_\mathrm{max}^w} \!\!
  \dd \mu \, P^\mathrm{(S)} (k, \mu) ,
\end{equation}
where $[\mu_\mathrm{min}^w, \mu_\mathrm{max}^w]$
is the range of the bin.
Figures~\ref{fig:multipoles_validation} and \ref{fig:wedges_validation}
demonstrate the performance of the response function approach for multipoles and wedges, respectively.
For multipoles, monopole ($\ell = 0$), quadrupole ($\ell = 2$), and hexadecapole ($\ell = 4$)
moments are considered.
For wedges, we consider two sets of wedge bins:
\begin{equation}
  [\mu_\mathrm{min}^w, \mu_\mathrm{max}^w] =
  \begin{cases}
  \left[0, \frac{1}{3}\right], \left[\frac{1}{3}, \frac{2}{3}\right],
  \left[\frac{2}{3}, 1\right] & (\text{3 wedges}), \\
  \left[0, \frac{1}{2}\right], \left[\frac{1}{2}, 1\right] & (\text{2 wedges}) .
  \end{cases}
\end{equation}
The response function approach has been employed to compute all 2-loop spectra
and TNS correction terms.
The accuracy of multipoles and wedges is well within $0.5\%$ up to $k = 0.3 \, \hMpcinv$.
Therefore, the fast scheme can be applied to practical analysis.

\section{Perturbation theory challenge in redshift space}
\label{sec:challenge}
Here, we describe the PT challenge analysis and
present details of $N$-body simulations for mock measurements of the redshift space power spectrum
and theoretical templates used in the analysis.

\subsection{Mock measurement of the power spectrum}
First, we prepare the realistic mock measurements of the power spectrum in the cosmological parameter inference.
The baseline $N$-body simulation is presented in Ref.~\cite{Nishimichi2017}
and here we briefly summarize specifications of the simulation.
The gravitational evolution of the matter density field is solved
by the Tree-PM code \texttt{L-Gadget-2} \cite{Springel2005}.
The initial condition is generated at the redshift $z_\mathrm{ini} = 15$
based on the second-order Lagrangian PT \cite{Scoccimarro1998,Crocce2006b,Nishimichi2009,Valageas2011}
with the transfer function calculated with the Boltzmann solver \texttt{CAMB} \cite{Lewis2000}.
Furthermore, the Fourier amplitude and phase of initial conditions are determined
based on the ``paired and fixed'' approach \cite{Angulo2016} to suppress the cosmic variance.
The employed cosmological parameters are the best-fit values of
\textit{Planck} 2015 results \cite{Planck2015XIII}, which are listed in Section~\ref{sec:introduction}.
The volume of the simulation box is $V_s = 2048^3 \, (\hMpc)^3$ with periodic boundary condition
and the number of particle is $N = 2048^3$.

Hereafter, we employ the particle snapshot at the redshift $z = 0.900902$.
We construct the density field with regular grids, where the number of grid is $1024$ on a side,
with cloud-in-cell mass assignment and then apply fast Fourier transform
to obtain the Fourier space density field.
For power spectrum measurements,
Fourier amplitude for each mode is summed up for the given range of bins:
for multipoles,
\begin{equation}
\hat{P}_\ell (k_i) = \frac{2\ell+1}{N_{k_i}} \sum_{k \in [k_i - \Delta k / 2, k_i + \Delta k / 2]}
|\tilde{\delta} (k_\parallel, \bm{k}_\perp)|^2 \mathcal{P}_\ell (\mu) ,
\label{eq:estim_multi}
\end{equation}
and for wedges,
\begin{equation}
\hat{P}_w (k_i) = \frac{1}{N_{k_i, w}}
\sum_{\substack{k \in [k_i - \Delta k / 2, k_i + \Delta k / 2] \\
|\mu| \in [\mu^w_\mathrm{min}, \mu^w_\mathrm{max}]}}
|\tilde{\delta} (k_\parallel, \bm{k}_\perp)|^2 ,
\label{eq:estim_wedge}
\end{equation}
where $\tilde{\delta}$ is the Fourier transform of the density field,
$\mu = k_\parallel/k$, $k = \sqrt{k_\parallel^2 + k_\perp^2}$\footnote{We take the $z$-direction
as the line-of-sight direction, and thus $k_\parallel = k_z$ and $\bm{k}_\perp = (k_x, k_y)$.}
and the number of modes is calculated by counting the modes for the three-dimensional regular grids:
\begin{align}
N_{k_i} &= \sum_{k \in [k_i - \Delta k / 2, k_i + \Delta k / 2]} 1 , \\
N_{k_i, w} &=
\sum_{\substack{k \in [k_i - \Delta k / 2, k_i + \Delta k / 2] \\ |\mu| \in [\mu^w_\mathrm{min}, \mu^w_\mathrm{max}]}} 1.
\end{align}
Here, we consider the linearly spaced bins with $\Delta k = 0.01 \, h\,\mathrm{Mpc}^{-1}$.
If the width is sufficiently small compared with $k$, the estimator converges.
However, on large scales, i.e., small $k$, the number of available modes is limited
and the discreteness of modes impacts the estimation of the power spectrum.
In order to incorporate the finite grid effect, in the model prediction,
we take the sum of modes as performed in the estimators
(Eqs.~\ref{eq:estim_multi} and \ref{eq:estim_wedge}) with replacing $|\tilde{\delta}|^2$
with the model template $P^{(\mathrm{S})}$ \cite{Taruya2013}.

We assume that the likelihood follows the multi-variate Gaussian distribution:
for multipoles,
\begin{align}
  & -2 \log \mathcal{L}_\text{multipoles} (\bm{\theta} ; k_\mathrm{max}) = \nonumber \\
  & \sum_{\ell, \ell'} \sum_{i, j}^{k_i, k_j < k_\mathrm{max}}
  \left[ \hat{P}_\ell (k_i) - P_\ell (\bm{\theta} ; k_i) \right]
  \mathrm{Cov}_{\ell, \ell'}^{-1} (k_i, k_j)
  \nonumber \\
  & \times \left[ \hat{P}_{\ell'} (k_j) - P_{\ell'} (\bm{\theta} ; k_j) \right]
  + \text{const.}
  \label{eq:lkl_multipoles} \\
  &\equiv \chi^2_\text{multipoles} + \text{const.}
  \label{eq:chi2_multipoles}
\end{align}
and for wedges,
\begin{align}
  & -2 \log \mathcal{L}_\text{wedges} (\bm{\theta} ; k_\mathrm{max}) = \nonumber \\
  & \sum_{w, w'} \sum_{i, j}^{k_i, k_j < k_\mathrm{max}}
  \left[ \hat{P}_w (k_i) - P_w (\bm{\theta} ; k_i) \right]
  \mathrm{Cov}_{w,w'}^{-1} (k_i, k_j)
  \nonumber \\
  & \times \left[ \hat{P}_{w'} (k_j) - P_{w'} (\bm{\theta} ; k_j) \right]
  + \text{const.}
  \label{eq:lkl_wedges} \\
  & \equiv \chi^2_\text{wedges} + \text{const.},
  \label{eq:chi2_wedges}
\end{align}
where $\bm{\theta}$ is the set of cosmological and nuisance parameters,
$k_\mathrm{max}$ is the maximum wave-number to exclude small-scale data points,
$\hat{P}_\ell (k_i)$ and $\hat{P}_w (k_i)$ are binned multipoles and wedges
measured from the simulation, and $P_\ell (\bm{\theta} ; k_i)$ and $P_w (\bm{\theta} ; k_i)$
are model predictions of multipoles and wedges,
and $\chi^2_\text{multipoles}$ and $\chi^2_\text{wedges}$ are
the chi-squares for multipoles and wedges, respectively.

\subsection{Covariance matrix}
In the actual observations, galaxies are used as a tracer of the matter distribution,
and thus, the galaxy power spectrum is employed in the analysis.
Though we are interested in how our approach performs for the \textit{matter} power spectrum
without uncertainty relevant to the galaxy bias,
it is important to take into account the shot noise due to the finite number of observed galaxies.
Hence, we include the effective shot noise term with angular dependence:
\begin{equation}
  \frac{1}{n_\mathrm{g}^\mathrm{eff}} (\mu) =
  \frac{(1+f \mu^2)^2}{(b_\mathrm{g}+f \mu^2)^2} \frac{1}{n_\mathrm{g}} ,
\end{equation}
where $b_\mathrm{g} = 1.41$ is the linear galaxy bias and
$n_\mathrm{g} = 8.4 \times 10^{-4} \, (\hMpc)^{-3}$ is the galaxy number density,
which are expected values in \textit{Euclid} survey at redshift $z = 1$ \cite{Amendola2018}.
Assuming furthermore that the field is Gaussian distributed,
the covariance matrix of multipoles is given by (e.g., \cite{Taruya2010,Grieb2016,Hand2017})
\begin{widetext}
\begin{equation}
  \mathrm{Cov}_{\ell, \ell'} (k_i, k_j) = \delta_{ij} \frac{2}{N_{k_i}}
  \frac{(2\ell+1)(2\ell'+1)}{2} \int_{k_i-\Delta k/2}^{k_i + \Delta k/2}
  \frac{4 \pi k^2 \dd k}{V_{k_i}}
  \int_{-1}^{+1} \!\! \dd \mu \,
  \mathcal{P}_{\ell} (\mu) \mathcal{P}_{\ell'} (\mu)
  \left[ P^{(\mathrm{S})}_\mathrm{L,FoG} (k, \mu) + \frac{1}{n_\mathrm{g}^\mathrm{eff}} (\mu) \right]^2 ,
\end{equation}
\end{widetext}
where $V_{k_i} = (4\pi/3) [(k_i + \Delta k/2)^3 - (k_i - \Delta k/2)^3]$,
$N_{k_i} = V_{k_i} V_s / (2\pi)^3$ is the number of modes within the bin.
The power spectrum $P^{(\mathrm{S})}_\mathrm{L,FoG} (k, \mu)$
is at linear order with the Lorentzian FoG damping function:
\begin{equation}
  P^{(\mathrm{S})}_\mathrm{L,FoG} (k, \mu) =
  D_\mathrm{FoG}^\text{Lorentzian} (k\mu f \sigma_\mathrm{v,L})
  (1 + f \mu^2)^2 P_\mathrm{L} (k) .
\end{equation}
Though the linear model underestimates the power spectrum at small scales
where non-linear evolution predominates, the damping due to the FoG effect
is strong on small scales and the shot noise is dominant in the covariance matrix.
Thus, the linear model yields reasonable estimates of the covariance matrix.
Similarly, the covariance matrix for wedges is given by \cite{Grieb2016,Hand2017}
\begin{widetext}
\begin{equation}
  \mathrm{Cov}_{w, w'} (k_i, k_j) =
  \delta_{ij} \delta_{w, w'} \frac{2}{N_{k_i} \Delta \mu^w}
  \int_{k_i-\Delta k/2}^{k_i + \Delta k/2} \frac{4 \pi k^2 \dd k}{V_{k_i}}
  \int_{\mu^w_\mathrm{min}}^{\mu^w_\mathrm{max}} \!\!
  \frac{\dd \mu}{\Delta \mu^w}
  \left[ P^{(\mathrm{S})}_\mathrm{L,FoG} (k, \mu) + \frac{1}{n_\mathrm{g}^\mathrm{eff}} (\mu) \right]^2 ,
\end{equation}
\end{widetext}
where $\Delta \mu^w \equiv \mu^w_\mathrm{max} - \mu^w_\mathrm{min}$
is the wedge bin width.
Note that these covariance matrices are calculated once with fiducial cosmological parameters
and thus, the cosmological dependence is not considered in the cosmological parameter inference.

\subsection{Markov chain Monte-Carlo analysis}
Our PT challenge for cosmological parameter estimation employs Markov chain Monte-Carlo (MCMC) technique to compute the posterior distribution of parameters.
To be specific, we use the Affine invariant ensemble sampler \texttt{emcee} \cite{Foreman-Mackey2013,Foreman-Mackey2019}.
As a baseline model, we adopt RegPT at 2-loop order with Lorentzian FoG,
the data vector consisting of 3 multipoles ($\ell = 0, 2, 4$), and the AP effect incorporated.
We also consider the various cases for PT modeling and data vectors in the MCMC analysis;
\begin{description}
  \item[PT model] RegPT (2-loop), RegPT+ (2-loop), IR-resummed EFT (1-loop)
  \item[Data vector] 3 multipoles ($\ell = 0, 2, 4$), 2 multipoles ($\ell = 0, 2$),
  3 wedges ($\left[0, 1/3 \right], \left[1/3, 2/3 \right], \left[2/3, 1 \right]$),
  2 wedges ($\left[0, 1/2 \right], \left[1/2, 1 \right]$)
  \item[FoG] Lorentzian, Gaussian, $\gamma$ FoGs
  \item[AP effect] Considered or ignored
\end{description}
Table~\ref{tab:models} summarizes the cases we examined
in the PT challenge analysis, in which the IR-resummed EFT at 1-loop order
is included as a reference model to clarify the performance of 2-loop PT models.
We vary the maximum wave-number $k_\mathrm{max}$, i.e., the smallest scale of the data points used in the analysis,
from $0.12 \, \hMpcinv$ to $0.30 \, \hMpcinv$ by the step of $0.03 \, \hMpcinv$.
The results based on 1-loop PT models (RegPT, RegPT+, and SPT) are discussed in Appendix~\ref{sec:1loop_results}.

In the subsequent MCMC analysis, we consider
five cosmological parameters ($\omega_\mathrm{cdm}, \omega_\mathrm{b}, h, A_\mathrm{s}, n_\mathrm{s}$)
plus nuisance parameters, which are described in Table~\ref{tab:models}.
Though the target power spectrum is the matter power spectrum,
all the models involve the linear bias parameter $b_1$ as a nuisance parameter.
That is because the linear bias is incorporated consistently in the models
and does not yield additional higher-order bias terms.
The true value of the bias parameter in this analysis is unity, which serves as a consistency check of the models.
Furtheremore, there is another implication about the linear bias.
In the real galaxy clustering analysis,
the amplitude information cannot be used directly
due to the bias uncertainties.
On the other hand, the degeneracy in the amplitude can be broken
through RSD since we assume there is no velocity bias.
Therefore, dropping the linear bias would lead to 
an unrealistic assessment of the model accuracy.
In order to incorporate the linear bias in the models,
$f$ and $P_{ab}$ in Eq.~\eqref{eq:RegPT_Pk_zspace}
are replaced with $\beta \equiv f/b_1$ and $b_1^2 P_{ab}$, respectively.
The TNS correction terms scale with the linear bias as
\begin{align}
  & A (k, \mu; f) \to b_1^3 A (k, \mu; \beta), \\
  & B (k, \mu; f) \to b_1^4 B (k, \mu; \beta).
\end{align}
The velocity dispersion parameter $\sigmav$ controls the scale of FoG damping and this parameter is required
by all the models with RegPT and RegPT+.
RegPT+ introduces the new parameter, the dispersion displacement parameter $\sigmad$,
which improves the modeling of small-scale power spectra.
If $\gamma$ FoG is selected,
the nuisance parameter $\gamma$ is included and this parameter determines the shape of FoG damping.
The IR-resummed EFT models introduce coefficients of counter terms
($\tilde{c}_0, \tilde{c}_2, \tilde{c}_4, \tilde{c}_{\nabla^4_z \delta}$) and the shot noise term $P_\mathrm{shot}$
as nuisance parameters.
Since the target power spectrum is the matter power spectrum,
the shot noise is expected to be close to zero
but included in the analysis for the consistency check
similarly to the linear bias.
In the cases of 2 multipoles and 2 wedges, we do not include the parameter $\tilde{c}_4$,
which is the coefficient of the counter term proportional to $\mu^4$,
because this term is less constrained due to the limited sampling in the $\mu$-direction. 

We add prior information for $\omega_\mathrm{b}$ and $n_\mathrm{s}$,
which are only poorly constrained with the redshift space power spectrum.
For both parameters, the prior distribution is Gaussian 
with mean values given by the fiducial ones
and standard deviations of $n_\mathrm{s}$ and $\omega_\mathrm{b}$ inferred respectively by the 
\textit{Planck} 2015 result and the constraints brought by big-bang nucleosynthesis
and observations of primordial deutrium abundance \cite{Cooke2018}.
Note that for the $\gamma$ FoG, the sampled parameter is $\gamma^{-1}$ instead of $\gamma$
because the $\gamma$ FoG asymptotes to the Gaussian FoG with $\gamma \to \infty$, i.e., $\gamma^{-1} \to 0$,
and the correspondence becomes clearer with $\gamma^{-1}$.
For other parameters, we assume flat prior distributions
and they are summarized in Table~\ref{tab:priors}.
Table~\ref{tab:derived} shows the fiducial values of the derived parameters.
All the chains are run with 80 walkers.
For convergence of the chains, the sampler is run
until the length of chains is 50 times longer than the auto-correlation time
for all cosmological parameters \footnote{For cases with RegPT+ with $\kmax = 0.12, 0.15, 0.18 \, \hMpcinv$,
$\sigmad$ and $\sigmav$ parameters converge slowly due to the parameter degeneracy.
We relax the convergence criterion only for $\sigmad$ and $\sigmav$
so that the length of chains is 10 times longer than the auto-correlation time after the burn-in process.
For higher $\kmax$, i.e. $\kmax = 0.21, 0.24, 0.27, 0.30 \, \hMpcinv$, the length of the chains is 50 times longer
than the auto-correlation time even for $\sigmad$ and $\sigmav$
because the parameter degeneracy is broken due to the small-scale data points.
The same problem occurs for $\gamma$ FoG and the convergence criterion is also relaxed for $\gamma^{-1}$.
For other nuisance parameters, i.e. $b_1$, and EFT parameters,
we adopt the same convergence criterion as cosmological parameters,
i.e. 50 times longer than the auto-correlation time.
In the presented analysis, nuisance parameters are always marginalized
and the relaxation of the convergence criterion does not have significant impacts on results.}.

\begin{table*}
  \caption{Descriptions of models of RegPT and RegPT+ at 2-loop order and IR-resummed EFT at 1-loop order.
  The label describes the components of the model:
  the PT model (RegPT, RegPT+, or IR-resummed EFT),
  the functional form of FoG damping (Lorentzian, Gaussian, or $\gamma$ FoGs),
  and the data vector (3/2 multipoles or 3/2 wedges).}
  \label{tab:models}
    \begin{tabular}{cccccc}
      \hline \hline
      Label & Model & Data vector & AP & FoG & Nuisance parameters \\
      \hline
      RPT-L-3$\ell$ & RegPT & 3 Multipoles ($\ell = 0, 2, 4$) & \checkmark & Lorentzian & $b_1, \sigmav$ \\
      RPT-G-3$\ell$ & RegPT & 3 Multipoles ($\ell = 0, 2, 4$) & \checkmark & Gaussian & $b_1, \sigmav$ \\
      RPT-$\gamma$-3$\ell$ & RegPT & 3 Multipoles ($\ell = 0, 2, 4$) & \checkmark & $\gamma$ FoG & $b_1, \sigmav, \gamma^{-1}$ \\
      RPT(no AP)-L-3$\ell$ & RegPT & 3 Multipoles ($\ell = 0, 2, 4$) & --- & Lorentzian & $b_1, \sigmav$ \\
      RPT-L-2$\ell$ & RegPT & 2 Multipoles ($\ell = 0, 2$) & \checkmark & Lorentzian & $b_1, \sigmav$ \\
      RPT-L-3$w$ & RegPT & 3 Wedges ($\left[0, 1/3 \right], \left[1/3, 2/3 \right], \left[2/3, 1 \right]$) & \checkmark & Lorentzian & $b_1, \sigmav$ \\
      RPT-L-2$w$ & RegPT & 2 Wedges ($\left[0, 1/2 \right], \left[1/2, 1 \right]$) & \checkmark & Lorentzian & $b_1, \sigmav$ \\
      RPT+-L-3$\ell$ & RegPT+ & 3 Multipoles ($\ell = 0, 2, 4$) & \checkmark & Lorentzian & $b_1, \sigmav, \sigmad$ \\
      RPT+-G-3$\ell$ & RegPT+ & 3 Multipoles ($\ell = 0, 2, 4$) & \checkmark & Gaussian & $b_1, \sigmav, \sigmad$ \\
      RPT+-$\gamma$-3$\ell$ & RegPT+ & 3 Multipoles ($\ell = 0, 2, 4$) & \checkmark & $\gamma$ FoG & $b_1, \sigmav, \gamma^{-1}, \sigmad$ \\
      RPT+(no AP)-L-3$\ell$ & RegPT+ & 3 Multipoles ($\ell = 0, 2, 4$) & --- & Lorentzian & $b_1, \sigmav, \sigmad$ \\
      RPT+-L-2$\ell$ & RegPT+ & 2 Multipoles ($\ell = 0, 2$) & \checkmark & Lorentzian & $b_1, \sigmav, \sigmad$ \\
      RPT+-L-3$w$ & RegPT+ & 3 Wedges ($\left[0, 1/3 \right], \left[1/3, 2/3 \right], \left[2/3, 1 \right]$) & \checkmark & Lorentzian & $b_1, \sigmav, \sigmad$ \\
      RPT+-L-2$w$ & RegPT+ & 2 Wedges ($\left[0, 1/2 \right], \left[1/2, 1 \right]$) & \checkmark & Lorentzian & $b_1, \sigmav, \sigmad$ \\
      EFT-3$\ell$ & IR-resummed EFT & 3 Multipoles ($\ell = 0, 2, 4$) & \checkmark & ---
      & $b_1, \tilde{c}_0, \tilde{c}_2, \tilde{c}_4, \tilde{c}_{\nabla^4_z \delta}, P_\mathrm{shot}$ \\
      EFT-2$\ell$ & IR-resummed EFT & 2 Multipoles ($\ell = 0, 2$) & \checkmark & ---
      & $b_1, \tilde{c}_0, \tilde{c}_2, \tilde{c}_{\nabla^4_z \delta}, P_\mathrm{shot}$ \\
      EFT-3$w$ & IR-resummed EFT & 3 Wedges ($\left[0, 1/3 \right], \left[1/3, 2/3 \right], \left[2/3, 1 \right]$) & \checkmark & ---
      & $b_1, \tilde{c}_0, \tilde{c}_2, \tilde{c}_4, \tilde{c}_{\nabla^4_z \delta}, P_\mathrm{shot}$ \\
      EFT-2$w$ & IR-resummed EFT & 2 Wedges ($\left[0, 1/2 \right], \left[1/2, 1 \right]$) & \checkmark & ---
      & $b_1, \tilde{c}_0, \tilde{c}_2, \tilde{c}_{\nabla^4_z \delta}, P_\mathrm{shot}$ \\
      \hline \hline
    \end{tabular}
\end{table*}

\begin{table}
  \caption{Priors and fiducial values for cosmological and nuisance parameters.
  The symbol $\mathcal{N} (\mu, \sigma)$ is the Gaussian distribution
  with the mean $\mu$ and the standard deviation $\sigma$ and the symbol $\mathcal{U} (a, b)$
  is the flat distribution in the range of $(a, b)$.}
  \label{tab:priors}
    \begin{tabular}{cccc}
      \hline \hline
      Parameter & Prior & Fiducial value & Unit \\
      \hline
      $\omega_\mathrm{cdm}$ & $\mathcal{U} (0, \infty)$ & $0.1198$ & --- \\
      $\omega_\mathrm{b}$ & $\mathcal{N} (0.02225, 0.0005)$ & $0.02225$ & --- \\
      $n_\mathrm{s}$ & $\mathcal{N} (0.9645, 0.0049)$ & $0.9645$ & --- \\
      $h$ & $\mathcal{U} (0, \infty)$ & $0.6727$ & --- \\
      $\ln (10^{10} A_\mathrm{s})$ & $\mathcal{U} (0, \infty)$ & $3.094$ & --- \\
      $b_1$ & $\mathcal{U} (0, \infty)$ & $1$ & --- \\
      $\sigmad$, $\sigmav$ & $\mathcal{U} (0, \infty)$ & --- & $\hMpc$ \\
      $\gamma^{-1}$ & $\mathcal{U} (0, \infty)$ & --- & --- \\
      $P_\mathrm{shot}$ & $\mathcal{U} (0, \infty)$ & --- & $(\hMpc)^3$ \\
      $\tilde{c}_0$, $\tilde{c}_2$, $\tilde{c}_4$
      & $\mathcal{U} (-\infty, \infty)$ & --- & $(\hMpc)^2$ \\
      $\tilde{c}_{\nabla^4_z \delta}$
      & $\mathcal{U} (-\infty, \infty)$ & --- & $(\hMpc)^4$ \\
      \hline \hline
    \end{tabular}
\end{table}

\begin{table}
  \caption{Fiducial values for derived parameters.}
  \label{tab:derived}
    \begin{tabular}{cc}
      \hline \hline
      Parameter & Fiducial value \\
      \hline
      $b_1 \sigma_8$ & $0.5273$ \\
      $f \sigma_8$ & $0.4524$ \\
      $\alpha_\perp$ & $1$ \\
      $\alpha_\parallel$ & $1$ \\
      $F_\mathrm{AP}$ & $1.195$ \\
      $D_V / r_d$ & $19.48$ \\
      \hline \hline
    \end{tabular}
\end{table}

\section{Results}
\label{sec:results}
In this Section, we present the results of the PT challenge analysis.
First, as a demonstration of the accuracy of PT schemes,
we show power spectra with fiducial and best-fit cosmological parameters.
Next, in order to quantify the performance of the PT schemes,
we introduce three measures: Figure of Bias (FoB), Figure of Merit (FoM), and reduced chi-square.
FoB corresponds to the normalized distance between the inferred and fiducial cosmological parameters,
and FoM indicates the constraining power of parameters.
The reduced chi-square is the goodness of fit,
i.e. how close the PT scheme predictions are to the data.

Here, we only show results for primary models with a part of maximum wave-numbers.
The complete results including all models are found in Supplemental Material
\footnote{See Supplemental Material at [URL will be inserted by publisher]
for bestfit and fiducial spectra with all $\kmax$ and
bestfit values and constraints of cosmological and nuisance parameters
of all examined models.}.

\subsection{Fiducial and best-fit power spectra}
\begin{figure}
  \includegraphics[width=\columnwidth]{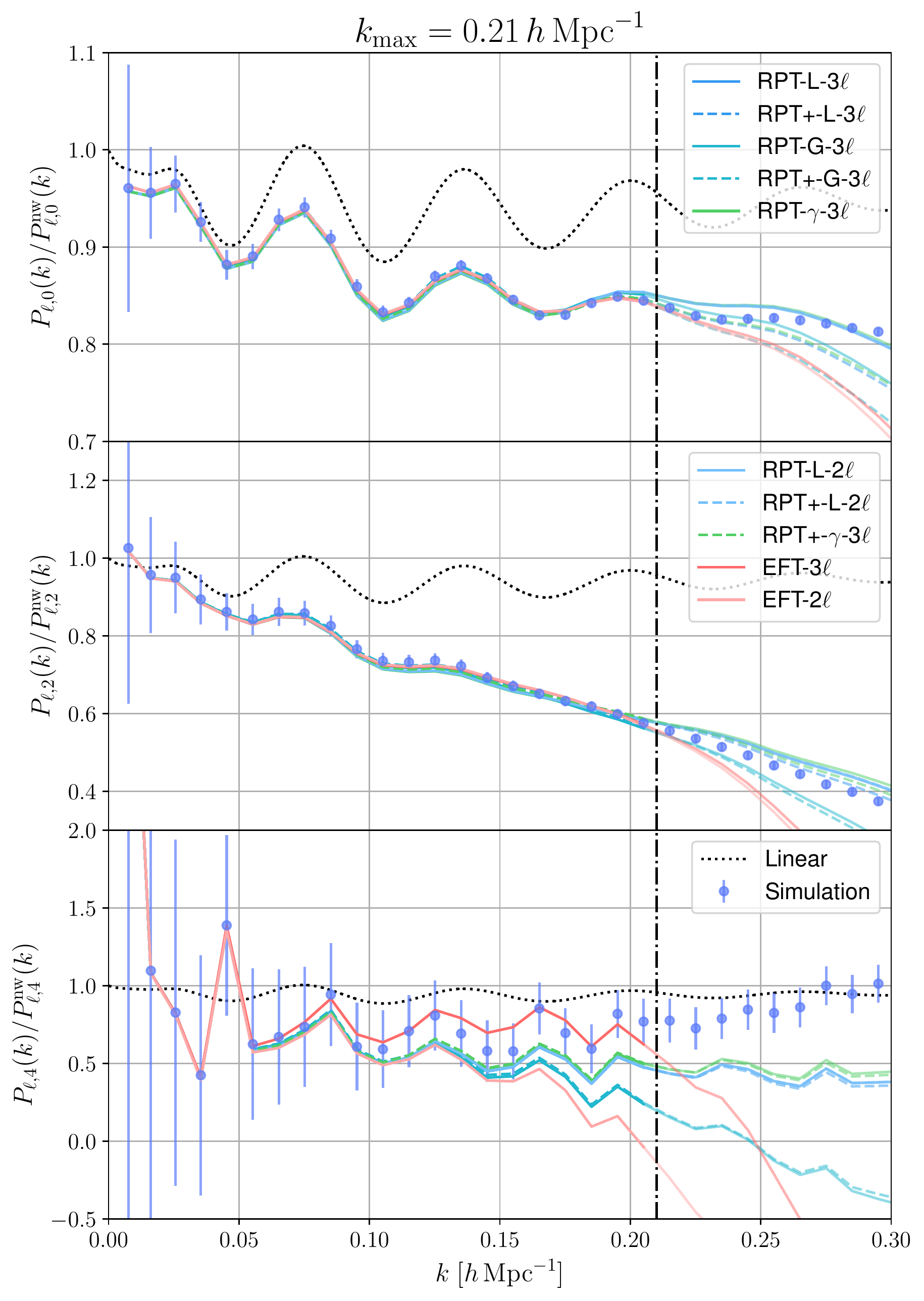}
  \caption{The monopole (upper panel), quadrupole (middle panel), and hexadecapole (lower panel) moments
  computed with fiducial cosmological parameters.
  The cyan points and black dotted lines correspond to the simulation result
  and the linear prediction, respectively.
  The nuisance parameters are fit to maximize the likelihood
  and the maximum wave-number of data points in the fitting is $\kmax = 0.21 \, \hMpcinv$,
  which is indicated as the black dot-dashed vertical line.
  Note that the hexadecapole moments of the models with 2 multipoles are not shown
  because they are not used for fitting.}
  \label{fig:fiducial_multipoles}
\end{figure}

\begin{figure}
  \includegraphics[width=\columnwidth]{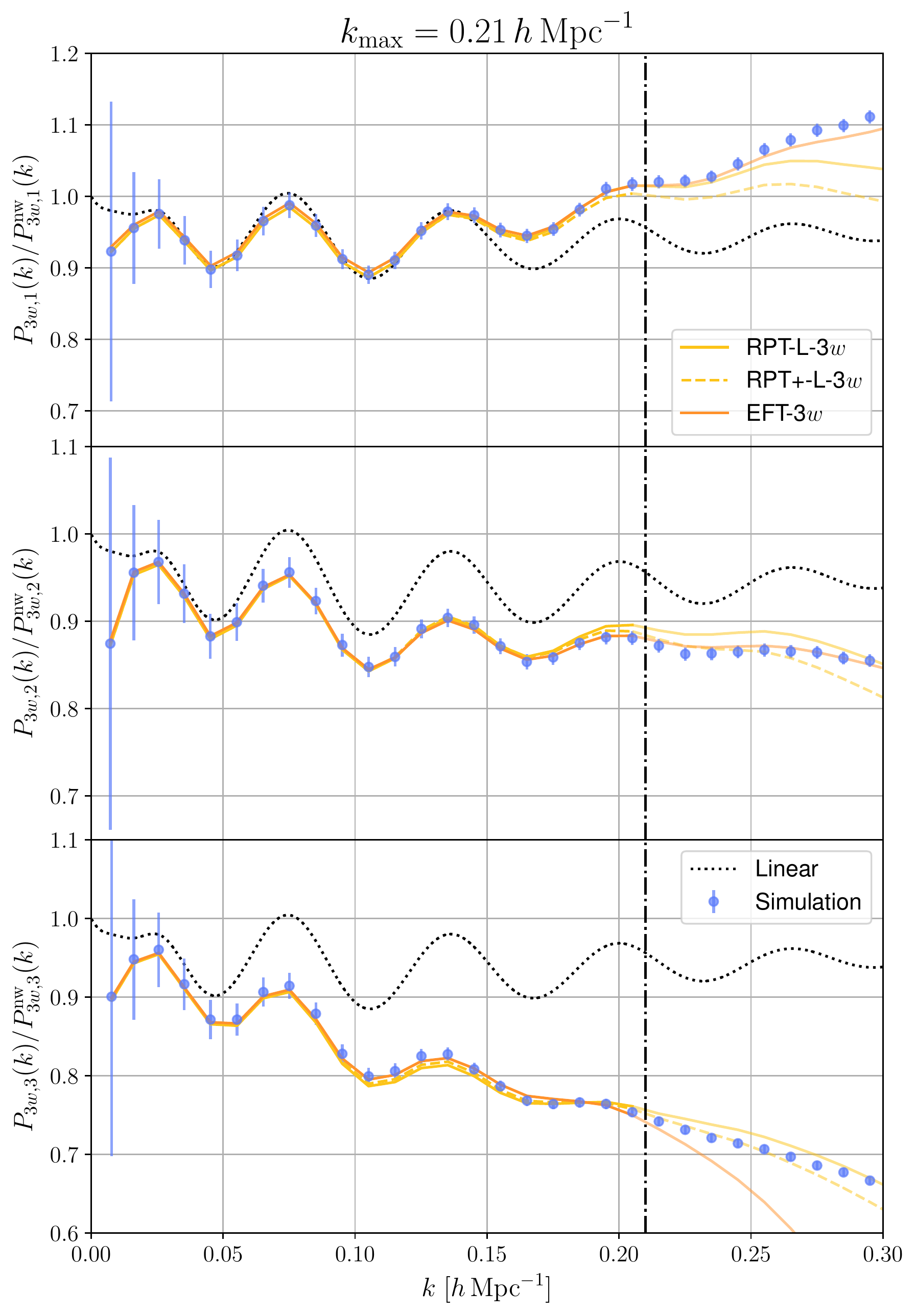}
  \caption{The same as Figure~\ref{fig:fiducial_multipoles} but for 3 wedges:
  $[0, 1/3]$ (upper panel), $[1/3, 2/3]$ (middle panel), and $[2/3, 1]$ (lower panel).}
  \label{fig:fiducial_3wedges}
\end{figure}

\begin{figure}
  \includegraphics[width=\columnwidth]{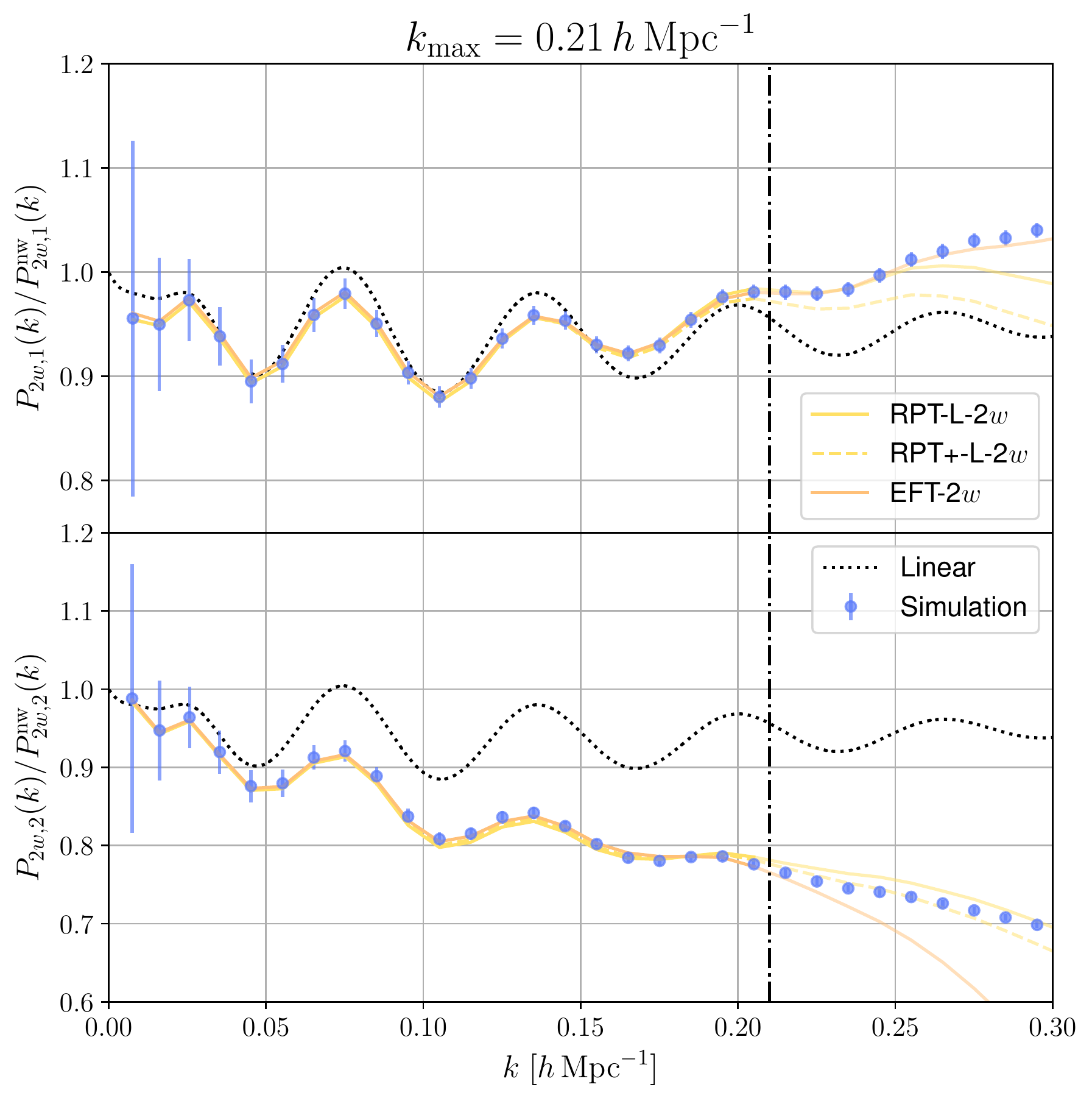}
  \caption{The same as Figure~\ref{fig:fiducial_multipoles} but for 2 wedges:
  $[0, 1/2]$ (upper panel) and $[1/2, 1]$ (lower panel).}
  \label{fig:fiducial_2wedges}
\end{figure}

Before the parameter inference, we present predictions with fiducial cosmological parameters,
which are used to generate the initial condition of the $N$-body simulations,
based on RegPT, RegPT+, and IR-resummed EFT.
The bias parameter is fixed as unity ($b_1 = 1$)
and other nuisance parameters are fit
with the likelihood functions defined in Eqs.~\eqref{eq:lkl_multipoles} and \eqref{eq:lkl_wedges}.
Figures~\ref{fig:fiducial_multipoles}, \ref{fig:fiducial_3wedges}, and \ref{fig:fiducial_2wedges}
show the fiducial multipoles, 3 wedges, and 2 wedges, respectively, in comparison with the simulation result.
The maximum wave-number is $\kmax = 0.21 \, \hMpcinv$ and the data points with $k < \kmax$
are used to fit nuisance parameters.
In general, all of the models yield good fits to the simulation spectra.
For multipoles, the difference is clear for hexadecapoles;
Gaussian FoG significantly underestimates the hexadecapole moment,
and the IR-resummed EFT can reproduce the hexadecopole the best among the models examined.
For 3 wedges and 2 wedges, there is an overshoot at large scale for IR-resummed EFT
because counter terms are adjusted to fit small-scale power, where the covariance is small,
and as a result, the accuracy at the large scale is degraded.

\begin{figure}
  \includegraphics[width=\columnwidth]{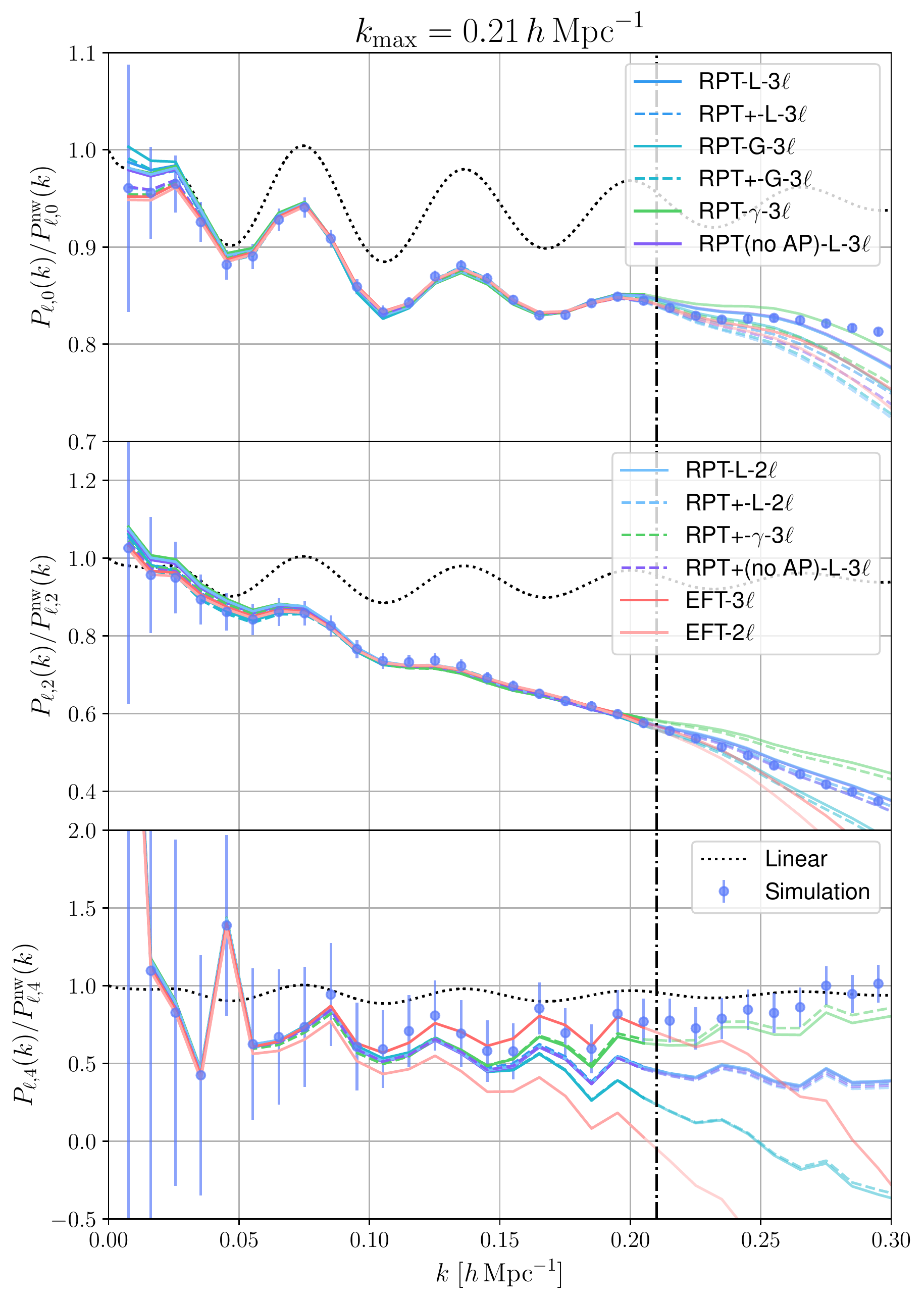}
  \caption{The monopole (upper panel), quadrupole (middle panel), and hexadecapole (lower panel) moments
  computed with best-fit parameters.
  The cyan points and black dotted lines correspond to the simulation result
  and the linear prediction, respectively.
  The cosmological and nuisance parameters are fit to maximize the posterior
  and the maximum wave-number of data points in the fitting is $\kmax = 0.21 \, \hMpcinv$,
  which is indicated as the black dot-dashed vertical line.
  Note that the hexadecapole moments of the models with 2 multipoles are not shown
  because they are not used for fitting.}
  \label{fig:bestfit_multipoles}
\end{figure}

\begin{figure}
  \includegraphics[width=\columnwidth]{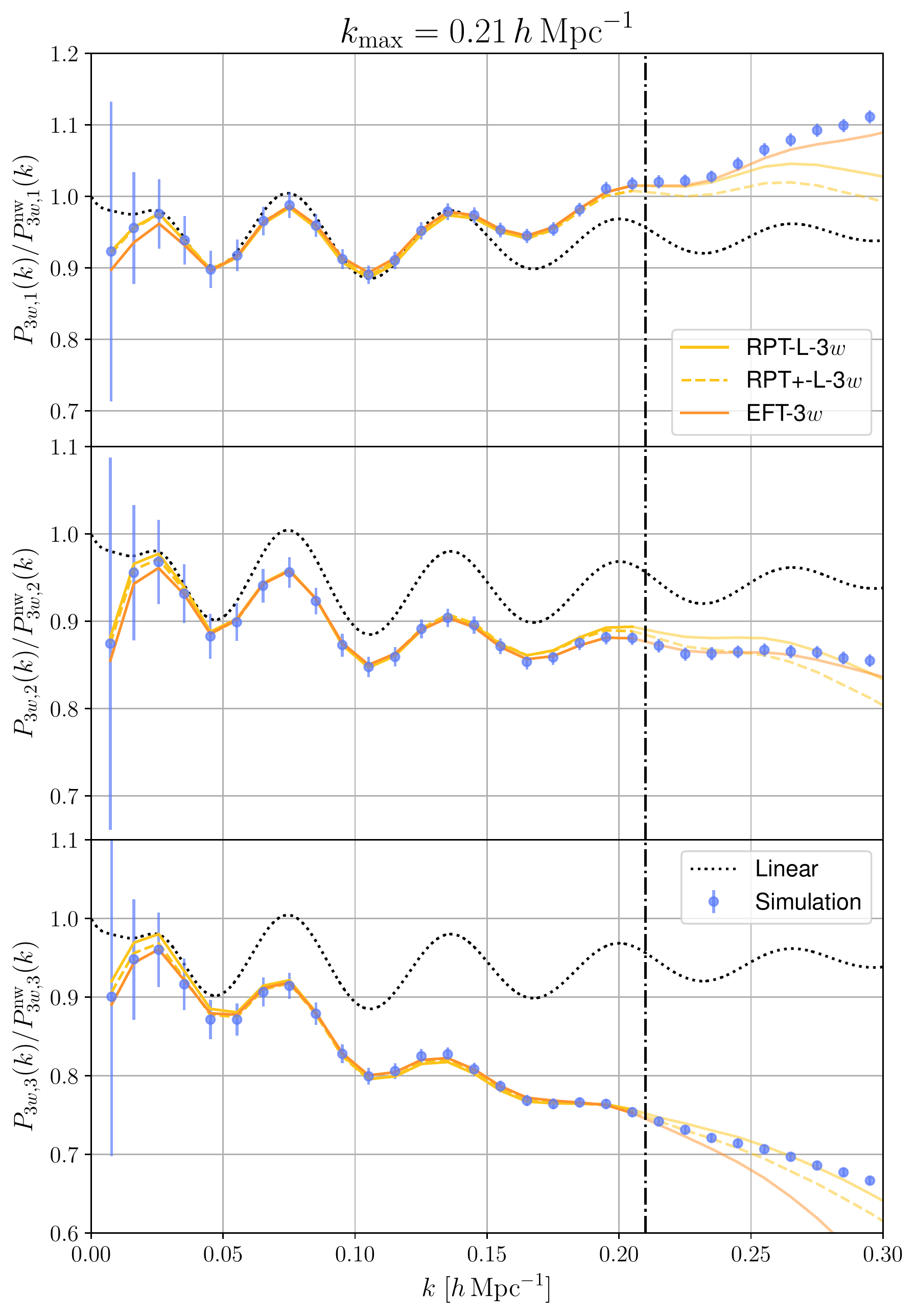}
  \caption{The same as Figure~\ref{fig:bestfit_multipoles} but for 3 wedges:
  $[0, 1/3]$ (upper panel), $[1/3, 2/3]$ (middle panel), and $[2/3, 1]$ (lower panel).}
  \label{fig:bestfit_3wedges}
\end{figure}

\begin{figure}
  \includegraphics[width=\columnwidth]{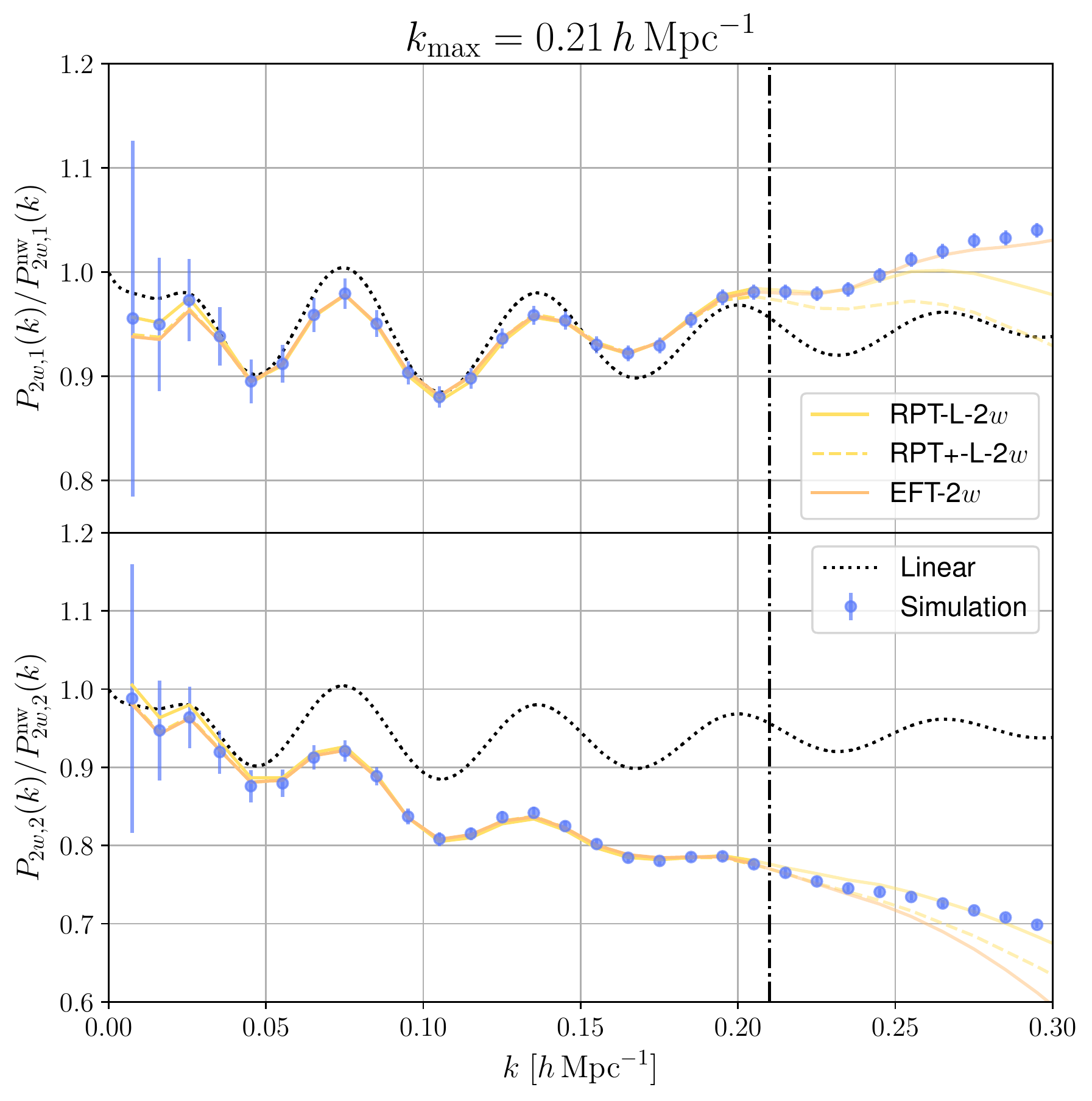}
  \caption{The same as Figure~\ref{fig:bestfit_multipoles} but for 2 wedges:
  $[0, 1/2]$ (upper panel) and $[1/2, 1]$ (lower panel).}
  \label{fig:bestfit_2wedges}
\end{figure}

Next, the cosmological parameters are varied and the posterior distributions
are inferred with the MCMC analysis.
We define the best-fit parameters as the ones which yield the maximum of the posterior
in the chain \footnote{We have confirmed that these parameters are consistent with
parameters derived by the optimization algorithm within $0.1\%$ level.}.
Figures~\ref{fig:bestfit_multipoles}, \ref{fig:bestfit_3wedges}, and \ref{fig:bestfit_2wedges}
show the best-fit multipoles, 3 wedges, and 2 wedges, respectively, in comparison with the simulation result.
The cosmological and nuisance parameters are determined to maximize
the posterior functions \footnote{In contrast to fiducial power spectra, the prior information on $\omega_\mathrm{b}$
and $n_\mathrm{s}$ is added in this fitting process.}.
At the cost of the large-scale power, most of models try to fit the small-scale power whose errors are small.
As a result, the best-fit power spectra are better matched with simulations
at the intermediate scale ($k = 0.1 \text{--} 0.2 \, \hMpcinv$)
compared with fiducial power spectra.

\subsection{Constraints on cosmological and derived parameters}
\begin{figure*}
  \includegraphics[width=0.8\textwidth]{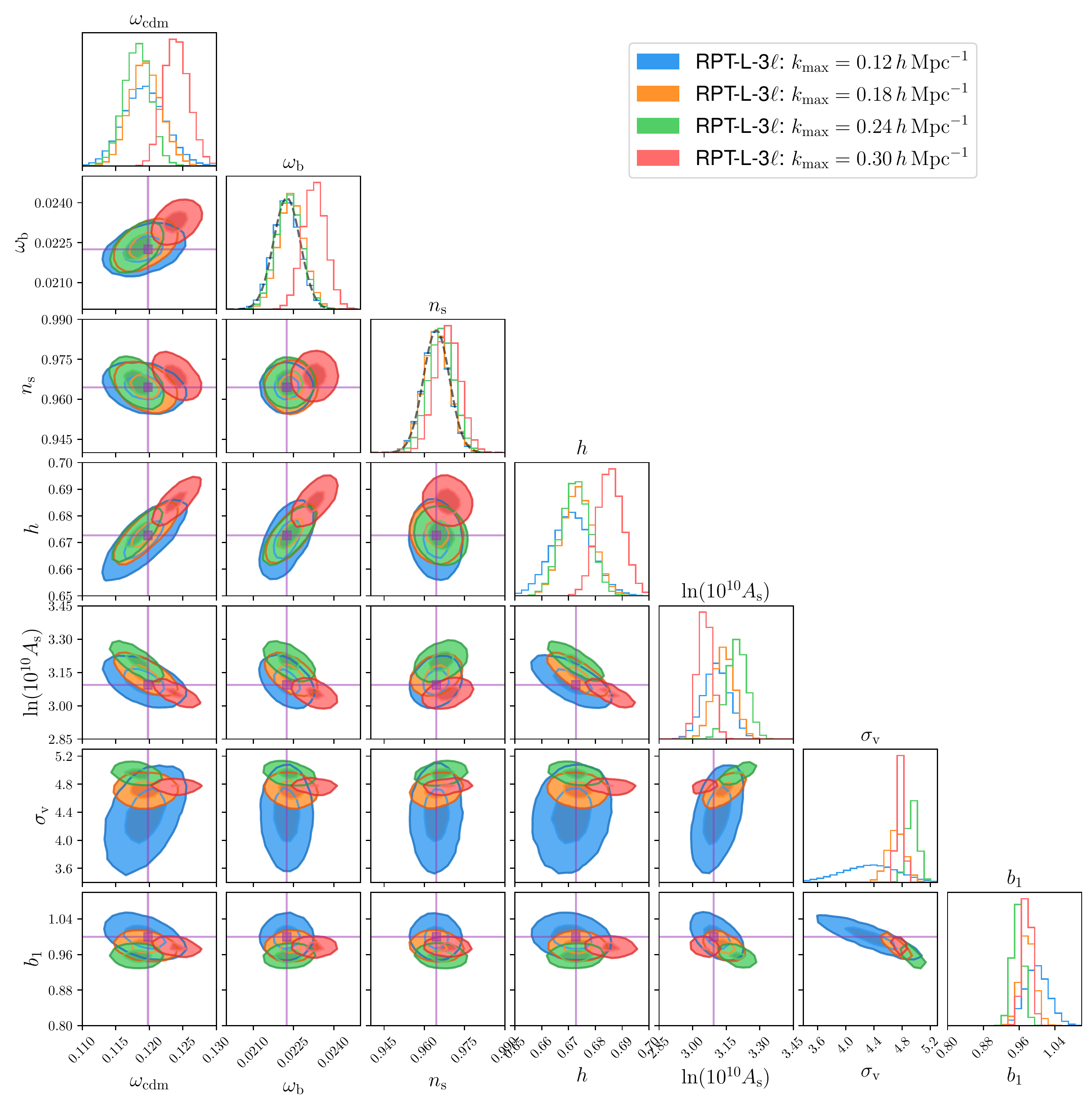}
  \caption{Constraints of cosmological and nuisance parameters
  inferred with RegPT with Lorentzian FoG and 3 multipoles
  for $\kmax = 0.12, 0.18, 0.24, 0.30 \, \hMpcinv$.
  The inner and outer ellipses correspond to the $1\text{-}\sigma$ and $2\text{-}\sigma$
  confidence levels, respectively.
  The purple lines correspond to fiducial values.
  The prior distributions of $\omega_\mathrm{b}$ and $n_\mathrm{s}$ are shown as black dashed lines.
  The unit of $\sigmav$ is $\hMpc$.}
  \label{fig:C1_triangle}
\end{figure*}

\begin{figure*}
  \includegraphics[width=0.8\textwidth]{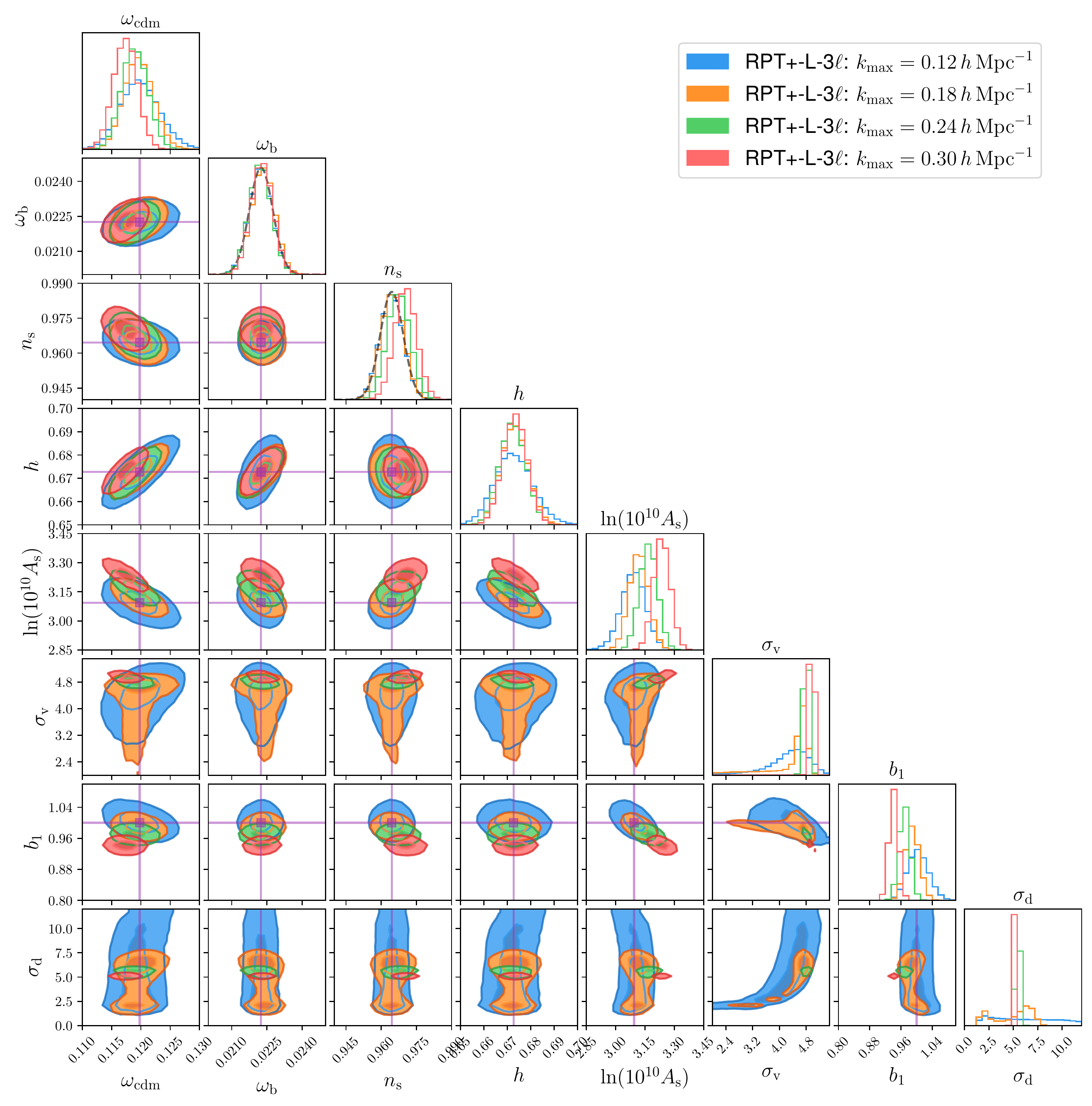}
  \caption{Same as Figure~\ref{fig:C1_triangle} but for RegPT+ with Lorentzian FoG and 3 multipoles.
  The unit of $\sigmav$ and $\sigmad$ is $\hMpc$.}
  \label{fig:C2_triangle}
\end{figure*}

\begin{figure*}
  \includegraphics[width=\textwidth]{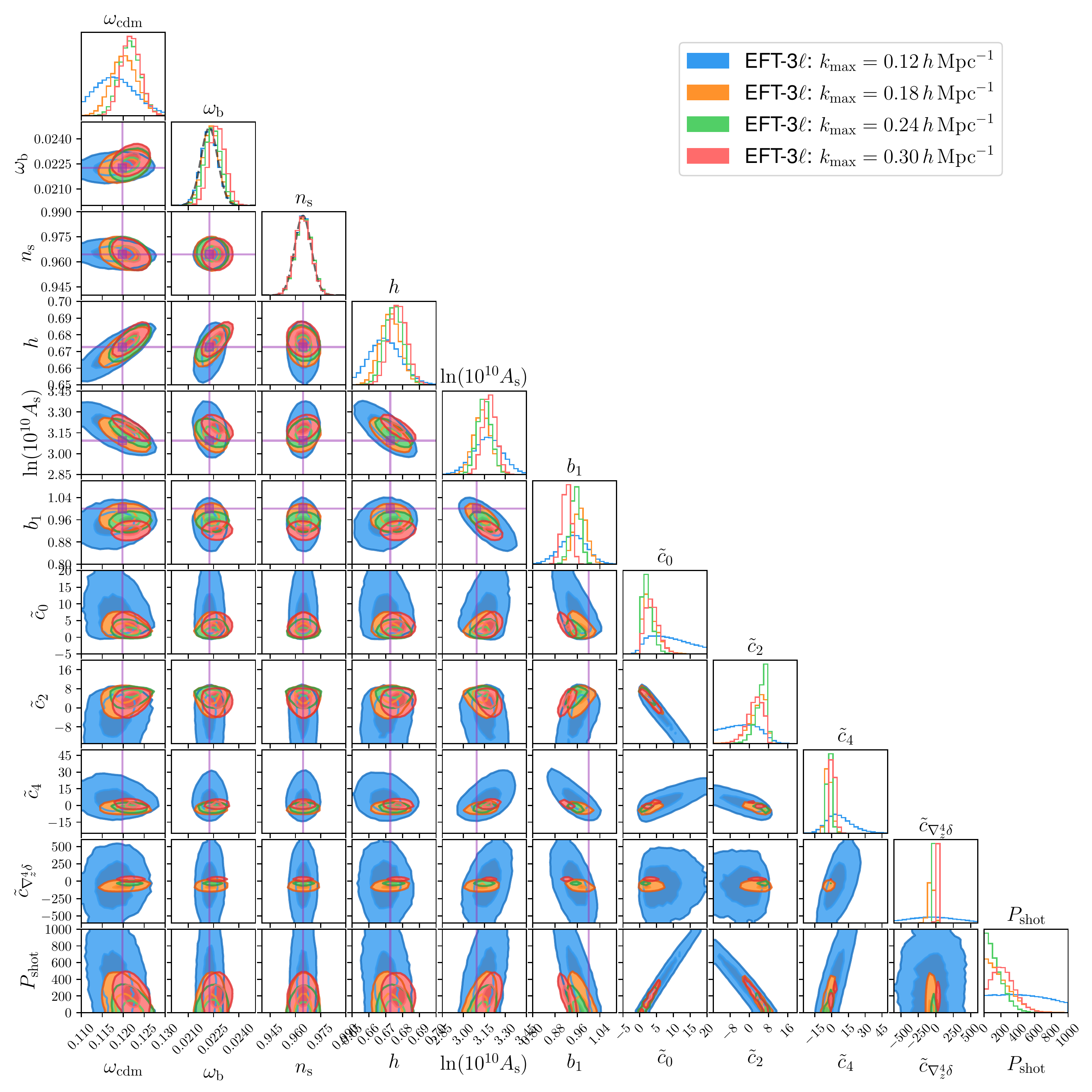}
  \caption{Same as Figure~\ref{fig:C1_triangle} but for IR-resummed EFT with 3 multipoles.
  For the units of $\tilde{c}_0$, $\tilde{c}_2$, $\tilde{c}_4$, $\tilde{c}_{\nabla^4_z \delta}$,
  and $P_\mathrm{shot}$, refer to Table~\ref{tab:priors}.}
  \label{fig:D4_triangle}
\end{figure*}

Here, we present examples of parameter inference results.
Figures~\ref{fig:C1_triangle}, \ref{fig:C2_triangle}, and \ref{fig:D4_triangle}
show constraints of cosmological and nuisance parameters for RegPT, RegPT+ and IR-resummed EFT, respectively.
The FoG damping function for RegPT and RegPT+ is Lorentzian and the data vector is 3 multipoles.
As the general trend, the parameter constraints become tighter
for larger $\kmax$ since more small-scale data are incorporated
in the parameter inference.
On the other hand, when the aggressive $\kmax$, e.g.,
$\kmax = 0.30 \, \hMpcinv$ for RegPT, is chosen,
the validity of the PT scheme ceases to be adequate
due to a stronger non-linearity in the power spectra, and the predictions are no longer reliable.
As a result, the inferred parameters are strongly biased.
This feature is common in all the models.

\begin{figure}
  \includegraphics[width=\columnwidth]{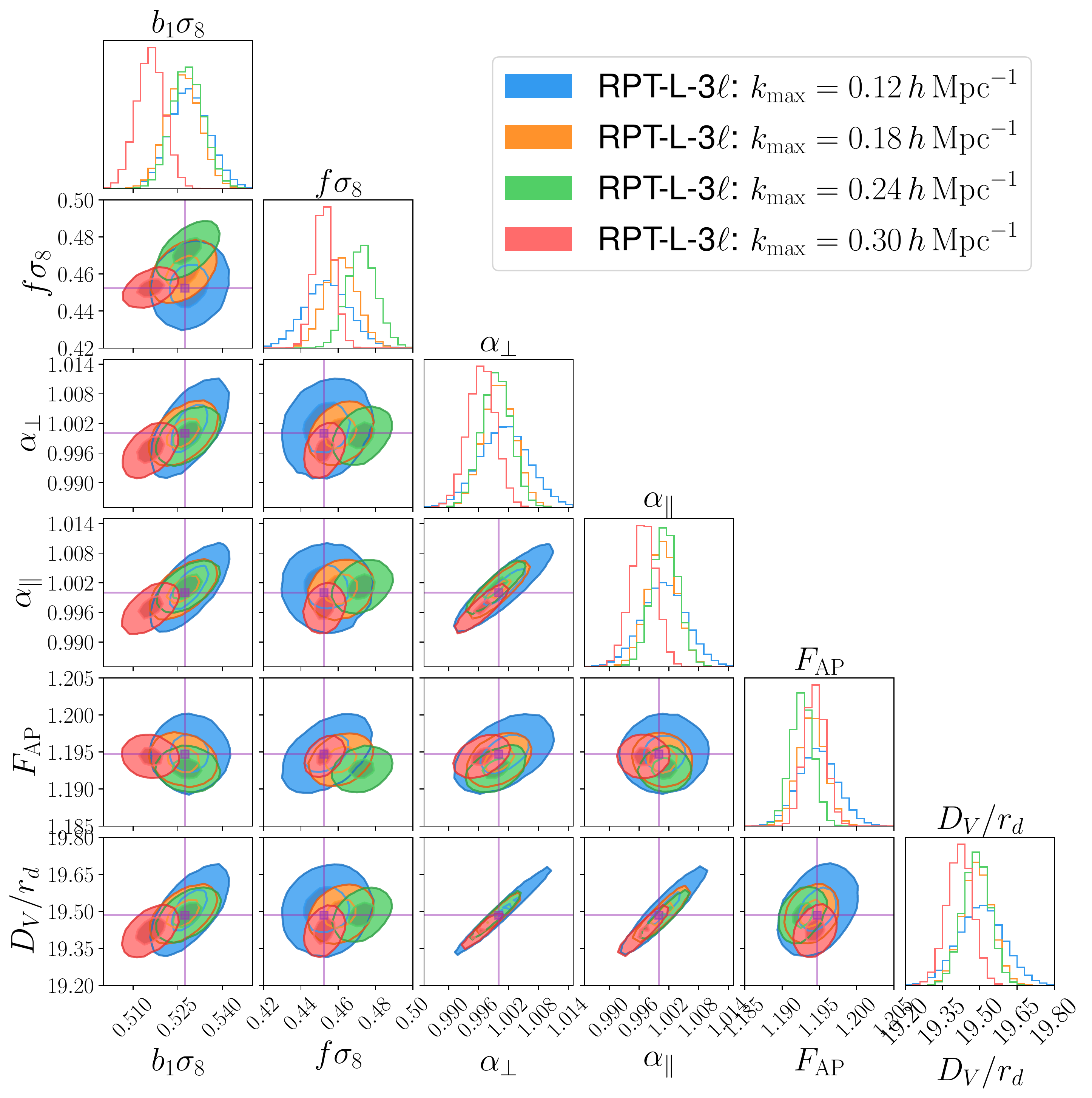}
  \caption{Constraints of derived parameters inferred with RegPT with 3 multipoles.
  for $\kmax = 0.12, 0.18, 0.24, 0.30 \, \hMpcinv$.
  The inner and outer ellipses correspond to the $1\text{-}\sigma$ and $2\text{-}\sigma$
  confidence levels, respectively.
  The purple lines correspond to fiducial values.}
  \label{fig:C1_triangle_derived}
\end{figure}

\begin{figure}
  \includegraphics[width=\columnwidth]{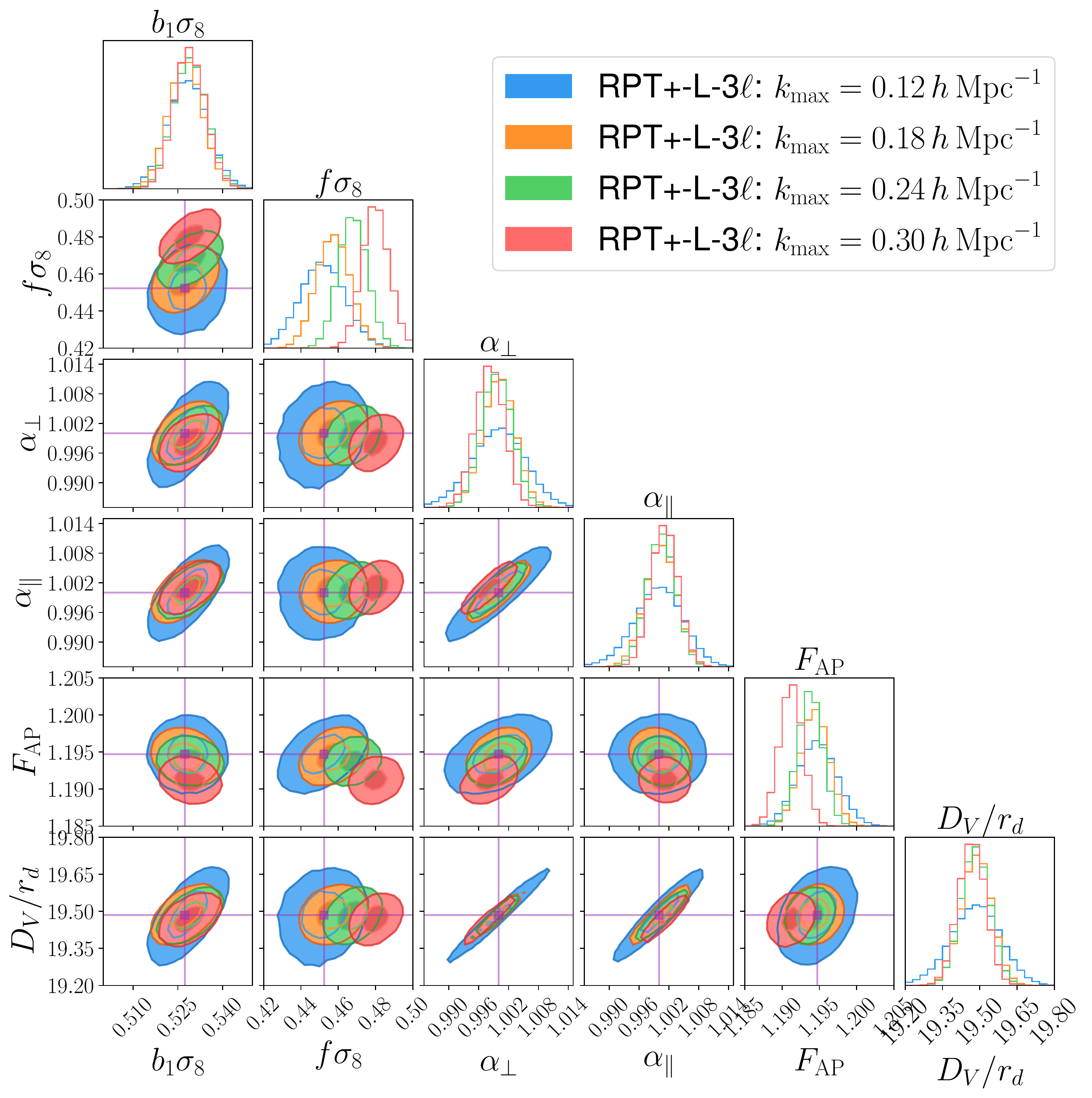}
  \caption{Same as Figure~\ref{fig:C1_triangle_derived} but RegPT+ with Lorentzian FoG and 3 multipoles.}
  \label{fig:C2_triangle_derived}
\end{figure}

\begin{figure}
  \includegraphics[width=\columnwidth]{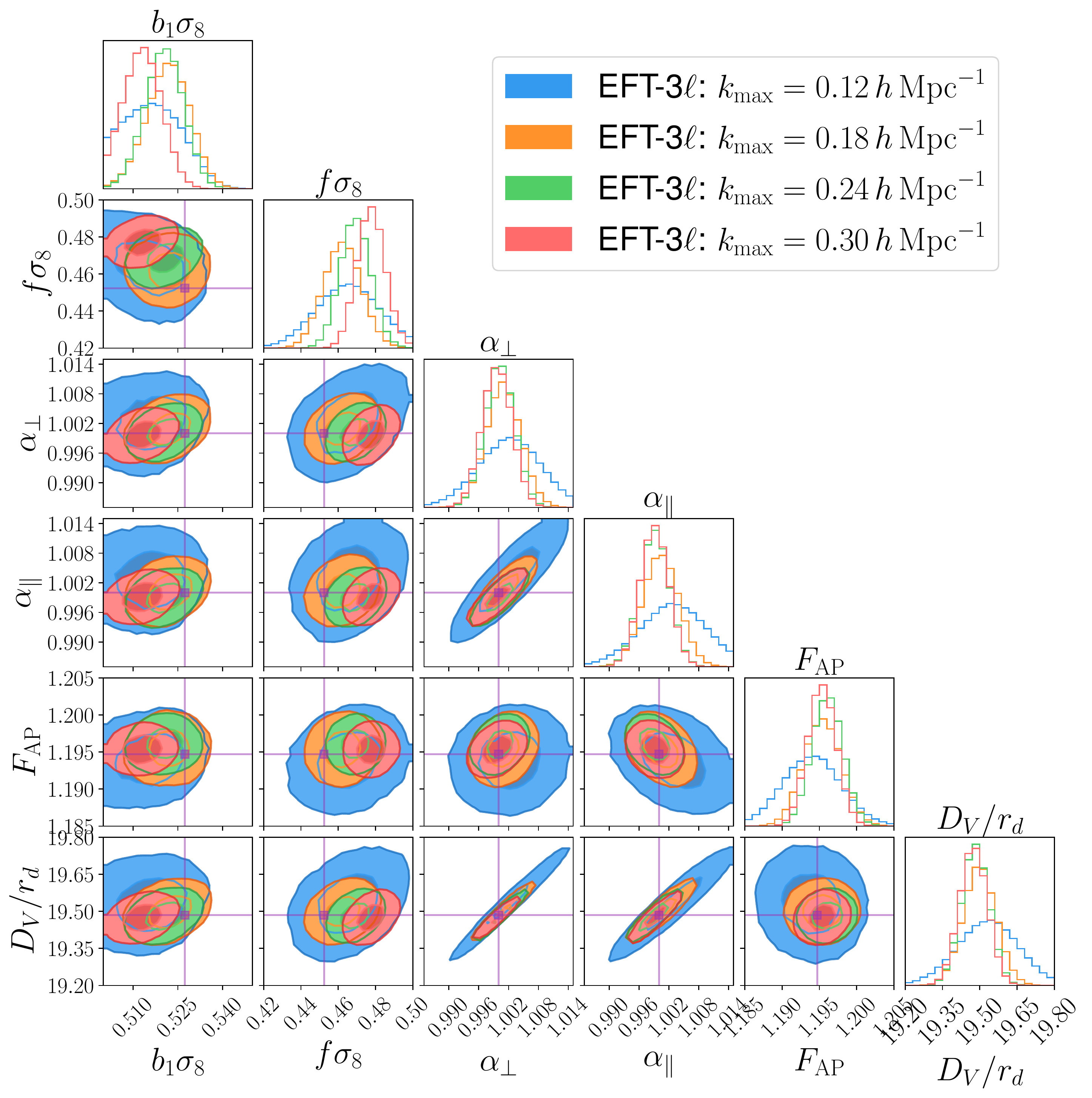}
  \caption{Same as Figure~\ref{fig:C1_triangle_derived} but for IR-resummed EFT with 3 multipoles.}
  \label{fig:D4_triangle_derived}
\end{figure}

Next, Figures~\ref{fig:C1_triangle_derived}, \ref{fig:C2_triangle_derived}, and \ref{fig:D4_triangle_derived}
show constraints on derived parameters.
Overall, the similar feature to cosmological parameters appears.
At low $\kmax$, the parameters are consistent with fiducial values but the constraining power is weak.
On the other hand, increasing high $\kmax$ leads to tight constraints but biased inference
because the PT schemes become less reliable.
This trend is clearer for $b_1 \sigma_8$ and $f \sigma_8$, which correspond to the amplitudes of
density and velocity power spectra. These parameters are sensitive to the small-scale power spectra,
and strongly biased with aggressive $\kmax$.
In contrast, AP parameters $(\alpha_\perp, \alpha_\parallel, F_\mathrm{AP})$ and the distance scale ($D_V / r_d$)
are more robustly determined even with high $\kmax$.
These parameters are constrained with the geometry information at large scales, and thus,
the parameter bias is less significant even for high $\kmax$.

\subsection{Figure of Bias, Figure of Merit, and reduced chi-square}
Here, we discuss the measures to quantify the parameter bias, the constraining power, and the goodness of fit.
First, we compute the covariance matrix of parameters $S$ from the chains:
\begin{equation}
S_{\alpha \beta} = \frac{1}{N-1} \sum_{i=1}^N (\theta_\alpha^i - \bar{\theta}_\alpha)
(\theta_\beta^i - \bar{\theta}_\beta) ,
\end{equation}
where $N$ is the number of samples in the chain and $\bar{\bm{\theta}}$ is the sample mean of the parameter.
In order to evaluate the parameter bias, we define Figure of Bias (FoB):
\begin{equation}
\mathrm{FoB} \equiv \left[ \sum_{\alpha, \beta} \delta \theta_\alpha
\left( \tilde{S} \right)^{-1}_{\alpha \beta} \delta \theta_\beta \right]^{\frac{1}{2}} ,
\end{equation}
where the covariance matrix $\tilde{S}$ is marginalized over all nuisance parameters
from the full parameter covariance matrix $S$
and $\delta \bm{\theta} = \bar{\bm{\theta}} - \bm{\theta}^\mathrm{fid}$ is
the difference between the sample mean and the fiducial parameter.
The probability distribution of FoB is the chi-squared distribution with the degree of freedom of $5$.
Then, we define the $1\text{-}\sigma$, $2\text{-}\sigma$, and $3\text{-}\sigma$
critical values of FoB, which are $68.3$, $95.5$, and $99.7$ percentiles, respectively.

Next, in order to evaluate the constraining power, we define the Figure of Merit (FoM):
\begin{equation}
\mathrm{FoM} \equiv \frac{V_\theta}{\sqrt{\det \tilde{S}}} ,
\end{equation}
where $V_\theta = \prod_\alpha \theta_\alpha^{\mathrm{fid}}$ is the normalization factor
and the subscript $\alpha$ runs over five cosmological parameters.
The FoM is roughly proportional to the hyper volume of $1\text{-}\sigma$ confidence region
and thus, larger FoM implies stronger constraining power.

The measure of the goodness of fit is the reduced chi-square $\chi^2 / N_\mathrm{dof}$,
where $\chi^2$ is the best-fit chi-square (Eqs.~\ref{eq:chi2_multipoles} and \ref{eq:chi2_wedges})
and $N_\mathrm{dof}$ is the effective degree of freedom.
We have introduced priors of several parameters and
the information content of the priors should be considered
in the degree of freedom.
The effective number of parameters $N_\mathrm{eff}$ is given as \cite{Raveri2019}
\begin{equation}
  N_\mathrm{eff} = N_p - \mathrm{tr} [\mathcal{C}_\mathrm{prior}^{-1} \mathcal{C}] ,
\end{equation}
where $N_p$ is the number of parameters (cosmological parameters plus nuisance parameters),
and $\mathcal{C}$ and $\mathcal{C}_\mathrm{prior}$ is the covariance matrix
of posterior and prior parameters, respectively.
Since all the adopted priors are Gaussian, the expression of the second term can be simplified as
\begin{equation}
  \mathrm{tr} [\mathcal{C}_\mathrm{prior}^{-1} \mathcal{C}]
  = \sum_{\alpha} \frac{\hat{\sigma}^2_\alpha}{\sigma^2_\alpha},
\end{equation}
where the summation runs over parameters with priors
($\alpha = \omega_\mathrm{b}, n_\mathrm{s}$),
$\hat{\sigma}^2_\theta$ is the sample variance of the parameter computed from the chain,
and $\sigma^2_\theta$ is the variance of the prior.
As a result, the degree of freedom is given as
\begin{equation}
  N_\mathrm{dof} = N_\mathrm{data} - N_\mathrm{eff},
\end{equation}
where $N_\mathrm{data}$ is the number of data points.
Note that the reduced chi-square in this analysis does not follow the chi-square distribution
because we have generated the initial condition of the $N$-body simulation
with the paired and fixed approach.
Therefore, the cosmic variance is strongly suppressed and the reduced chi-square is always close to zero.

\begin{figure*}
  \includegraphics[width=\columnwidth]{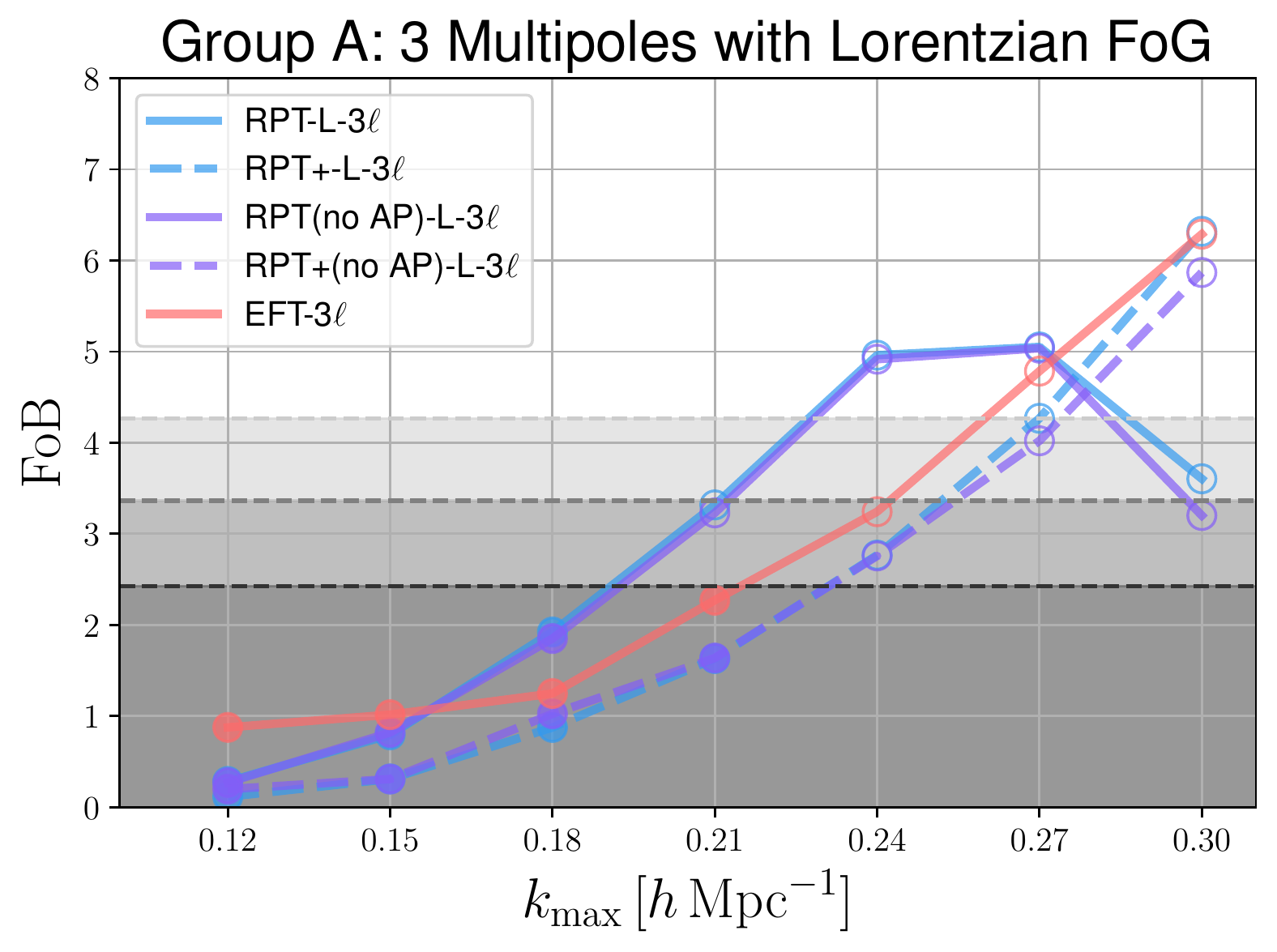}
  \includegraphics[width=\columnwidth]{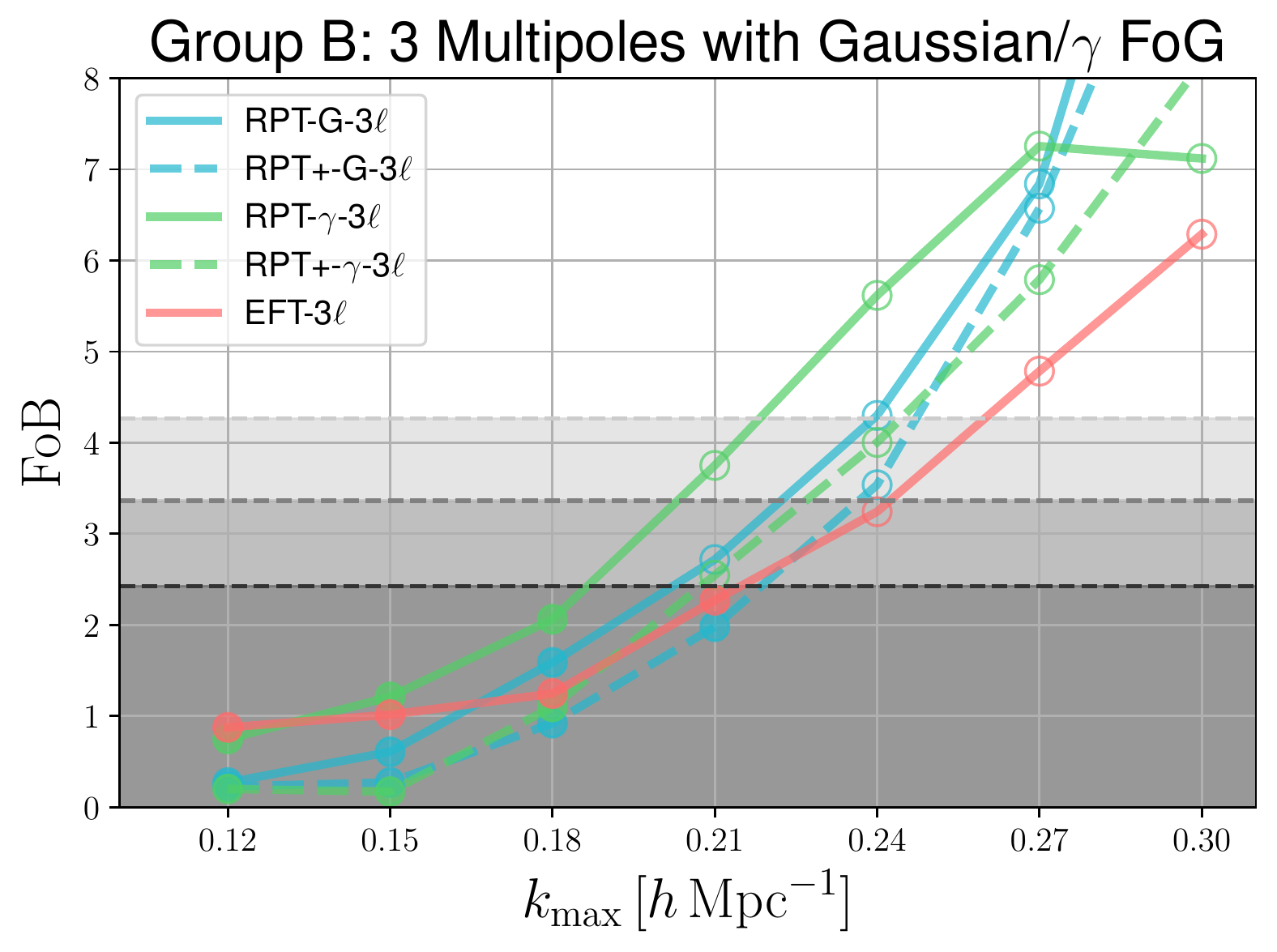}
  \includegraphics[width=\columnwidth]{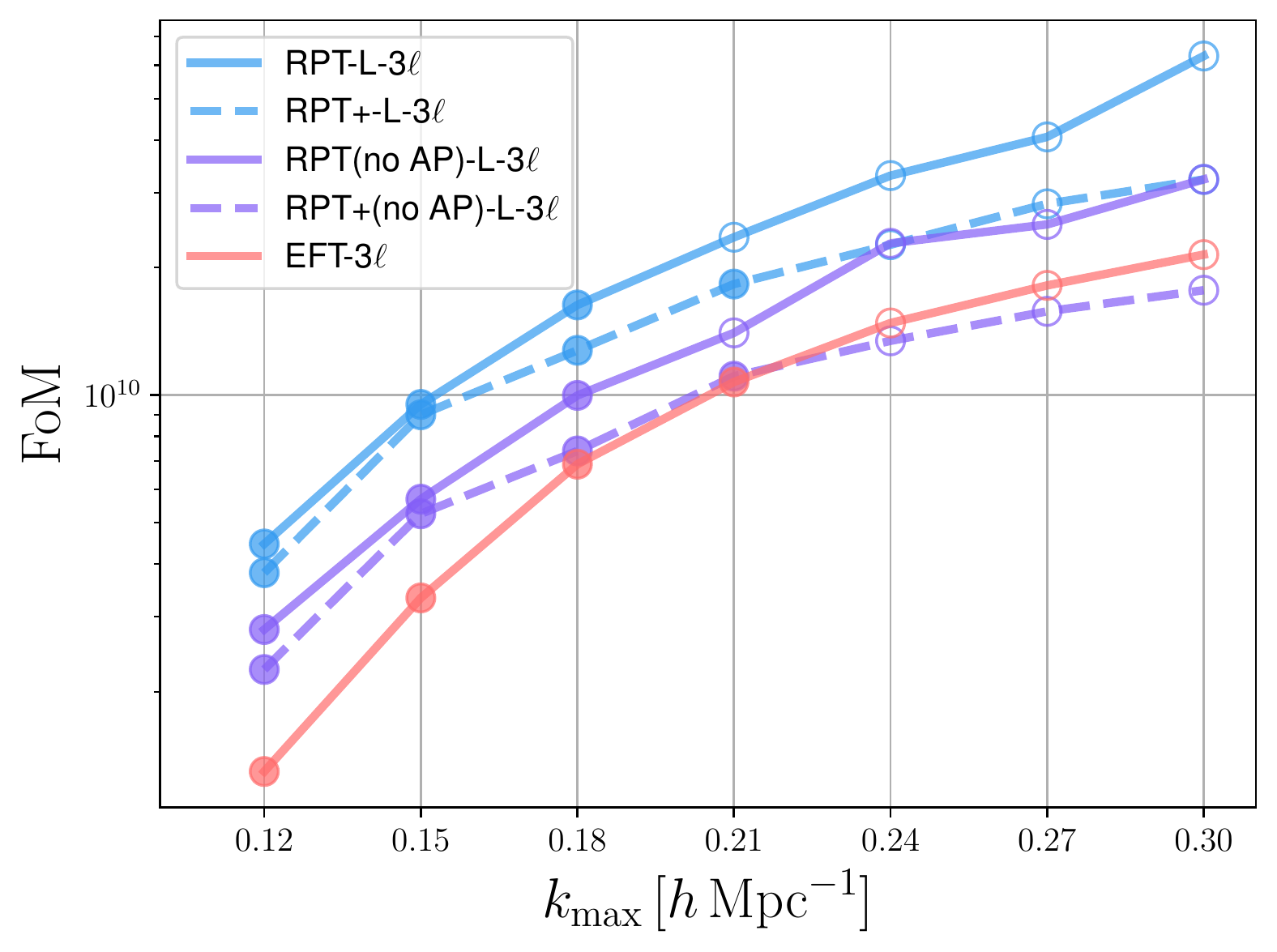}
  \includegraphics[width=\columnwidth]{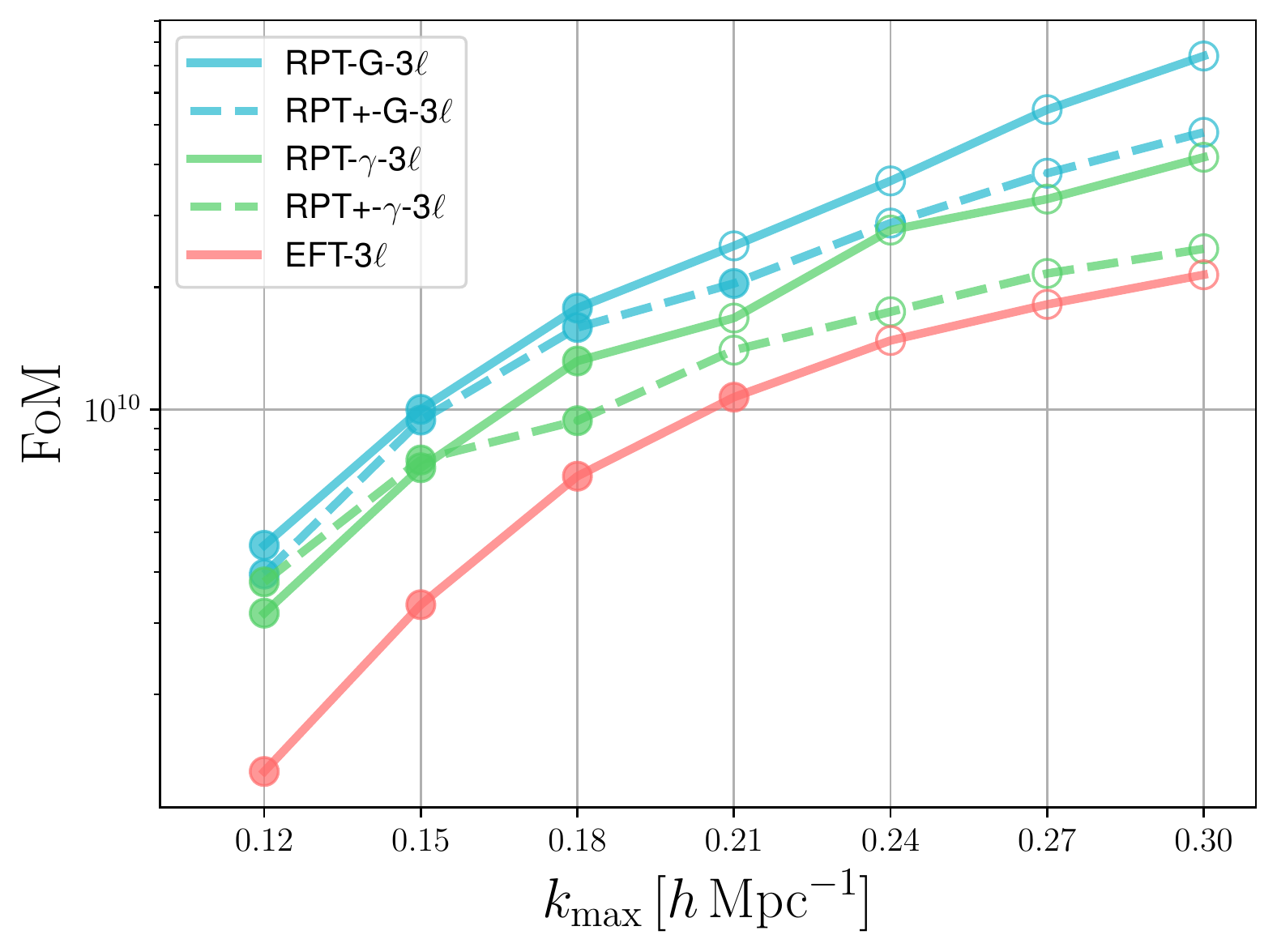}
  \includegraphics[width=\columnwidth]{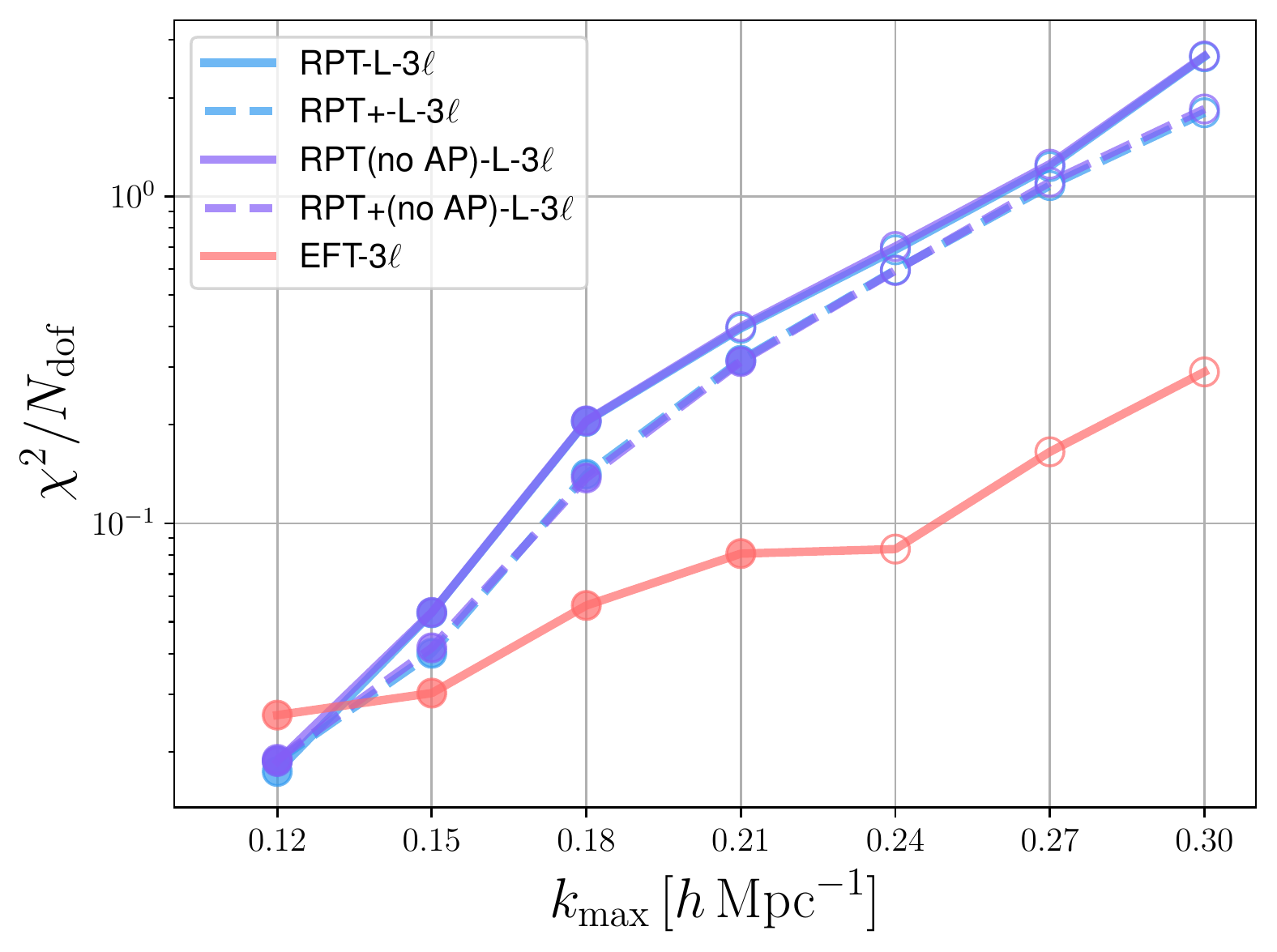}
  \includegraphics[width=\columnwidth]{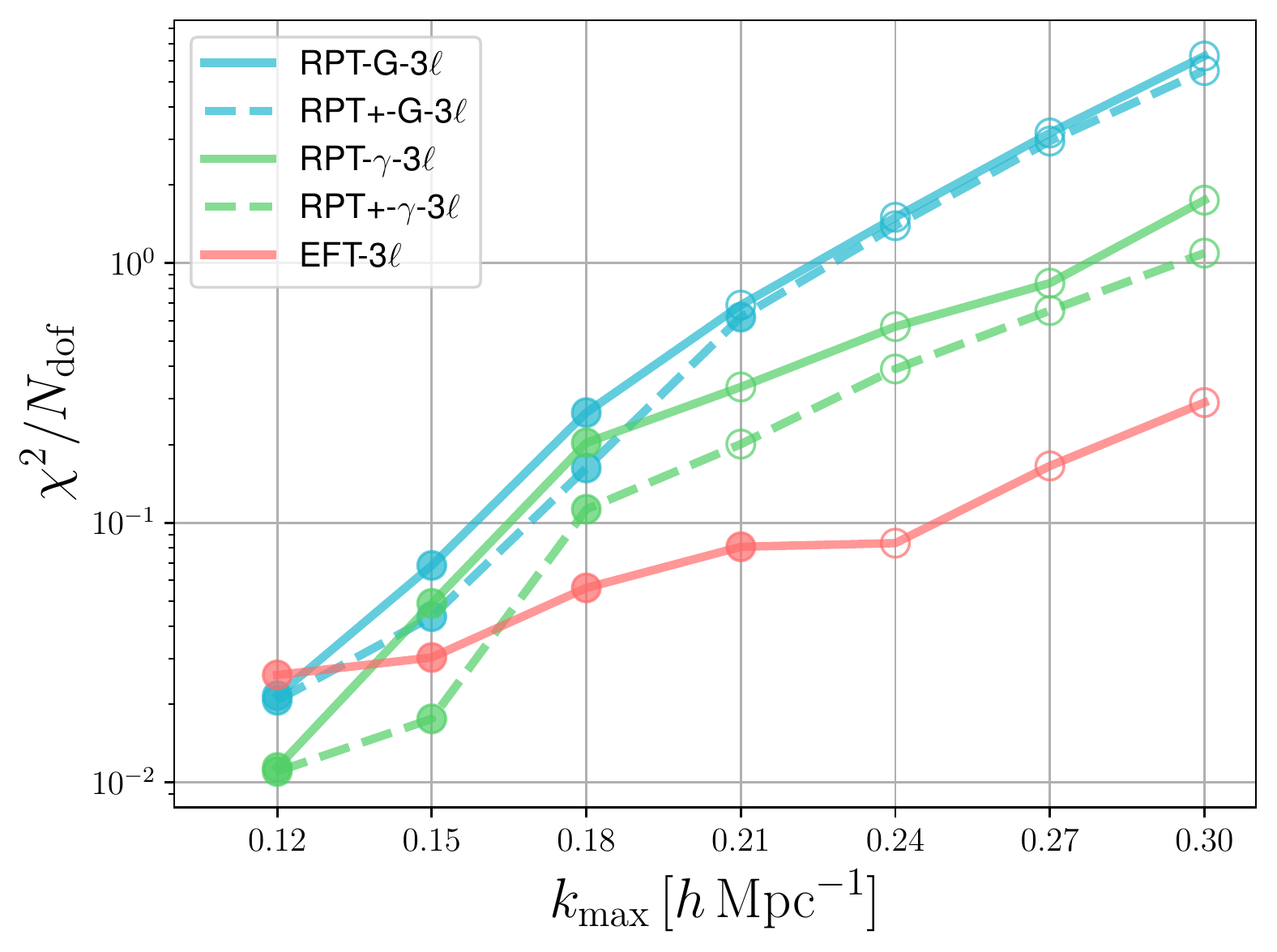}
  \caption{The FoB (upper panels), FoM (middle panels), and reduced chi-square (lower panels)
  for Group A (left panels) and Group B (right panels).
  For FoB, the gray regions correspond to $1\text{-}\sigma$, $2\text{-}\sigma$, and $3\text{-}\sigma$
  critical values from bottom to top.}
  \label{fig:measures_GroupAB}
\end{figure*}

\begin{figure*}
  \includegraphics[width=\columnwidth]{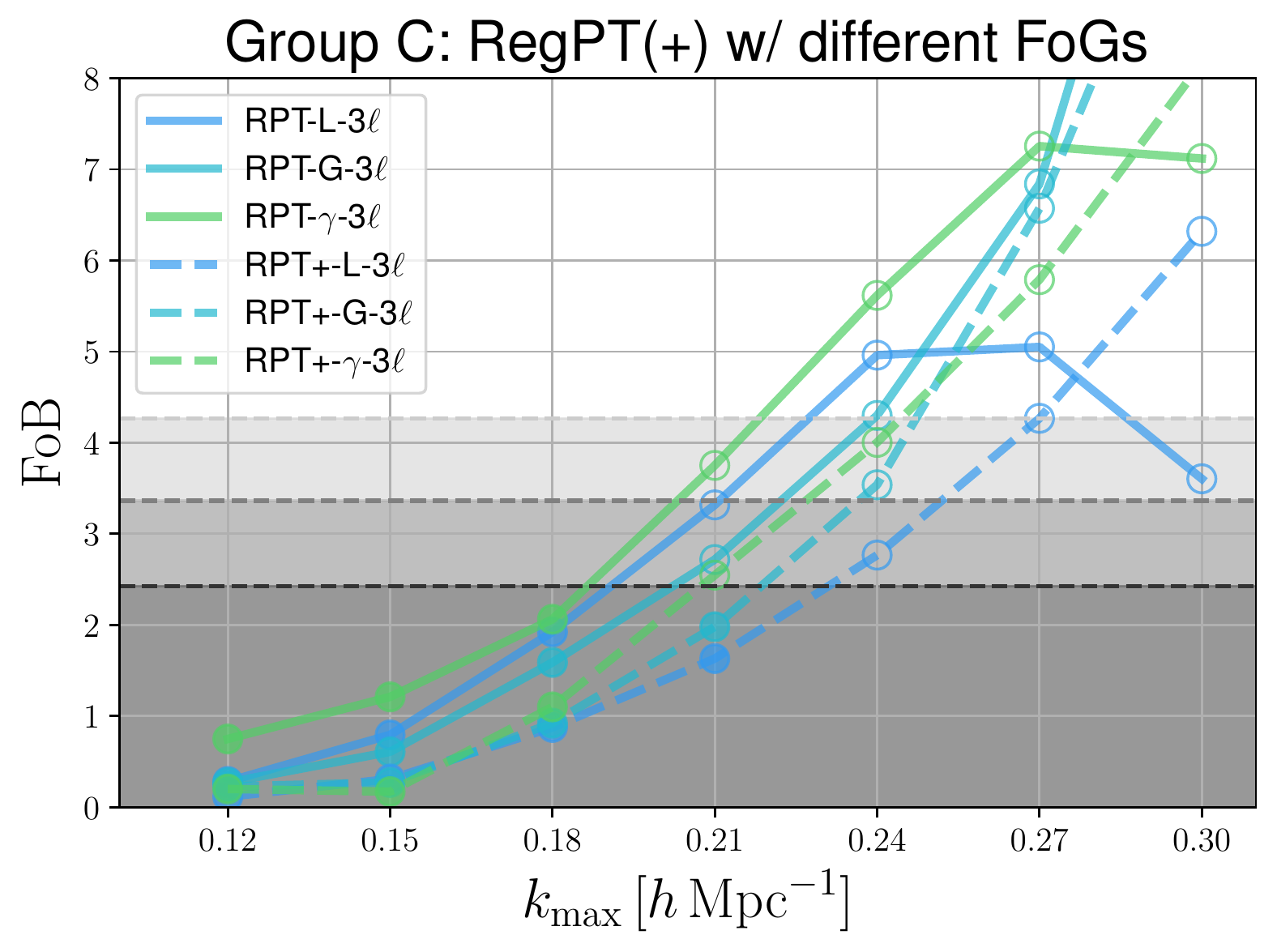}
  \includegraphics[width=\columnwidth]{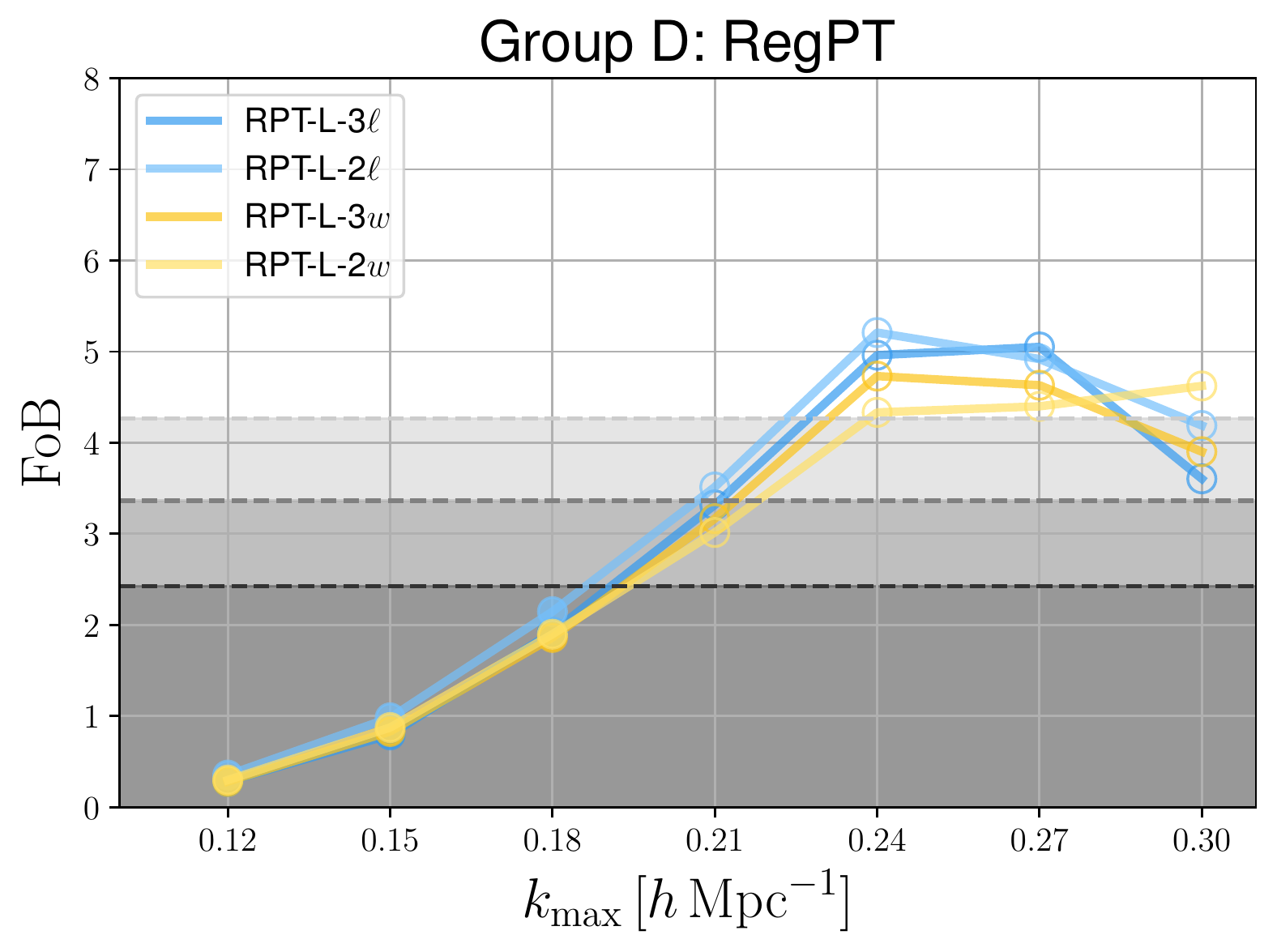}
  \includegraphics[width=\columnwidth]{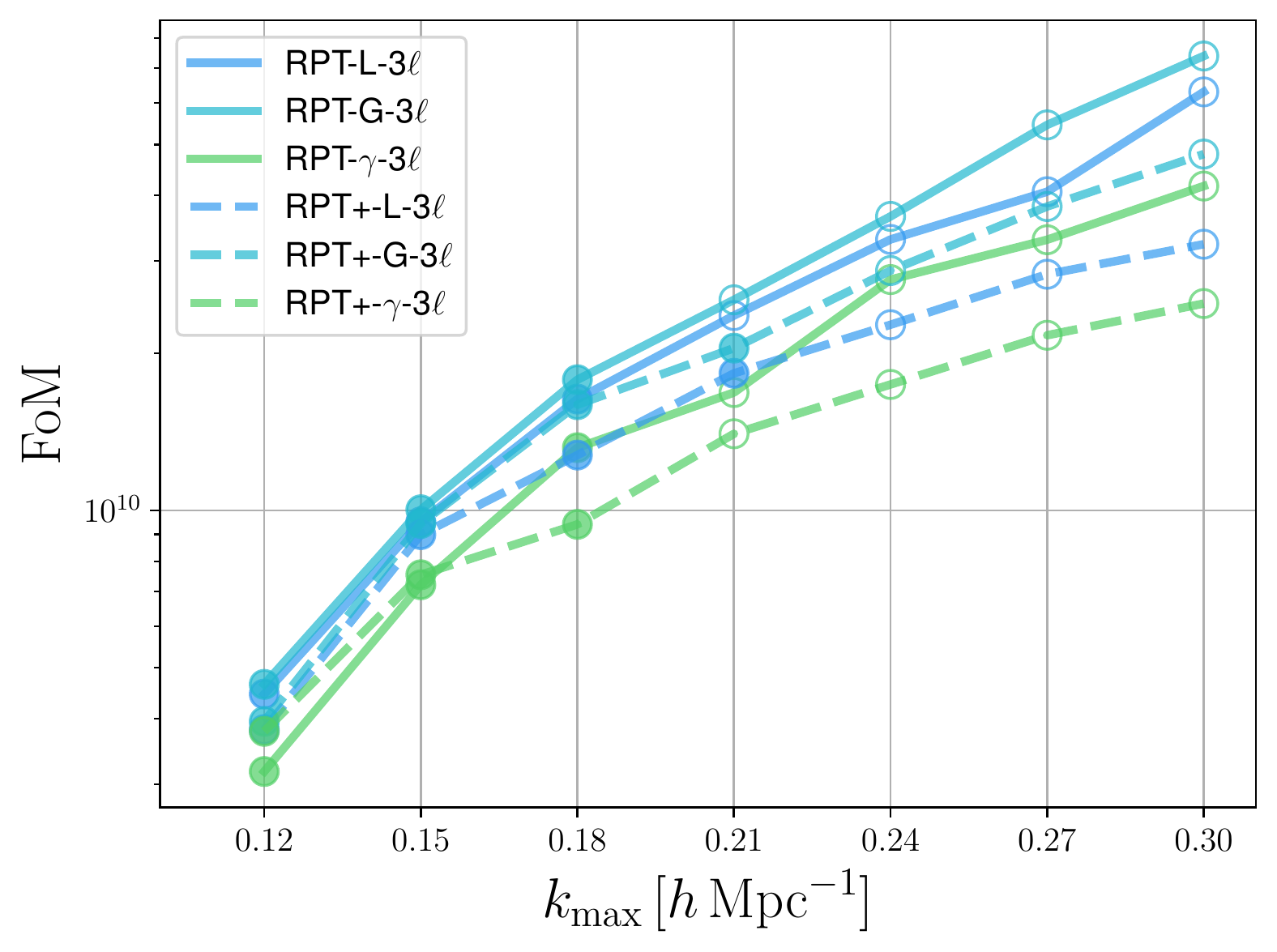}
  \includegraphics[width=\columnwidth]{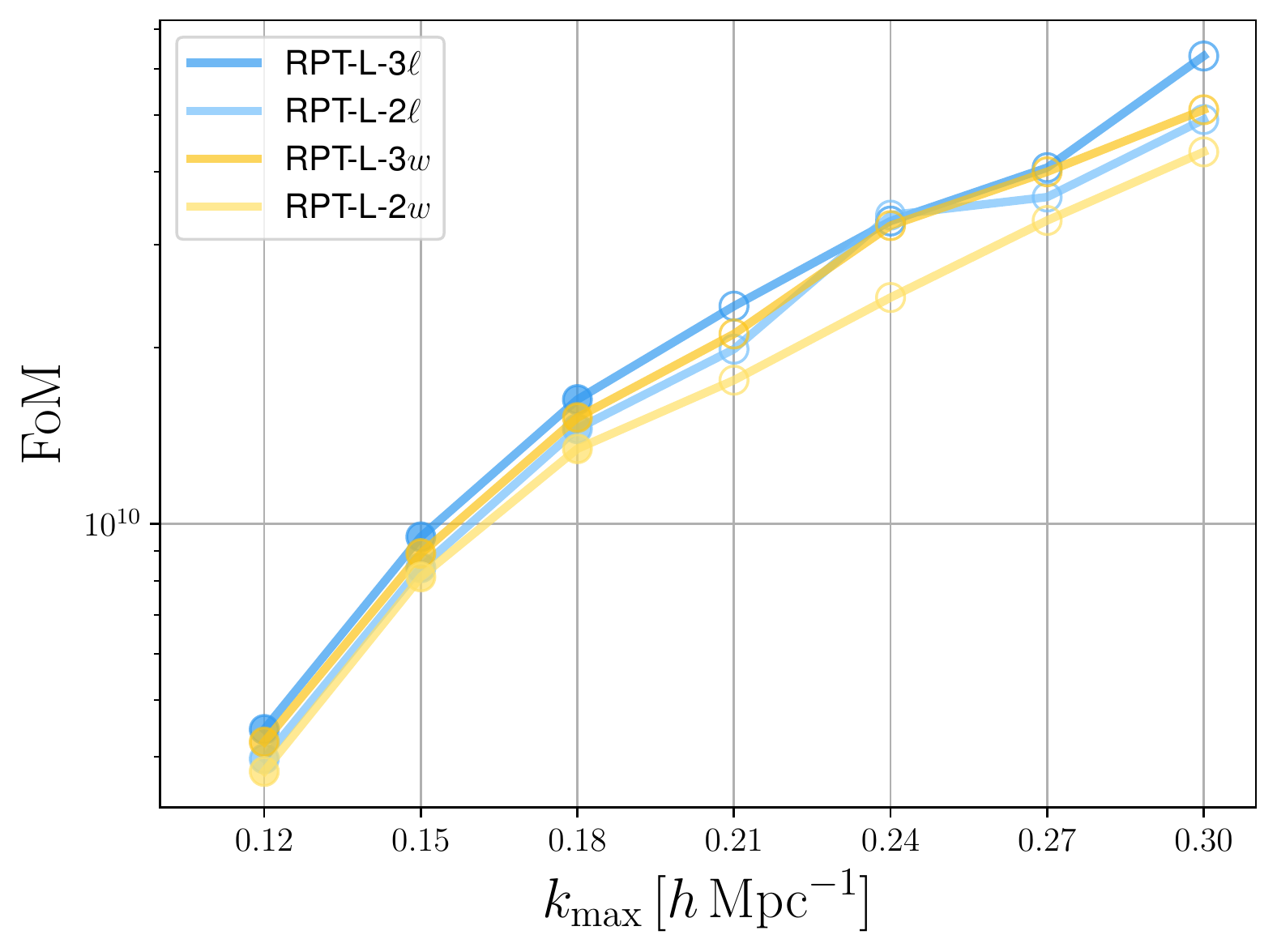}
  \includegraphics[width=\columnwidth]{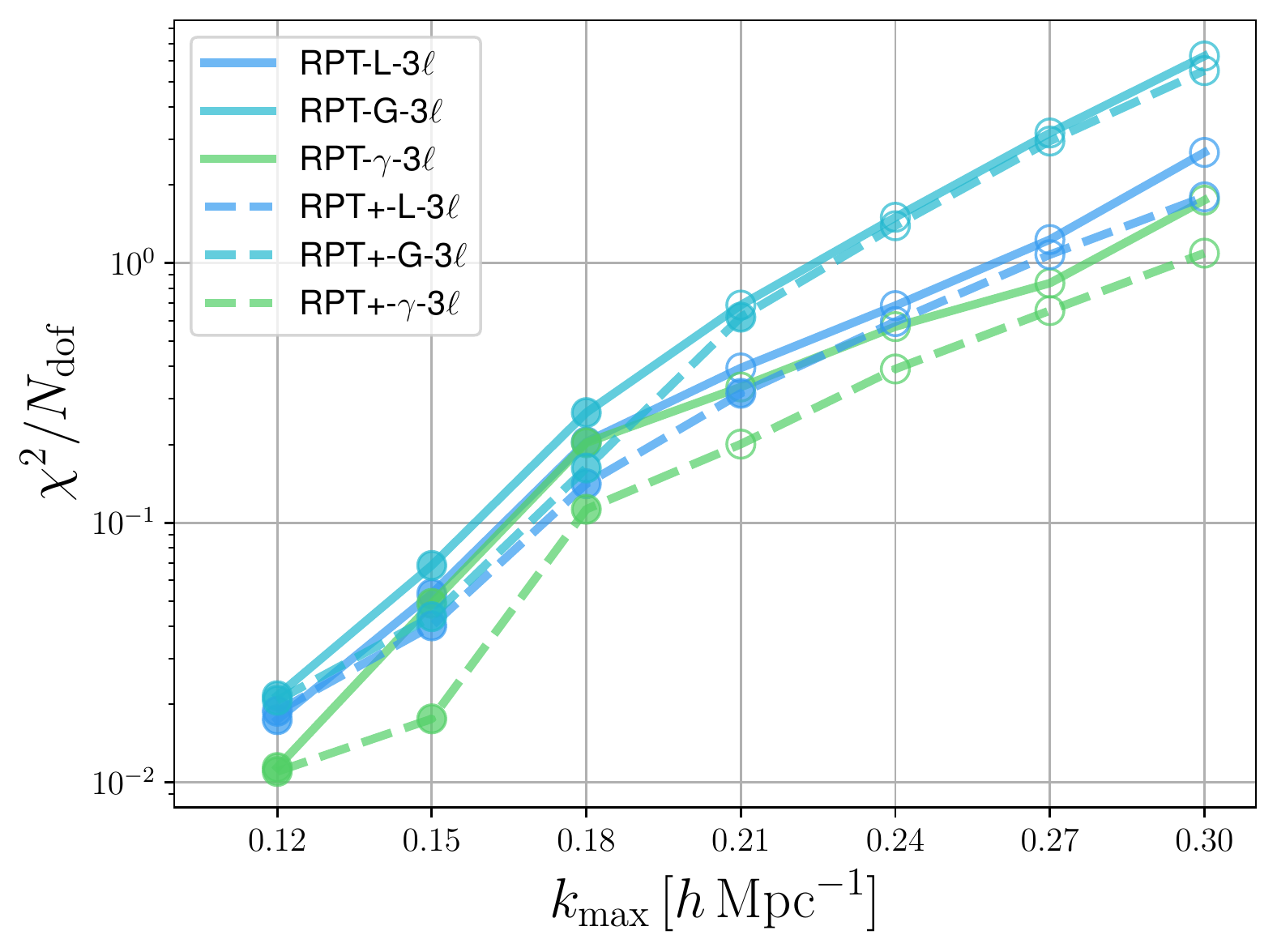}
  \includegraphics[width=\columnwidth]{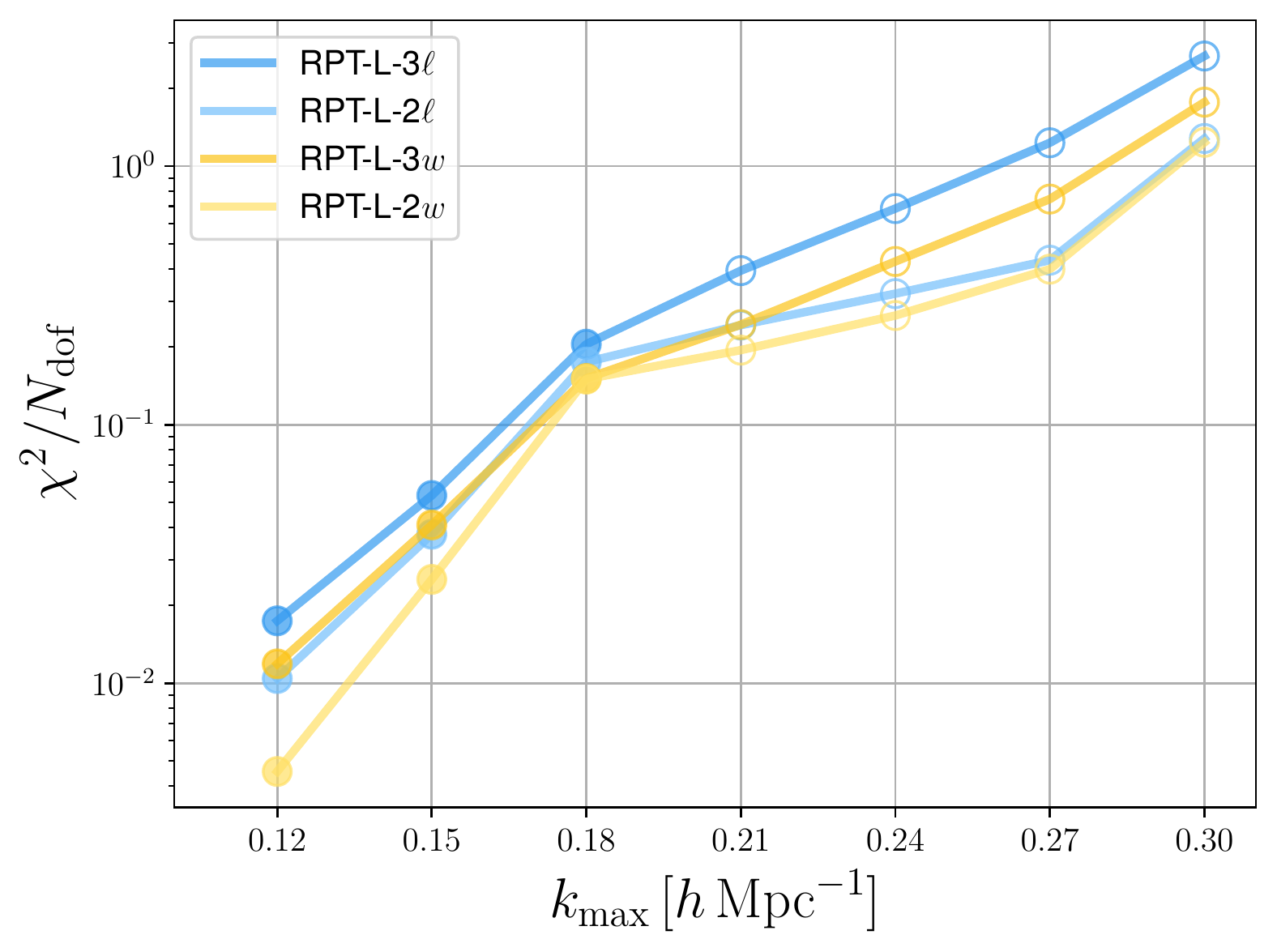}
  \caption{Same as Figure~\ref{fig:measures_GroupAB} but
  for Group C (left panels) and Group D (right panels).}
  \label{fig:measures_GroupCD}
\end{figure*}

\begin{figure*}
  \includegraphics[width=\columnwidth]{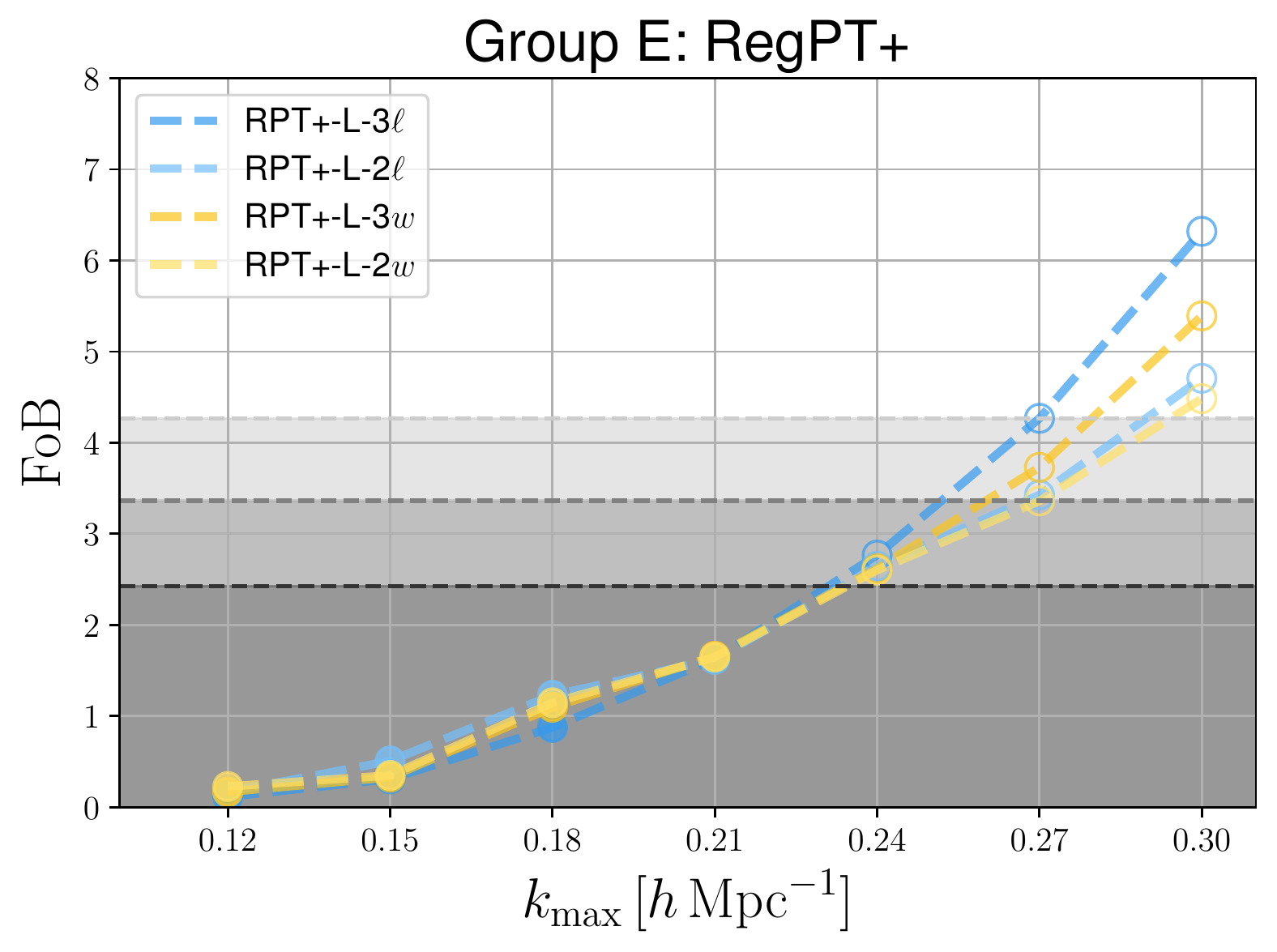}
  \includegraphics[width=\columnwidth]{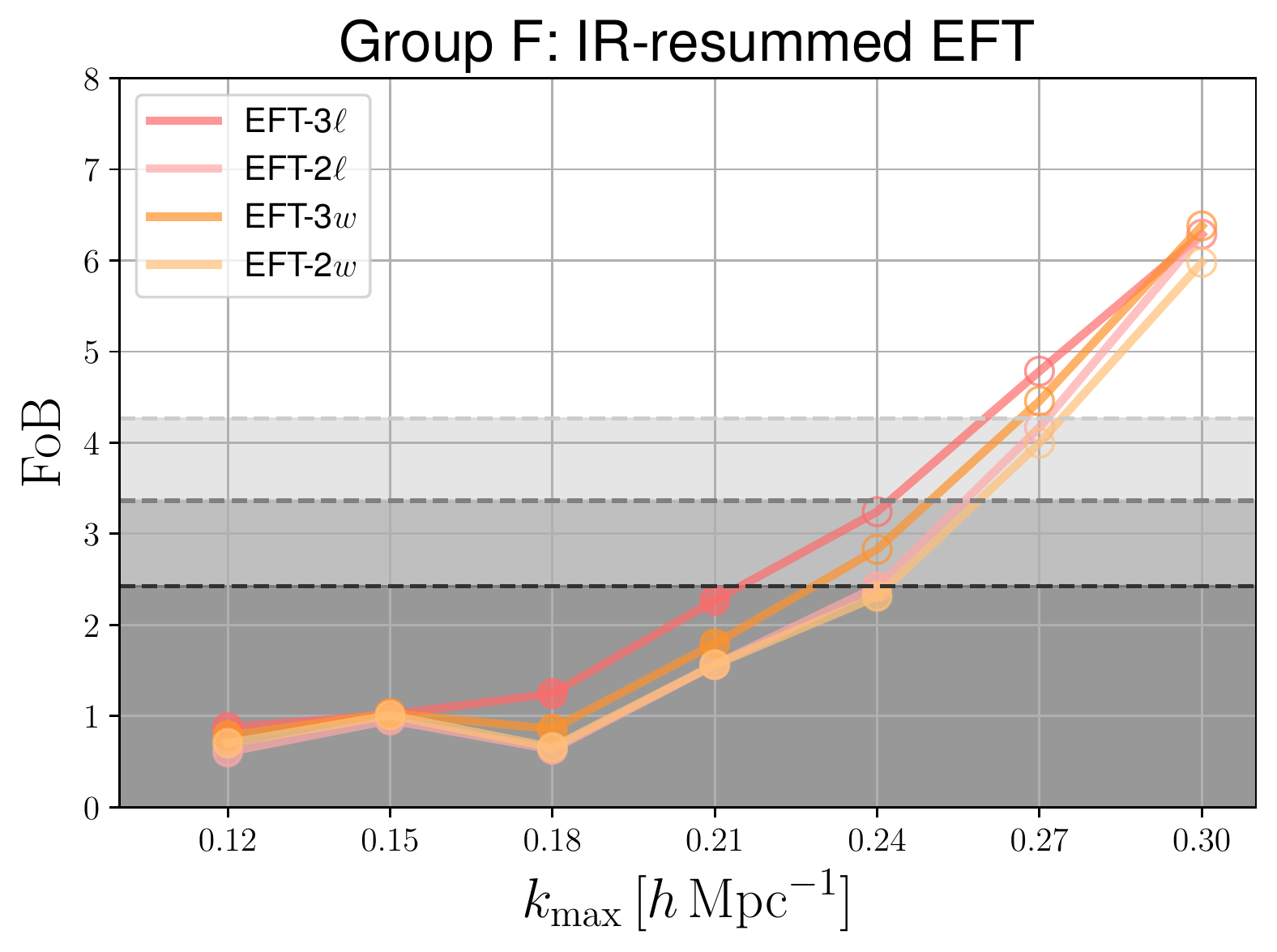}
  \includegraphics[width=\columnwidth]{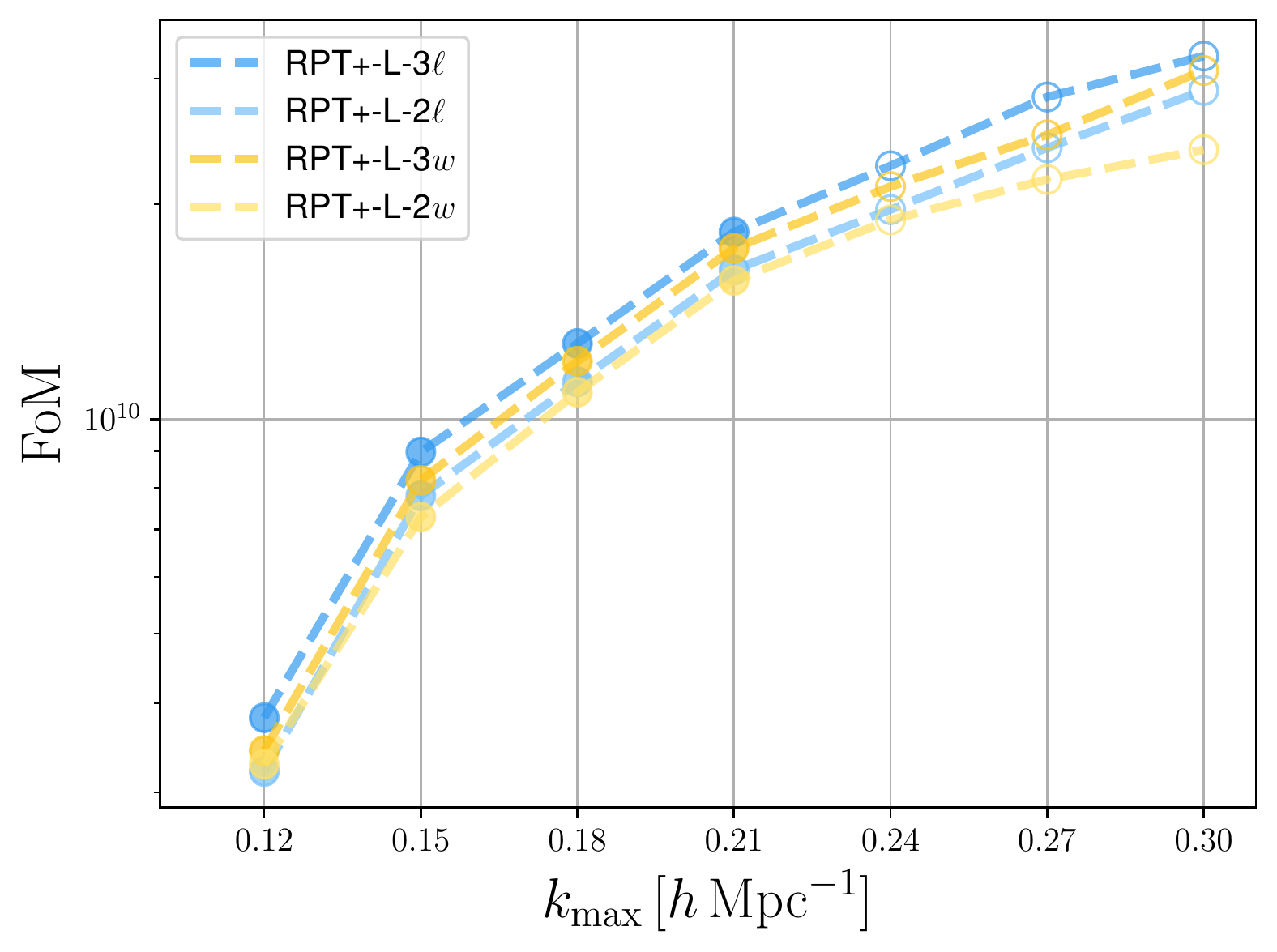}
  \includegraphics[width=\columnwidth]{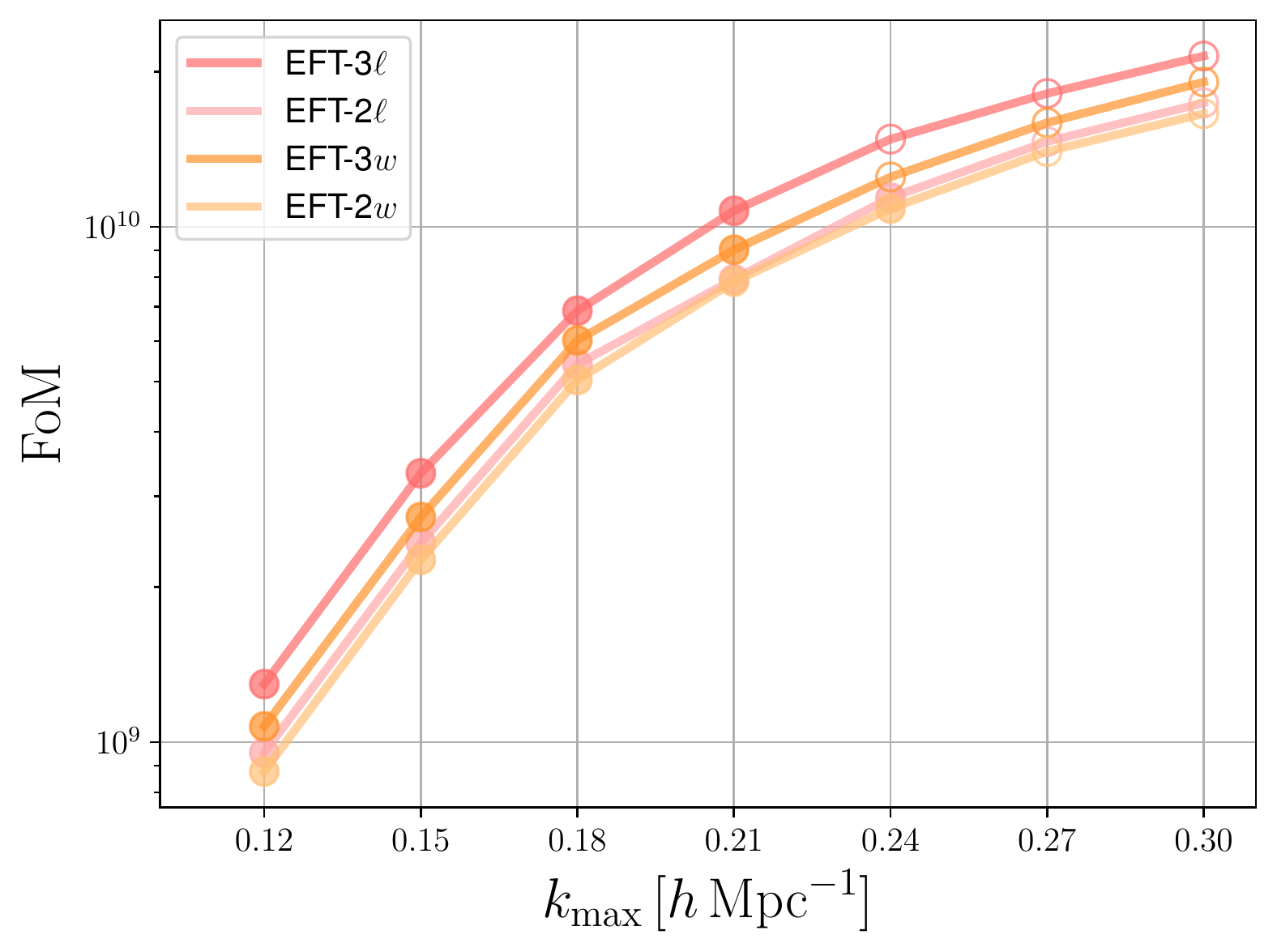}
  \includegraphics[width=\columnwidth]{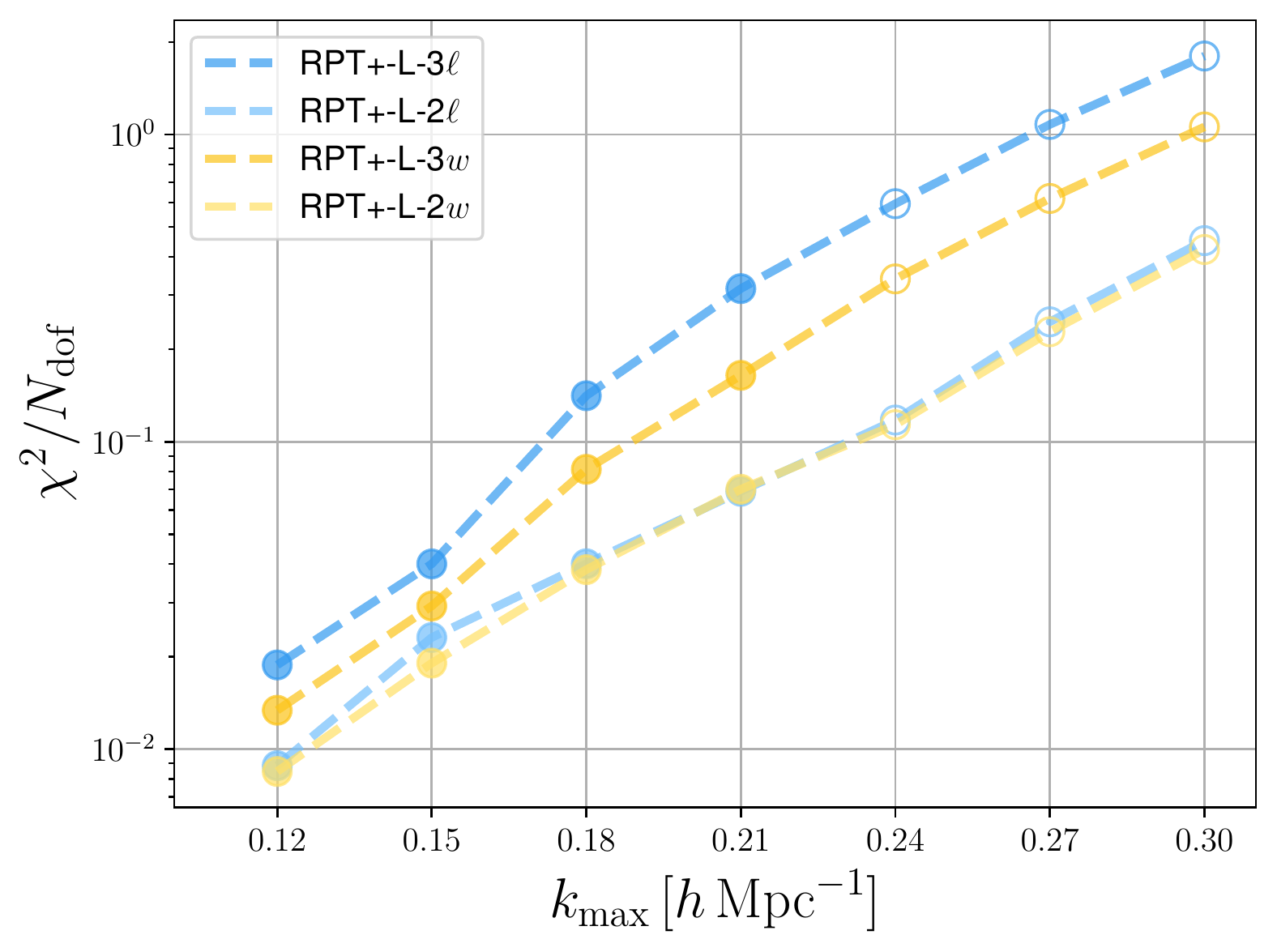}
  \includegraphics[width=\columnwidth]{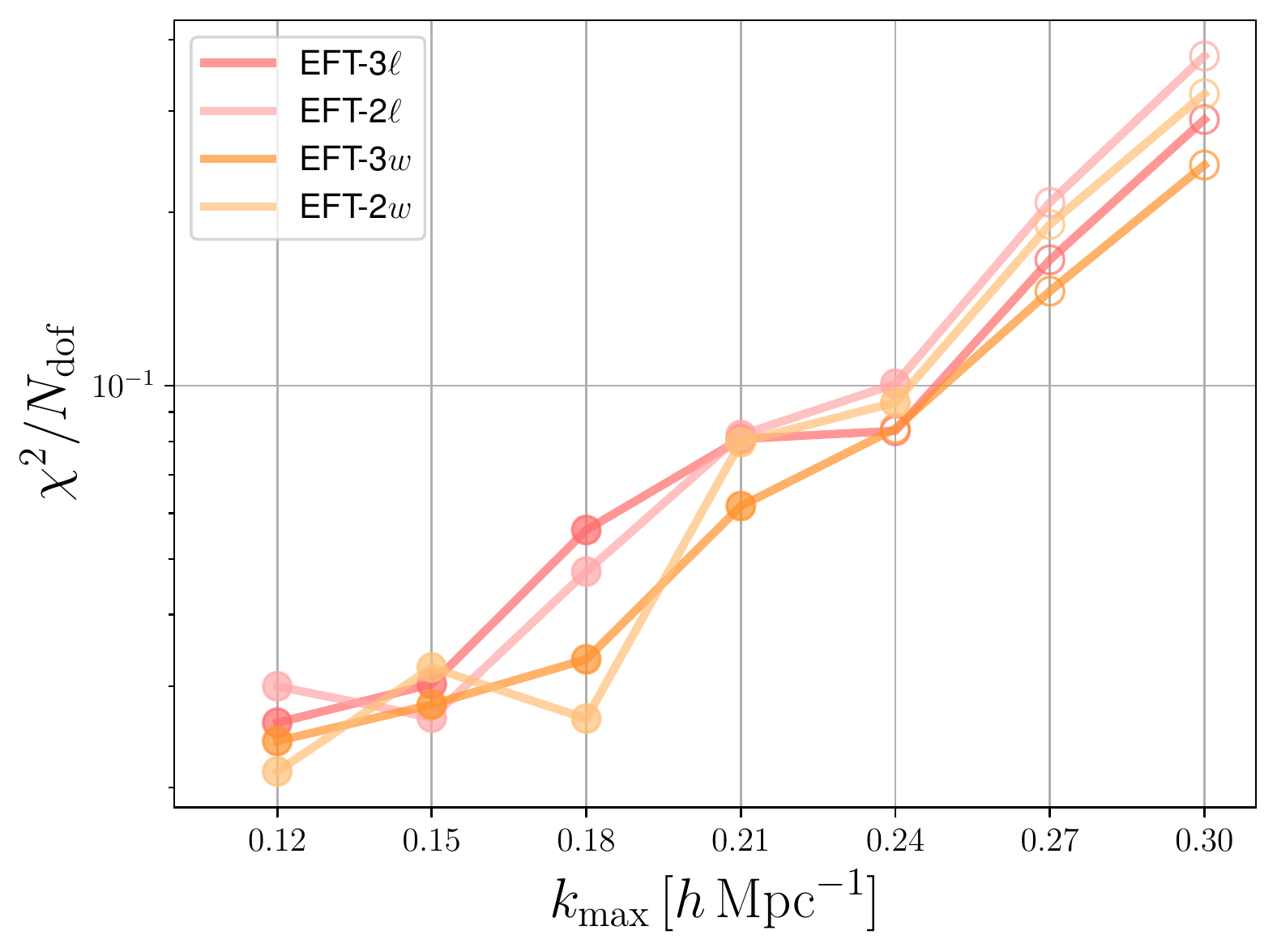}
  \caption{Same as Figure~\ref{fig:measures_GroupAB} but
  for Group E (left panels) and Group F (right panels).}
  \label{fig:measures_GroupEF}
\end{figure*}

In order to highlight the difference between models, we discuss the results for following groups:
\begin{description}
\item[Group A] RegPT and RegPT+ with Lorentzian FoG with and without AP effect, and IR-resummed EFT with 3 multipoles,
\item[Group B] RegPT and RegPT+ with Gaussian FoG, $\gamma$ FoG, and IR-resummed EFT with 3 multipoles,
\item[Group C] RegPT and RegPT+ with Lorentzian, Gaussian, $\gamma$ FoGs with 3 multipoles,
\item[Group D] RegPT with 3/2 multipoles and 3/2 wedges,
\item[Group E] RegPT+ with 3/2 multipoles and 3/2 wedges,
\item[Group F] IR-resummed EFT with 3/2 multipoles and 3/2 wedges.
\end{description}
Figures~\ref{fig:measures_GroupAB}, \ref{fig:measures_GroupCD}, and \ref{fig:measures_GroupEF}
show the FoB, FoM, and reduced chi-square for each group.

\subsubsection{Group A}
This group consists of RegPT, RegPT+, and IR-resummed EFT with the 3 multipoles data vector.
First, FoBs of RegPT+ are the lowest among the models presented
and they do not exceed the $1\text{-}\sigma$ critical value up to $\kmax = 0.21 \, \hMpcinv$
whereas FoB for RegPT with $\kmax = 0.21 \, \hMpcinv$ is around the $2\text{-}\sigma$ level.
In terms of FoM, the FoM of RegPT is always the highest
because RegPT contains only two nuisance parameters (three for RegPT+),
which can be degenerate with cosmological parameters and as a result, weaken parameter constraints.
However, the highest FoM with unbiased parameter estimates, i.e. FoB less than the $1\text{-}\sigma$ value,
can be achieved by RegPT+ though there is only a slight difference between RegPT and RegPT+.
Second, FoB of IR-resummed EFT is comparable with RegPT+ but
FoM is much smaller than RegPT and RegPT+ because the number of nuisance parameters of IR-resummed EFT
is six and these many nuisance parameters lead to weak parameter constraints.
On the other hand, reduced chi-squares of IR-resummed EFT
are the smallest except $\kmax = 0.12 \, \hMpcinv$.
This feature has an important implication that the good fit to the measured power spectrum
does not guarantee the tight or unbiased parameter constraints
and potentially, over-fitting occurs due to the large degrees of freedom.
Finally, FoB and reduced chi-squares are not much affected by the AP effect but
FoM with the AP effect considered is higher than that without the AP effect.
The AP effect induces additional dependence on cosmological parameters relevant to the geometry,
and thus, more information can be accessible from the AP effect.
Therefore, when the AP effect is considered,
the parameter constraints become tighter and the resultant FoM becomes larger.

\subsubsection{Group B}
In this group, we investigate whether the performance of RegPT and RegPT+ in comparison with IR-resummed EFT,
which is addressed in Group A, is affected by different FoG functional forms: Gaussian and $\gamma$ FoGs.
The general trend is quite similar to the results with fiducial Lorentzian FoG;
though the reduced chi-square is larger compared with IR-resummed EFT,
FoM and FoB for both models with Gaussian or $\gamma$ FoG are better.
That is because these models contain less nuisance parameters.

\subsubsection{Group C}
Here, we address how the choice of functional form of FoG function affects the measures.
The $\gamma$ FoG has one additional free parameter which determines the shape of the FoG function
and includes Loretzian and Gaussian forms as a special case.
The best-performing model in terms of FoB and FoM is Gaussian for RegPT and RegPT+.
Though the fitting results for hexadecapole moments with Gaussian FoGs are worse than Lorentzian FoG,
the models with Gaussian FoG can perform better for monopole and quadrupole moments.
The reduced chi-square with the $\gamma$ FoG is the best compared with other FoG models.
However, the FoM and FoB are significantly worse
and thus, the free parameter of the $\gamma$ FoG leads to over-fitting to the data.

\subsubsection{Group D}
This group addresses the choice of data vectors with RegPT: 3/2 multipoles and 3/2 wedges.
The 3 (2) multipoles and 3 (2) wedges contain the same number of data points
but different projection of the anisotropic power spectrum is adopted.
For FoB, the results for the pair of results of 3 (2) multipoles and 3 (2) wedges are quite similar.
On the other hand, FoM is better for multipoles and reduced chi-square is smaller for wedges.
Originally, the concept of wedges has been proposed in Ref.~\cite{Kazin2012}
to efficiently constrain the geometry of the Universe.
The transverse wedge ($0 < \mu < 1/2$) and radial wedge ($1/2 < \mu < 1$)
are sensitive to $H (z)$ and $D_A (z)$, respectively.
However, our analysis incorporates the full shape information of the power spectrum,
and in this case, the multipole expansion can constrain parameters better
because it weights the anisotropic part of the power spectrum more.
We have investigated only the equally spaced wedges but
the different spacing of wedges, e.g. weighing more on $\mu \simeq 0$
to avoid the region where FoG damping is eminent,
has the potential to yield performance similar to multipoles.

\subsubsection{Group E}
This group presents results similar to those in Group D but with RegPT+.
For FoB, the results are almost the same as Group D
but the difference for FoM and chi-square is clearer.
The RegPT+ has better flexibility of the FoG damping
and thus, can extract more information from the power spectrum at directions
where strong FoG damping happens, i.e., $\mu \simeq 1$.

\subsubsection{Group F}
This group also presents results for different data vectors with IR-resummed EFT.
The general trend is the same as Group D and Group E but the difference due to data vectors
is comparable with RegPT but less significant than RegPT+.
Note that for all data vectors the reduced chi-squares are the lowest among all PT models
but in compensation, FoMs are the lowest and thus,
the constraining power of cosmological parameters is weak.
It should be noted that the reduced chi-square with small $\kmax$ is unstable
because there are too many free parameters compared with the number of data points
and over-fitting may occur in this case.

\begin{figure}
  \includegraphics[width=\columnwidth]{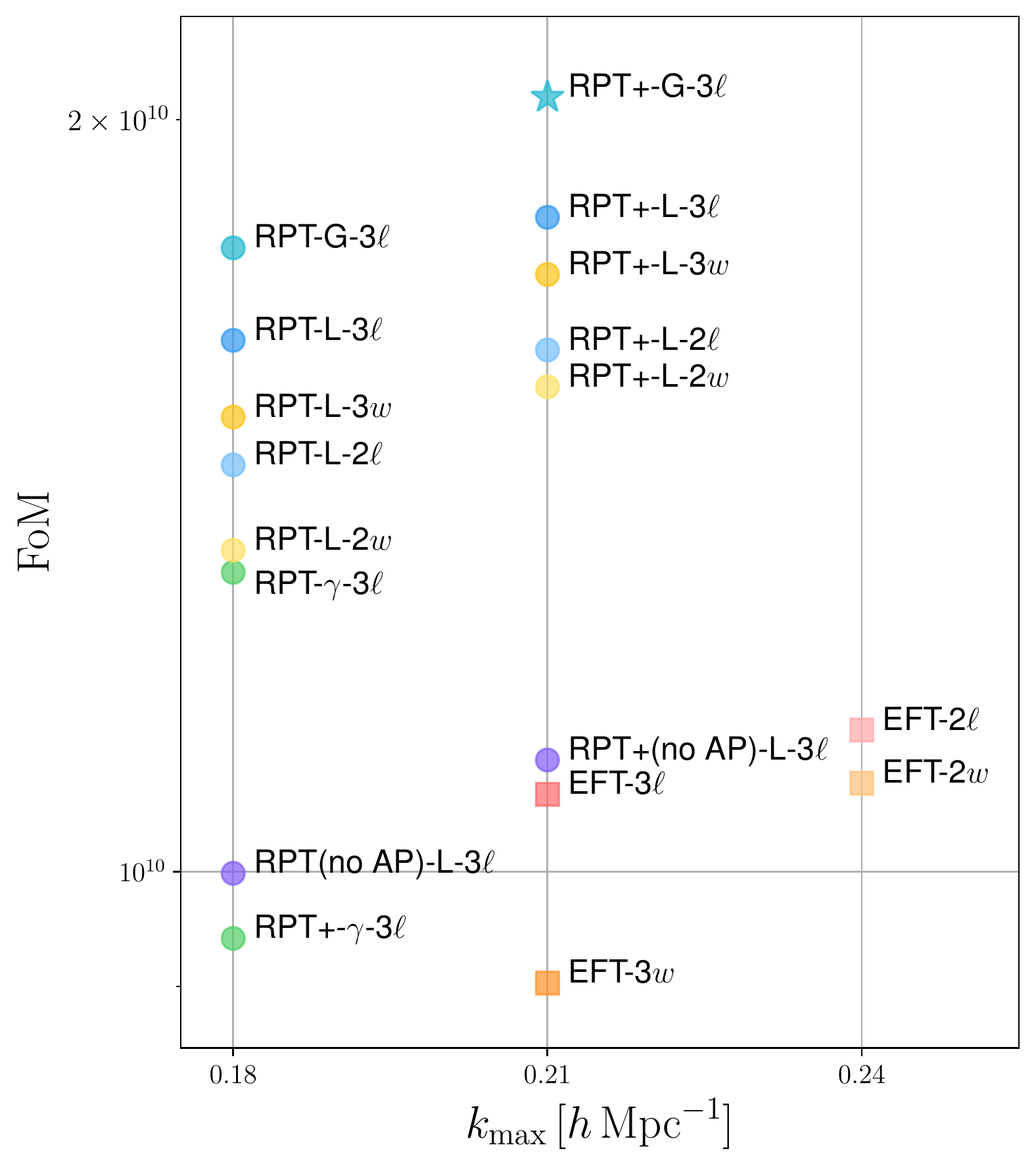}
  \caption{The highest FoM with the constraint that FoB does not exceed
  the $1\text{-}\sigma$ critical value. RegPT+ with Gaussian FoG and 3 multipoles yields the highest FoM,
  which is displayed as a star symbol.
  Note that IR-resummed EFT is at 1-loop order and the FoMs of IR-resummed EFT
  are shown as square symbols.
  For reference, a factor of 2 better FoM corresponds to $14.8\%$ better constraint for each parameter on average.}
  \label{fig:bestFoM}
\end{figure}

To summarize, Figure~\ref{fig:bestFoM} illustrates the highest FoM achieved by each model
with the constraint that FoB does not exceed the $1\text{-}\sigma$ critical value.
RegPT+ generally performs better than RegPT and the best-performing model among all the examined models
is RegPT+ with Gaussian FoG and 3 multipoles.
IR-resummed EFT with 2 multipoles or 2 wedges can utilize the small-scale power spectra
up to $\kmax = 0.24 \, \hMpcinv$.
However, due to many nuisance parameters introduced in IR-resummed EFT, the parameter constraining power, i.e. FoM,
is weaker than RegPT and RegPT+.
There is a caveat about IR-resummed EFT.
In the presented results, the order of IR-resummed EFT is 1-loop in contrast to 2-loop for RegPT and RegPT+
since the implementation of IR-resummed EFT at 2-loop order is still challenging.
Thus, the comparison between 1-loop IR-resummed EFT and 2-loop RegPT and RegPT+ is not fair
and IR-resummed EFT at 2-loop order has potential to achieve the lower FoB with higher $\kmax$ than RegPT+.

\section{Conclusions}
\label{sec:conclusions}
The galaxy clustering analysis has been playing a central role in observational cosmology
to constrain cosmological parameters.
The theoretical model to predict the statistics, e.g., power spectra,
given a cosmological model is an essential ingredient in cosmological parameter inference,
and the accuracy of the model is critical to place stringent constraints on cosmological parameters.
For accurate predictions on small scales,
the non-linearity due to gravitational evolution must be incorporated into the model.
In addition, the peculiar motion of galaxies distorts the distance estimate in the line-of-sight direction,
which is referred to as the RSD effect, and the effect needs to be considered in the theoretical model.
A variety of theoretical models based on PT have been proposed and
different assumptions are employed depending on the model.

The PT challenge analysis proposed here follows our precedent work \cite{Osato2019}, where the real space power spectrum is employed.
In this work, we extend the PT challenge analysis with the redshift space power spectrum.
This was made possible with the help of the implementation of the response function method \cite{Osato2021}
to accelerate the PT calculations of power spectra with RSD effects
and the calculation module is integrated into the framework of \texttt{Eclairs} \cite{Eclairs}.
In the challenge analysis, we first perform the $N$-body simulation with fiducial cosmological parameters
and measure the power spectrum from the matter density field.
Then, we regard the measured power spectrum as a given data vector
and carry out parameter inference with the PT models described above.
The inferred cosmological parameters can then be directly compared with the fiducial values.
We can assess the constraining power and the bias of cosmological parameters
as the function of the maximum wave-number scale of data points, which we denote $\kmax$.
More precisely, $\kmax$ is determined with the help of the bias parameter, the Figure of Bias (FoB),
defined as the difference between the inferred and fiducial values normalized by the covariance.
Parameter inference is then tagged as ``biased'', and therefore excluded,
if FoB exceeds the $1\text{-}\sigma$ critical value
as the probability distribution of FoB follows the multi-variate Gaussian distribution.
The constraining power of the parameters is defined with a Figure of Merit (FoM),
which is the inverse of the square root of the parameter covariance matrix,
which corresponds to the inverse of the hyper-volume of confidence regions.
The key objective of the PT challenge analysis is then first,
for each model to assign a $\kmax$ for which its FoB remains below the $1\text{-}\sigma$ critical value,
and then to identify the models which yield the highest FoM.

The examined PT models are RegPT, RegPT+, and IR-resummed EFT.
RegPT is the extended version of SPT by reorganizing the SPT expansion and
the accuracy and convergence have been enhanced compared with SPT. The
RegPT+ model contains one additional free parameter giving
the dispersion of displacement $\sigmad$ that controls
the small-scale damping feature.
Previous studies have shown that taking this parameter as a free parameter can significantly improve
the model \cite{Osato2019}.
The last model is IR-resummed EFT, where the small-scale spectra are described by effective counter terms. 
This class of models can accurately account for the small-scale power spectra but with the introduction of many nuisance parameters.
In addition to different theoretical models, we also address
how FoG functional forms, the AP effect, and sampling of data vectors (multipoles or wedges) affect the parameter inference.
In this analysis, however, we use a simple, perhaps naive, model of linear galaxy bias and assume Poisson noise.
In particular, we assume a simple linear scale independent bias,
ignoring the possibility that the galaxy bias is likely to have non-trivial scale dependencies
\cite[for a review, see][]{Desjacques2018}.
It should be kept in mind that this is likely to be an important limitation
on the scope of the analysis we present below.
In particular, the performance of the models we consider is likely to be affected differently
when additional nuisance parameters are introduced.

It is also to be noted that the reference survey we consider, in terms of volume and number density of galaxies,
reproduces the raw characteristics of the \textit{Euclid} spectroscopic survey.
The conclusions we reach are then a priori valid for such a setting only
but we do not expect the results to be very sensitive to those assumptions.

In the PT challenge analysis,
five cosmological parameters $(\omega_\mathrm{b}, \omega_\mathrm{cdm}, h, \ln (10^{10} A_\mathrm{s}), n_\mathrm{s})$
and the linear bias parameter $b_1$ are inferred from the redshift space power spectrum
at the redshift $z = 1$ assuming the Gaussian covariance matrix with the survey volume
and the galaxy number density expected for the \textit{Euclid} mission.
In order to quantify the goodness of fit and the induced parameter bias,
we have introduced three measures: FoM, FoB, and reduced chi-squares.
The reduced chi-square is irrelevant to the parameter inference
but expresses the performance of the fitting of the power spectrum.
The findings derived from the PT challenge analysis are summarized below:
\begin{itemize}
  \item RegPT yields the highest FoM but the largest reduced chi-square.
  IR-resummed EFT delivers the opposite results: the lowest FoM but the smallest reduced chi-square.
  RegPT+ is in between. This feature can be explained by the number of nuisance parameters.
  RegPT has no nuisance parameter, RegPT+ has one, and IR-resummed EFT has five for 3 multipoles case.
  The nuisance parameters add flexibility especially at small scales,
  and thus, introducing nuisance parameters improves the fitting at small scales,
  i.e., smaller reduced chi-square.
  However, the nuisance parameters are degenerate with cosmological parameters
  and thus degrade constraints on cosmological parameters, i.e., lower FoM.
  The trade-off is also observed in the previous analysis of the real-space power spectrum \cite{Osato2019}.
  \item Under the condition that the FoB does not exceed the $1\text{-}\sigma$ critical value,
  the PT model which yields the highest FoM is RegPT+ with $\kmax = 0.21 \, \hMpcinv$.
  This model has only one free parameter $\sigmad$, which controls the small-scale damping feature.
  Therefore, we can conclude that RegPT+ is the most competent without significant parameter bias
  among the examined PT models.
  \item We have examined the different functional forms of FoG damping:
  Lorentzian, Gaussian, $\gamma$ FoGs.
  The $\gamma$ FoG has one additional parameter, which determines the shape of the damping,
  and contains Lorentzian and Gaussian forms as special cases.
  Among the three models, Gaussian FoG yields the best FoM and FoB.
  On the other hand, for the $\gamma$ FoG,
  over-fitting occurs due to the free parameter introduced in the model.
  Hence, the reduced chi-square is the lowest but the FoM and FoB are the worst.
  \item The redshift space power spectrum has two arguments:
  the magnitude of wave-number $k$ and the directional cosine $\mu$ in the line-of-sight direction.
  In the practical analysis, the variable $\mu$ is projected in two different manners:
  multipoles and wedges.
  Overall, for 3 multipoles and 3 wedges, FoMs are slightly better than
  2 multipoles and 2 wedges, respectively, because there are more data points
  and more information is accessible.
  The difference between 3 (2) multipoles and 3 (2) wedges is quite small
  but overall, multipoles lead to slightly better FoM but worse reduced chi-square.
\end{itemize}

These conclusions on the relative performances of the models we explore should not however be considered definitive.
In particular, as mentioned before, we made a simplifying assumption on galaxy bias behavior.
The introduction of a more elaborate model is likely to affect the result and the relative performances of the models.
This is precisely what we would like to explore in the subsequent paper of this series,
exploiting the fact that the computation of higher-order contributions in
galaxy bias expansion can also be accelerated with the response function approach.

\begin{acknowledgments}
K.O. was supported by JSPS Research Fellowships for Young Scientists.
This work was supported in part by MEXT/JSPS KAKENHI Grant Number
JP20H05861 and JP21H0108 (A.T., T.N.), JP21J00011 and JP22K14036 (K.O.), JP19H00677 and JP22K03634 (T.N.), 
and JSPS Core-to-Core Program (grant number: JPJSCCA20210003). This work was also supported in part by Japan Science and
Technology Agency AIP Acceleration Research grant No. JP20317829 (A.T., T.N.). 
Numerical simulations were carried out on Cray XC50
at the Center for Computational Astrophysics, National Astronomical Observatory of Japan.
\end{acknowledgments}

\appendix

\section{1-loop SPT power spectrum in the redshift space}
\label{sec:1loop_SPT}
There is another formalism to derive the expression of
the power spectrum in the redshift space based on SPT
through the mapping of coordinates $\bm{s} = \bm{x} + \frac{v_z (\bm{x})}{aH} \hat{z}$,
where $\bm{s}$ and $\bm{x}$ are the coordinates in the redshift space and the real space, respectively,
and $\hat{z}$ is the unit vector along the line-of-sight direction.
Thus, one can obtain the expression of the power spectrum
in the redshift space \cite{Heavens1998,Scoccimarro1999,Matsubara2008}
and the expression at 1-loop order can be reorgarnized with TNS correction terms \cite{Taruya2010}:
\begin{align}
  P^{\mathrm{(S)}}_\mathrm{SPT} (k, \mu) =& [1-(k \mu f \sigma_{\mathrm{v},\mathrm{L}})^2 ]
  \nonumber \\
  & \times [P_{\delta \delta} (k) + 2 f \mu^2 P_{\delta \theta} (k) + f^2 \mu^4 P_{\theta \theta} (k) ]
  \nonumber \\
  & + A(k, \mu) + B(k, \mu) + C(k, \mu) ,
\end{align}
where the velocity dispersion at linear order $\sigma^2_{\mathrm{v, L}}$ is defined in Eq.~\eqref{eq:sigmav_L}.
The power spectra $P_{\delta \delta} (k)$, $P_{\delta \theta} (k)$,
and $P_{\theta \theta} (k)$ are computed based on SPT at 1-loop order.
For the correction terms at 1-loop order,
the bispectra in the $A$-term and the power spectra in the $B$-term
should be computed at the tree level, i.e., $P_{ab} (k, z) = P_\mathrm{L} (k, z)$.
Then, the $C$-term is given as
\begin{align}
  C (k, \mu) =& (k \mu f)^2 \int \frac{\dd[3] p}{(2\pi)^3} \frac{\dd[3] q}{(2\pi)^3}
  \delta_\mathrm{D} (\bm{k}-\bm{p}-\bm{q}) \frac{\mu_p^2}{p^2} P_{\theta \theta} (p)
  \nonumber \\
  & \times \left\{ P_{\delta \delta} (q) + 2 f \mu_q^2 P_{\delta \theta} (q) +
  f^2 \mu_q^4 P_{\theta \theta} (q) \right\} ,
\end{align}
where $\mu_p = p_z/p$ and $\mu_q = q_z/q$.
To keep the consistency of the order,
the power spectra should be computed at linear order.
Then, the $C$-term is given as
\begin{align}
  C(k, \mu) =& \sum_{n=1}^{3} \sum_{a, b=1}^{2} \mu^{2n} (-f)^{a+b} \frac{k^3}{(2 \pi)^2}
  \int_0^{\infty} \!\! \dd r \int_{-1}^{+1} \!\! \dd x
  \nonumber \\
  & \times \left[ C^n_{ab} (r, x) P_{ab} \left( k \sqrt{1+r^2-2rx} \right) P_{22} (kr) \right.
  \nonumber \\
  & \left. + \tilde{C}^n_{ab} (r, x) P_{22} \left( k \sqrt{1+r^2-2rx} \right) P_{ab} (kr) \right] .
\end{align}
The non-vanishing components of $C^n_{ab} (r, x)$ and $\tilde{C}^n_{ab} (r, x)$ are
\begin{widetext}
  \begin{align}
    C^1_{11} (r,x) =& -\frac{x^2-1}{4} , \\
    C^1_{12} (r,x) =& -\frac{3 r^2 (x^2-1)^2}{8(1+r^2-2rx)} , \\
    C^1_{22} (r,x) =& -\frac{5 r^4 (x^2-1)^3}{16(1+r^2-2rx)^2} , \\
    C^2_{11} (r,x) =& \frac{3 x^2-1}{4} , \\
    C^2_{12} (r,x) =& \frac{(x^2-1) (2-12rx-3r^2+15 r^2 x^2)}{4 (1+r^2-2rx)} , \\
    C^2_{22} (r,x) =& \frac{3 r^2 (x^2-1)^2 (7-30rx-5r^2+35 r^2 x^2)}{16(1+r^2-2rx)^2} , \\
    C^3_{12} (r,x) =& -\frac{-4+12x^2+8rx (3-5x^2) + r^2 (3-30x^2+35x^4)}{8 (1+r^2-2rx)} , \\
    C^3_{22} (r,x) =& -\frac{(x^2-1) [4+3r\{ -16x + r(-14+70x^2 + 20rx (3-7x^2) + 5r^2 (1-14x^2+21x^4))\}]}
    {16 (1+r^2-2rx)^2} , \\
    C^4_{22} (r,x) =& \frac{-4 + 12x^2 + 16rx(3-5x^2) + 7r^2 (3-30x^2+35x^4) - 6r^3x (15-70x^2+63x^4)}{16 (1+r^2-2rx)^2}
    \nonumber \\
    & + \frac{r^4 \{ -5+21x^2 (5-15x^2+11x^4) \}}{16 (1+r^2-2rx)^2} , \\
    \tilde{C}^1_{11} (r,x) =& -\frac{r^4 (x^2-1)}{4 (1+r^2-2rx)^2} , \\
    \tilde{C}^1_{12} (r,x) =& -\frac{3 r^4 (x^2-1)}{8 (1+r^2-2rx)^2} , \\
    \tilde{C}^2_{11} (r,x) =& \frac{r^2 (2-4rx-r^2+3r^2 x^2)}{4 (1+r^2-2rx)^2} , \\
    \tilde{C}^2_{12} (r,x) =& \frac{r^2 (x^2-1) (2-12rx-3r^2+15r^2x^2)}{4 (1+r^2-2rx)^2} , \\
    \tilde{C}^3_{12} (r,x) =& -\frac{r^2 \{ -4+12x^2+8rx (3-5x^2) +r^2 (3-30x^2+35x^4) \}}{8 (1+r^2-2rx)^2} .
  \end{align}
\end{widetext}
The scaling of $C$-terms with respect to the linear bias $b_1$ is given as
\begin{equation}
  C (k, \mu ; f) \to b_1^4 C (k, \mu ; \beta) .
\end{equation}

\section{PT Challenge results with models at 1-loop order}
\label{sec:1loop_results}
\begin{figure}
  \includegraphics[width=\columnwidth]{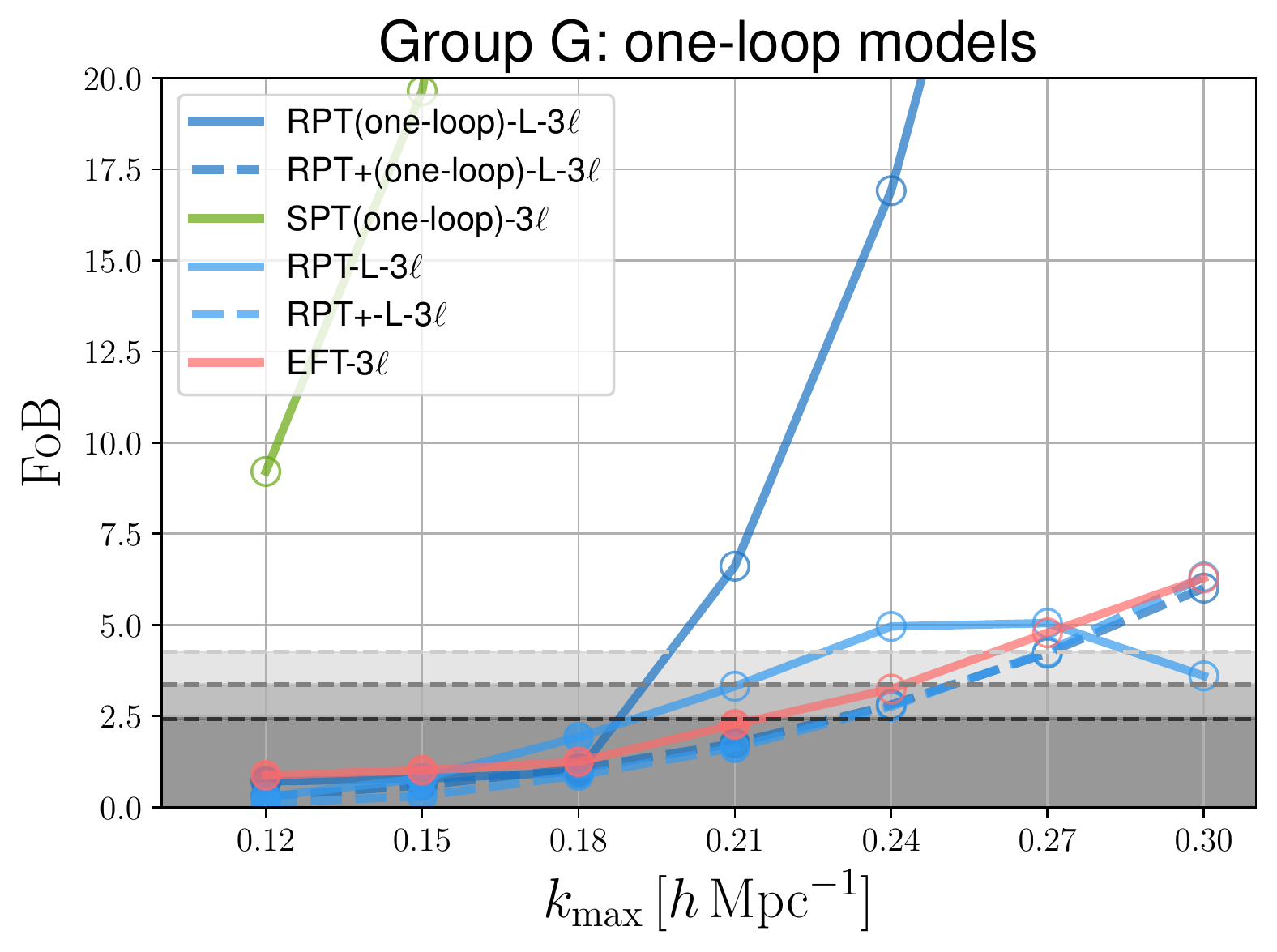}
  \includegraphics[width=\columnwidth]{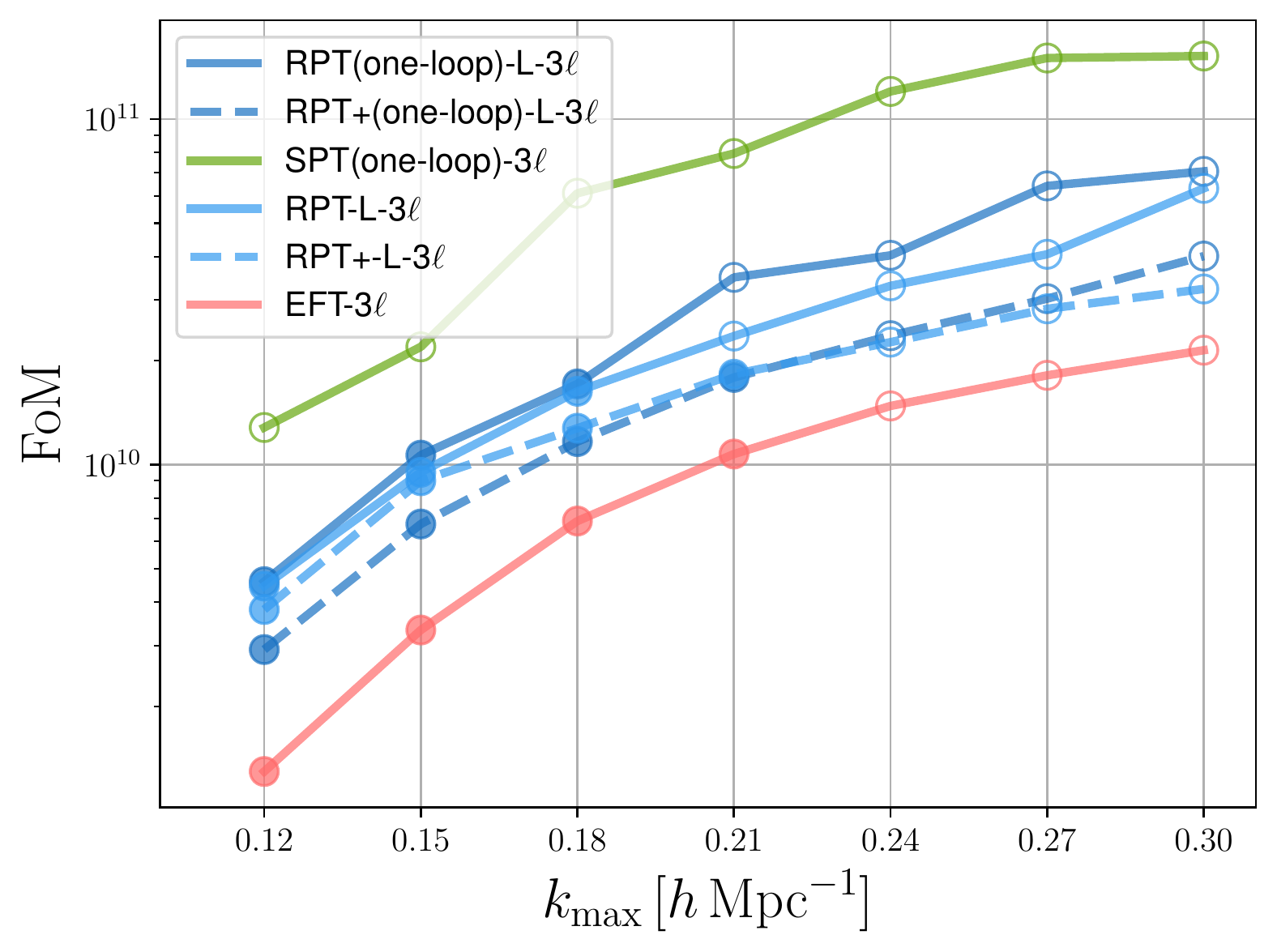}
  \includegraphics[width=\columnwidth]{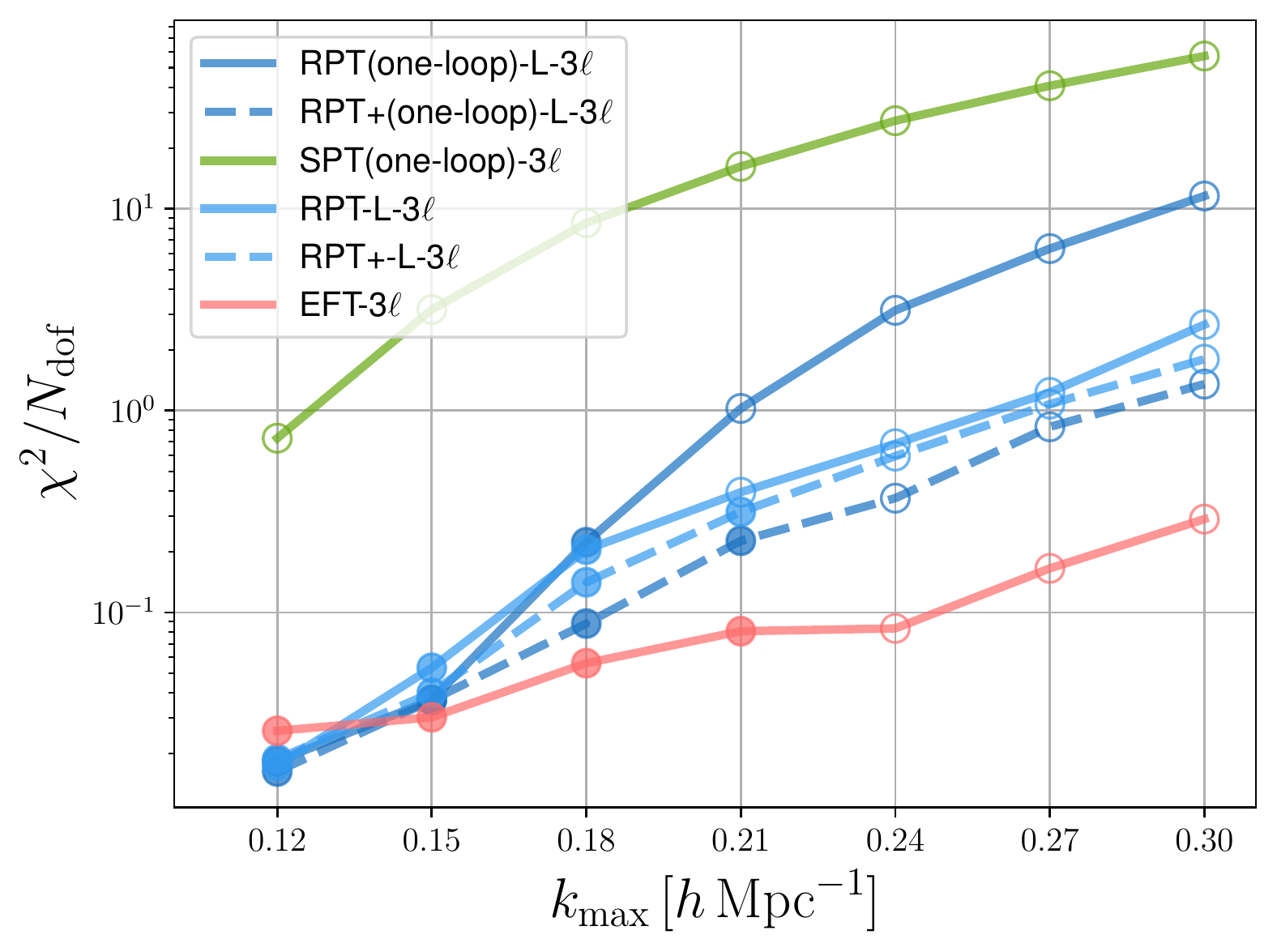}
  \caption{The FoB (upper panel), FoM (middle panel), and reduced chi-square (lower panel)
  for Group G.}
  \label{fig:measures_GroupG}
\end{figure}

Here, we carry out the PT challenge analysis for SPT, RegPT, and RegPT+ at 1-loop order
to investigate the effect of the order of PT models.
For these 1-loop PT models, the computational cost is not problematic and thus,
we use the direct integration for computations instead of the response function approach.
Table~\ref{tab:models_1loop} summarizes the examined models
and we define \textit{Group G}, which includes RegPT and RegPT+ at 1-loop and 2-loop orders
and SPT and IR-resummed EFT at 1-loop order to investigate measures of FoB, FoM, and the reduced chi-square.
Figure~\ref{fig:measures_GroupG} shows the results for Group G.
In this group, the FoG damping function and the data vector
are fixed as Lorentzian and 3 multipoles, respectively,
and the AP effect is included.
First, SPT demonstrates poor performance even at the small $\kmax$.
The FoB is already larger than the $1\text{-}\sigma$ critical value at $\kmax = 0.12 \, \hMpcinv$.
As SPT has no nuisance parameters, it yields high FoM.
However, the FoB and reduced chi-square are too high and thus,
SPT at 1-loop order cannot be used to robustly constrain cosmological parameters.
In terms of RegPT, FoB is better for the 2-loop model and
FoB of the 1-loop model suddenly increases at $\kmax = 0.21 \, \hMpcinv$
because the accuracy at the small scale of this model is worse.
A similar feature is found in the reduced chi-square.
The FoM is quite similar at lower $\kmax$ because the number of the nuisance parameter is the same.
In contrast, the performance of RegPT and RegPT+ is comparable but
FoM is better for the 2-loop PT model. The free parameter specific to RegPT+,
i.e., the dispersion of displacement,
has considerable flexibility to fit the small-scale power spectrum.
IR-resummed EFT yields the lowest reduced chi-square and FoM is slightly higher
than RegPT+.
Similarly to the 2-loop case, FoM of IR-resummed EFT is quite suppressed
due to many nuisance parameters.

\begin{table*}
  \caption{Descriptions of models with RegPT, RegPT+, and SPT at 1-loop order.}
  \label{tab:models_1loop}
    \begin{tabular}{cccccc}
      \hline \hline
      Label & PT Model & Data vector & AP & FoG & Nuisance parameters \\
      \hline
      RPT(one-loop)-L-3$\ell$ & RegPT & 3 Multipoles ($\ell = 0, 2, 4$) & \checkmark & Lorentzian & $b_1, \sigmav$ \\
      RPT+(one-loop)-L-3$\ell$ & RegPT+ & 3 Multipoles ($\ell = 0, 2, 4$) & \checkmark & Lorentzian & $b_1, \sigmav, \sigmad$ \\
      SPT(one-loop)-3$\ell$ & SPT & 3 Multipoles ($\ell = 0, 2, 4$) & \checkmark & --- & $b_1$ \\
      \hline \hline
    \end{tabular}
\end{table*}

\bibliographystyle{apsrev4-2}
\bibliography{main}

\end{document}